\documentclass[aip,preprint]{revtex4-1}
\usepackage{graphicx,bm,amsmath}
\usepackage{color}

\newcommand{\p}{{\partial}}

\draft 

\begin{document}

\title{Extensional Flow of a Free Film of Nematic Liquid Crystal with Moderate Elasticity} 

\author{M.J. Taranchuk}%
\affiliation{Department of Mathematical Sciences, University of Delaware, USA}%

\author{L.J. Cummings}%
\affiliation{ Department of Mathematical Sciences, New Jersey Institute of Technology,USA}%

\author{T.A. Driscoll}%

\author{R.J. Braun}%
\email{rjbraun@udel.edu.}%
\affiliation{Department of Mathematical Sciences, University of Delaware, USA}%

\date{24 March 2023}

\begin{abstract}
Motivated by problems arising in tear film dynamics, we present a model for the extensional flow of thin sheets of nematic liquid crystal. The rod-like molecules of these substances impart an elastic contribution to its response. We rescale a weakly elastic model due to Cummings et al. [European Journal of Applied Mathematics 25 (2014): 397-423] to describe a case of moderate elasticity. The resulting system of two nonlinear partial differential equations for sheet thickness and axial velocity is nonlinear and fourth order in space, but still represents a significant reduction of the full system. We analyze solutions arising from several different boundary conditions, motivated by the underlying application, with particular focus on dynamics and underlying mechanisms under stretching. We solve the system numerically, via collocation with either finite difference or Chebyshev spectral discretization in space, together with implicit time stepping. 
At early times, depending on the initial film shape, pressure either aids or opposes extensional flow, which changes the shape of the sheet and may result in the loss of a minimum or maximum at the moving end. We contrast this finding with the cases of weak elasticity and Newtonian flow, where the sheet retains all extrema from the initial condition throughout time.
\end{abstract}

\pacs{}

\maketitle 

\section{\label{sec:Intro} Introduction}
\begin{figure}[b]
\centering
\includegraphics[width=.6\textwidth]{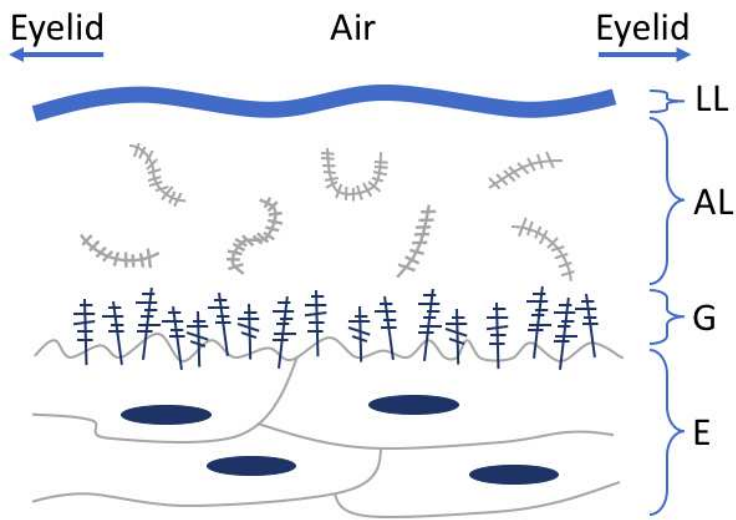}
\caption{\label{fig:tf_xsection}A sketch of the tear film on the ocular surface.  Here LL denotes the lipid layer, AL the aqueous layer, G the glycocalyx, and C is the outermost part of the corneal epithelium.  The objects in the interior of the aqueous layer represent large mucin and protein molecules.} \label{fig:tf_ocul_surf} 
\end{figure}

The tear film of the eye is a thin multi-layer protective liquid film lying over the cornea.  It is painted onto the ocular surface during the upstroke of the blink, and is re-formed rapidly after each blink.\cite{braun2015dynamics} Proper function of the tear film is essential for eye health and clear vision.\cite{willcox2017tfos}  The most abundant component of the tear film is the aqueous layer, sandwiched between a mucin layer called the glycocalyx, that is bound to the ocular surface, and a thin lipid layer that floats on it. A sketch of a cross section of a small part of the tear film is shown in Fig.~\ref{fig:tf_ocul_surf}.
Proceeding toward the eye from the surrounding air, the first layer encountered is the lipid layer, which averages on the order of tens of nanometers thick.\cite{KSOculSurfRev11}  Next comes the aqueous layer, which is typically a few microns thick\cite{King-SmithFink04}, and which contains large molecules such as soluble mucins and proteins.\cite{BronTiffRev04}  The glycocalyx is a forest of membrane-bound mucins and associated molecules that form a protective barrier for the ocular surface.\cite{Gipson04,Gipson10,BronEtal15,fini2020membranemucins}  Finally, the outer surface of the corneal epithelium is the beginning of the ocular surface itself.\cite{hogan1971histology}  

The normal tear film structure can fail to form initially, or sometime after a blink
develop tear breakup, where the tear film fails to coat
the ocular surface.\cite{EwenTBUreview17,yokoi2017classification}  Tear breakup and associated hyperosmolarity (excessive saltiness of the local tears) is thought to play an important role in the development of dry eye disease, which affects millions of people.\cite{GilbardFarris78,BaudouinEtal13,craig2017tfos} The tear film lipid layer is of interest because it plays an important role in preventing tear breakup.  Simultaneous imaging of the lipid layer and the aqueous layer\cite{King-SmithIOVS13a} shows a strong correlation between lipid layer dynamics and tear breakup.  The lipid layer is typically thought to be a barrier to evaporation, thus providing an important function to preserve the tear film between blinks.\cite{MishimaMaurice61,KSHinNic10}
However, the lipid layer composition \cite{butovich2014biophysical} and structure \cite{Leiske11,Leiske12,rosenfeld2013structural} are complex and not yet fully understood. Meibum, an oily secretion from meibomian glands in the eyelids,\cite{KnopKnop2011} is the primary component of the lipid layer; it is not uncommonly used as a model for the lipid layer.  X-ray scattering methods applied to \emph{in vitro} meibum films have suggested that there are ordered particles in the meibum films with layered structures;\cite{rosenfeld2013structural} these particles may have liquid crystal structure.  Hot-stage imaging of meibum droplets have shown birefringence,\cite{butovich2014biophysical} another sign of order within the meibum.  And in the meibomian glands\cite{KnopKnop2011} in the human eyelid which produce meibum, freeze fracture with electron microscopy shows a layered structure of the lipids inside the cells that are the source of the meibum.\cite{sirigu92freeze}  We interpret this evidence to suggest that the tear film lipid layer could be an extended liquid crystalline layer with (possibly many) defects.\cite{King-SmithOS13}   It is not known whether the entire lipid layer has these qualities, or whether isolated chunks of structured particles float in the layer; however, there is general agreement that the lipid layer has non-Newtonian properties.\cite{PanNagBT99,Leiske12,rosenfeld2013structural,butovich2014biophysical,georgiev2019lipidsat}   These areas of structure in the lipid layer are thought to provide the barrier against evaporation of the aqueous layer. \cite{rosenfeld2013structural, butovich2014biophysical}  
 In addition, cooling of liquid crystals facilitates orientation of the molecules in the same direction.\cite{YangWu15} The cooling of the lipid layer may encourage the formation of liquid crystal structure \emph{in vivo}.\cite{Leiske11}   

\begin{figure}[htbp]
\centering
\includegraphics[width=.6\textwidth]{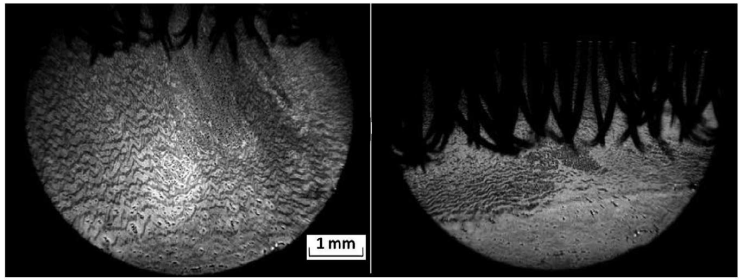}
\caption{\label{fig:ripples} Ripples in the lipid layer of the tear film before a blink (left), and after a blink (right). The ripples become compressed after a blink, and may not extend to cover the cornea once the eye is open.\cite{braun2015dynamics} }
\end{figure}

As the eye reopens during a blink, the lipid layer undergoes extensional flow as the tear film is painted across the surface of the eye. \cite{braun2015dynamics}  Rather than spreading smoothly and uniformly over the eye, imaging of the tear film reveals stripes or ripples in the lipid layer (see Fig.~\ref{fig:ripples}). \cite{braun2015dynamics} The goal of this paper is to model extensional flow of thin sheets of liquid crystal using both weak and moderate elasticity limits, and to lay the foundation to explore whether we can replicate the type of rippling seen in the tear film.  

Theoretical modeling of extensional flow was developed quite extensively in the twentieth century \cite{petrie2006one} and continues to be an active area of study, in part because of industrial applications such as optical fiber drawing\cite{howell1994extensional} and the use of polymers for a wide range of industrial purposes.   Thus, much work has been done on extensional flow of both Newtonian and non-Newtonian fluids, especially thin sheets or fibers. We do not attempt a comprehensive review here, but simply highlight a few studies of relevance to our problem. Evolution of Newtonian fibers under extensional flow has been studied extensively, from axisymmetric viscous fibers with one-dimensional flow by Schultz and Davis,\cite{schultz1982one} to more complicated three-dimensional models for non-axisymmetric fibers by Dewynne et al. \cite{dewynne1992systematic, dewynne1994slender} Wylie et al.\cite{wylie2011stretching} discuss the role of inertia and surface tension in the extensional flow of viscous fibers, and find that, while effects of surface tension are higher order and can be neglected, there are times when inertia plays an important role in the evolution.  Howell\cite{howell1994extensional} also presented exact solutions for the extensional flow of both sheets and fibers of primarily Newtonian fluid (and also provides a good overview of earlier extensional flow modeling).  Such flows are relevant for glass manufacturing, printing, and other applications; see \textcite{dewynne1994slender} for further discussion and references. Non-Newtonian fluids have also received attention; for example, the development of beads on a string has been described by Clasen et al.\cite{clasen2006beads} for polymer fluids in a jet or liquid bridge,  and by Sostarecz and Belmonte\cite{sostarecz2004beads} for micellar fluid under stretching,  while Smolka et al.\cite{smolka2004exact} presented an exact solution for the extensional flow of a thread of fluid under both weakly and strongly viscoelastic limits.  Most relevant to our application, however, Cummings et al. \cite{cummings2014extensional} studied extensional flow of nematic liquid crystals, and this is the scenario on which we now focus, as we hope it can help to explain the dynamics of the tear film of the eye during the blink cycle.

 As a starting point, we use the model of Cummings et al.,\cite{cummings2014extensional}  which uses the Ericksen-Leslie equations to describe extensional flow of a thin sheet of nematic liquid crystal. The main focus of that paper is the response of the liquid crystal film to an applied electric field (relevant to many technological applications, such as electronic displays). In a biological setting such as the human eye, however, no electric field is present, thus, we neglect this aspect of the modeling but follow the same asymptotic approach. We rescale the governing equations and consider a new limit for the case of moderate elastic effects. We analyze a range of boundary conditions, which are found to strongly affect the shape of the evolving sheet under stretching. We then investigate rippling in the sheet by introducing more waves into the initial condition.

The paper is organized as follows. In Section \ref{sec:Model} we describe the problem formulation, and the mathematical models for both weak and moderate elasticity. Section \ref{sec:Solving} provides details of the numerical methods used to solve the models. In Section \ref{sec:Results} we present our results. These include profiles of the sheet thickness, fluid velocity, and film pressure that result from the different boundary conditions. We track the location of minimum sheet thickness under these different scenarios. We compare and contrast the solutions under weak and moderate elasticity, when either the surface tension or speed of the moving end is varied. Then we present results when multiple waves are added to the initial condition, and the mechanisms for the observed dynamics. Finally, in Section \ref{sec:Discussion} we discuss the results and outline our conclusions. 

\section{\label{sec:Model}Models}

\begin{figure}[htbp]
\centering
\includegraphics[width=.4\textwidth]{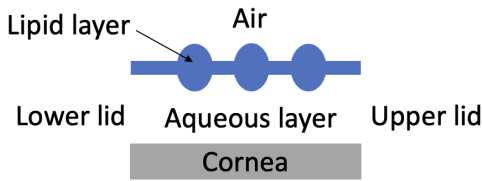}
\caption{\label{fig:beads}Simplified sketch of a cross section of the tear film.}
\end{figure}

 To reduce the complexity of the lipid layer geometry seen in Fig.~\ref{fig:tf_xsection}, we simplify the cross section of the tear film (the sagittal plane) to the geometry shown in Fig.~\ref{fig:beads}, where the ripples in the lipid layer of the tear film appear in a 2D configuration analogous to beads on a string. In this work, we neglect the aqueous layer, and consider the lipid layer alone, in two dimensions. Thus, as a first step, we consider it to be a thin free film in a sheet configuration, with multiple waves on the fluid/air interfaces in the initial condition.

\begin{figure}[htbp]
\centering
\includegraphics[width=.6\textwidth]{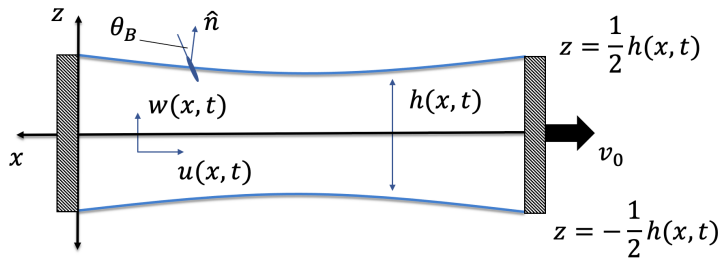}
\caption{\label{fig:schematic}Schematic of a sheet of nematic liquid crystal stretched between two plates. The left end is fixed, while the right end is moved with a prescribed velocity $v_0$. The sheet thickness is $h(x,t)$ and the axial fluid velocity, $u(x,t)$. The transverse velocity is $w(x,t)$. Molecules on the surface lie at an angle $\theta_B$ relative to $\hat{n}$, the outward-facing normal vector.} 
\end{figure}

The sheet of fluid is assumed fixed at the left end, while the right end moves with a prescribed constant speed $v_0$, providing a simple model of the opening eyelid following a blink. A sketch is shown in Fig.~\ref{fig:schematic}. As a further simplification, the lipid sheet is assumed symmetric about its midline, and the midline is assumed to be straight. We denote the thickness of the sheet by $h(x,t)$, the axial fluid velocity by $u(x,t)$, and the transverse velocity by $w(x,t)$. The liquid crystal molecules in the lipid sheet are assumed to have a preferred angle of $\theta_B$ relative to $\hat{n}$, the outward-facing unit vector normal to the sheet surface. The angle of the molecules within the sheet is described by the director field $\bf{n}=(\sin \theta,\cos \theta)$; the director field is discussed further in the appendix.

\subsection{Weak elasticity}
Our approach follows that of Cummings et al., \cite{cummings2014extensional} who used multiple scale perturbation methods to simplify the Ericksen-Leslie equations \cite{leslie1979theory} governing nematic liquid crystal dynamics. The Ericksen-Leslie equations (see Eqs.~(\ref{eq:ELwhole}) of the Appendix) are nondimensionalized using the scalings given below, where primes denotes dimensional quantities. The coordinates $(x', z')$ and velocity components $(u',w')$ correspond to the axial and transverse directions respectively, $h'$ represents the sheet thickness, $t'$ is time, $p'$ is pressure, and $\gamma'$ is the surface tension at the film/air interface:
\begin{align}
    x'&=Lx, \hspace{25pt} z'=\delta Lz, \hspace{25pt} u'=Uu, \hspace{25pt} w'=\delta Uw,  \\
    h'&=\hat{h}h, \hspace{27pt} t'=\frac{L}{U}t, \hspace{30pt}  p'= \frac{\mu U}{L}p, \hspace{18pt} \gamma'= \frac{\mu U}{\delta}\gamma.
\end{align}

The dimensional parameters used in the model are defined in Table \ref{tab:dim_param}, along with the non-dimensional parameters that result from the chosen scalings.

\begin{table}
\caption{\label{tab:dim_param}Parameters used in the model scalings for weak elasticity. Different scales for $\gamma$ and $\hat{N}$ are used in the moderate elasticity model; see Eq.~(\ref{eq:moderate_scales}).  Note that $\hat{N}$, $\delta$ and $\gamma$ are dimensionless.}
\begin{ruledtabular}
\begin{tabular}{cl}
Parameter     & Description \\
\hline
$\mu$         & dynamic viscosity  \\
$U$           & typical axial velocity  \\
$L$           & typical sheet length  \\
$\hat{h}$     & typical initial sheet thickness\\
$\gamma'$     & surface tension of air/sheet interface \\
$K$           & elastic constant of the liquid crystal      \\
$\delta=\frac{\hat{h}}{L} \ll 1$  &  aspect ratio \\
$\gamma = \frac{\delta}{\mu U} \gamma'$  & surface tension/viscosity comparison \\
$\hat{N}=\frac{K}{\mu U \delta L}$  & inverse Ericksen number
\end{tabular}
\end{ruledtabular}
\end{table}

Asymptotic expansion of the dependent variables in the small parameter $\delta =\hat{h}/L$ (see section \ref{sec:ModElasticity} of the Appendix), yields a closed system of equations for the (leading order) sheet thickness $h$ and axial velocity $u$:
\begin{align}
    h_t+(hu)_x&=0,\label{eq:1}\\
    \frac{F(\theta_B)}{G(\theta_B)}(hu_x)_x+\frac{\gamma}{2}hh_{xxx}&=0.\label{eq:2}
\end{align}
Eq.~(\ref{eq:1}) represents conservation of mass, and Eq.~(\ref{eq:2}) is the axial force balance. The coefficient of the axial gradient term is formed from functions $F(\theta_B)$ and $G(\theta_B)$, which depend on material properties of the fluid as well as the leading order solution for the director angle, $\theta_0$ (see Eqs.~(\ref{eq:G}) and (\ref{eq:F}) in the Appendix).  However, in the situation considered here, $\theta_0=\theta_B$ is a fixed angle, and $F$ and $G$ are themselves also constant. If the properties of a Newtonian fluid are used, then $F/G=4$, and Eq.~(\ref{eq:2}) simplifies to
\begin{align}
    4(hu_x)_x+\frac{\gamma}{2}hh_{xxx}&=0. \label{eq:2n}
\end{align}
For the remainder of this paper, we use this coefficient value of 4 when presenting weak elasticity solutions. We note that for the weak elasticity scalings chosen here, the pressure is defined as\cite{cummings2014extensional} 
\begin{align}
    p = -2 u_x - \frac{\gamma}{2} h_{xx}.
    \label{eq:p_weak}
\end{align}
Whenever the pressure is shown for solutions to the weak elasticity model, we make use of Eq.~(\ref{eq:p_weak}). The Newtonian limit, with zero surface tension $\gamma=0$, becomes the Trouton model, \cite{trouton1906coefficient} considered extensively within a Newtonian framework by Howell\cite{howell1994extensional} (see also references therein).

The tension $T(t)$ in the sheet is found by taking the first integral of the axial force balance in the Newtonian case, Eq.~(\ref{eq:2n}), which gives
\begin{align}
    T(t)&=4hu_x + \frac{\gamma}{2}\left(hh_{xx}-\frac{1}{2}h_x^2\right).
\end{align}
The tension is spatially uniform throughout the sheet (independent of $x$).\cite{howell1994extensional}

Since we specify the speed of the moving end, we impose the following boundary conditions (BCs), where $s(t)=1+v_0t$ denotes the location of the moving end,
\begin{align}
    u(0,t)&=0, \quad u(s(t),t)=v_0,\label{eq:bc1}\\
    h_x(0,t)&=0, \quad h_x(s(t),t)=0.\label{eq:bc2}
\end{align}
Typically, we take $v_0=1$, with the exception of Section \ref{sec:speed}, where we explore varying the speed of the moving end. Neumann BCs on $h$ specify the contact angle of the film with end plates; the plates are assumed to have no effect on the director field.

For the weak elasticity case, we solve the system of partial differential equations (PDE) found in Eqs.~(\ref{eq:1}) and (\ref{eq:2n}), subject to the BCs in Eqs. (\ref{eq:bc1}) and (\ref{eq:bc2}), as well as given initial conditions (ICs) for $h(x,0)$ and $u(x,0)$ discussed below.

\subsection{Moderate elasticity}
To consider the case of moderate elasticity, we rescale the inverse Ericksen number, the pressure, and the surface tension as follows, while keeping the other scalings the same:
\begin{align}
    \hat{N}=\frac{K}{\mu U L},\quad p'=\frac{\mu U}{\delta L}p,\quad \gamma'=\frac{\mu U}{\delta^2}\gamma.
    \label{eq:moderate_scales}
\end{align} 
Here primes denote dimensional quantities.  For $O(1)$ $p$ and $\gamma$, the dimensional values are both scaled to be larger than the weak elasticity case.  Following the derivation outlined in \ref{sec:ModElasticity} of the Appendix, we find the the leading order pressure
\begin{align}
\label{eq:moderate_pressure_def}
    p=-\frac{\gamma}{2}h_{xx},
\end{align}
and obtain the following system
\begin{align}
    h_t+(hu)_x&=0, \label{eq:masscon}\\
    \left(hu_{x}\right)_x+\tilde{\gamma}(h^2h_{xxx})_x&=0, \label{eq:axforcebal}
\end{align}
where $\tilde{\gamma}=\gamma C_2(\theta_B)/B_2(\theta_b)$ is the scaled surface tension with the scale factors $B_2$ and $C_2$ given in Eqs.~(\ref{eq:BigB_2}) and (\ref{eq:BigC_2}) of the Appendix. 
For simplicity, we take $\tilde{\gamma}=\gamma$ in our computational solutions. In this case of moderate elasticity, the tension in the sheet is now given by
\begin{align}
    T(t)&=hu_x+\gamma h^2h_{xxx}.\label{eq:T}
\end{align}

Although the surface tension at the lipid layer/air interface of the tear film is unknown, we use a value based on surface tension measurements for the nematic liquid crystal 5CB at a range of temperatures surrounding $35^\circ$C, \cite{tarakhan2006determination} which is close to the temperature at the surface of the eye.\cite{peng2014evaporation} Unless otherwise noted, we take $\gamma = 0.025$. 

We note that Eq.~(\ref{eq:axforcebal}) is higher order than the weak elasticity or Newtonian cases; this change will be consequential for the dynamics of the film. This higher order system requires more boundary conditions on $h$. To determine the number of boundary conditions needed we use Eq.~(\ref{eq:T}) to eliminate  $hu_x$ from Eq.~(\ref{eq:masscon}), yielding
\begin{align}
    h_t + uh_x + T(t)-\gamma h^2h_{xxx}=0 \label{countbc}.
\end{align}
The highest derivative in this equation is third order, implying that we need three boundary conditions on $h$ to solve the system; thus, we will need an additional boundary condition apart from those given in Eqs.~(\ref{eq:bc1}) and (\ref{eq:bc2}).

\subsubsection{Reducing the order}
To solve the model numerically, it is preferable to reduce the order of the system by adding a dependent variable.  
We can add the pressure, $p$, shown in Eq.~(\ref{eq:moderate_pressure_def}) to our system of PDEs as an additional dependent variable, and substitute into the axial force balance Eq.~(\ref{eq:axforcebal}) to reduce the order of the highest derivative appearing in the system. We obtain:
\begin{align}
    h_t+(hu)_x&=0,\label{eq:mod1}\\
    \left(hu_{x}\right)_x-(h^2p_{x})_x&=0, \label{eq:mod2}\\
    p+\frac{\gamma}{2}h_{xx}&=0.\label{eq:mod3}
\end{align}
Using this substitution, we can write the equation for tension in the moderate elasticity case (see Eq.~(\ref{eq:T})) as
\begin{align}
    T(t)=hu_{x}-h^2p_{x}. \label{eq:T_mod_using_p}
\end{align}
The space-dependent terms on the right hand side of this equation combine to be independent of $x$.  

\subsubsection{Boundary and initial conditions}

The boundary conditions for the axial velocity $u$ are as in Eq.~(\ref{eq:bc1}) for the weak elasticity case: $u(0,t)=0$ and $u(s(t),t)=v_0$.
For the sheet thickness $h$, we consider four sets of boundary conditions for the moderate elasticity model that are summarized in Table~\ref{tab:bc_summary}. In all cases, the third (additional) boundary condition on $h$ is enforced by setting  $p_x(0,t)=0$ on the fixed end.  

Turning to Table~\ref{tab:bc_summary}, Cases I and II specify Neumann conditions (homogeneous and non-homogenous) on $h$. Cases III and IV specify a Robin condition on the right (moving) or left (fixed) end respectively. The parameter $\nu$ may vary between 0 and 1; a smaller value for $\nu$ results in a boundary condition that is close to a pure Dirichlet condition at that end. In the physical sense, the Robin boundary conditions model capillarity on one end of the sheet. A Dirichlet condition would represent fluid pinned to the plate, with the slope free to vary. The Neumann conditions specify the contact angle formed by the liquid crystal fluid and the plate, but the thickness of the film is free to vary. Homogeneous Neumann conditions represent a contact angle of $\pi/2$.

\begin{figure}[htbp]
\centering
\includegraphics[width=.5\textwidth]{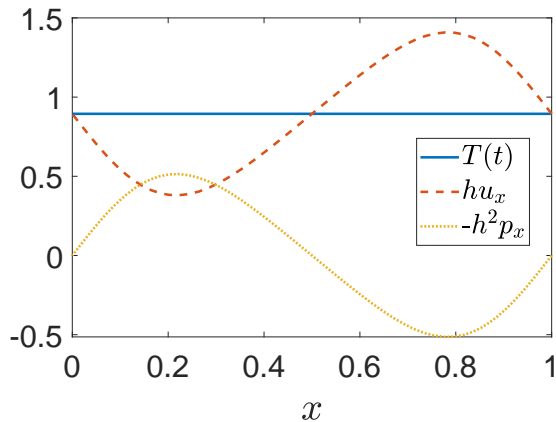}
\caption{The tension $T(t)$ and its component terms from Eq.~(\ref{eq:T_mod_using_p}) are plotted at $t=0$ for $h(x,0)=0.9+0.1 \cos (2 \pi x)$. At $x=0$, all the tension comes from $hu_x \neq 0$, and none from $h^2p_x=0$. Because $h(0,t) \neq 0$, this justifies the choice of $p_x(0,t)=0$ as a third boundary condition in the case of moderate elasticity.}
\label{fig:tension}
\end{figure}

The tension equation Eq.~(\ref{eq:T_mod_using_p}) can be used to determine the remaining boundary condition that is needed.  
To evaluate the individual terms in Eq.~(\ref{eq:T_mod_using_p}), we use the initial condition $h(x,0)=0.9+0.1 \cos (2 \pi x)$ and find $u(x,0)$ by solving Eq.~(\ref{eq:axforcebal}) subject to $u(0,0)=0$ and $u(1,0)=v_0$ with $\gamma=0.025$ and $v_0=1$. 
To find $p(x,0)$, the definition in Eq.~(\ref{eq:mod3}) is used. We then plot the individual terms from Eq.~(\ref{eq:T_mod_using_p}) (or equivalently, Eq.~(\ref{eq:T})). These curves result from valid initial conditions for which we present solutions below.  We see that one component of the 
tension, $h^2 p_x$, is zero at the left end, while the other is not.  This is important because it suggests that $p_x(0,t)=0$, and that we can enforce it as an additional boundary condition at $x=0$ for the moderate elasticity model.  In physical terms, Fig.~\ref{fig:tension} shows that, at $x=0$, all of the tension is in the extensional term while none is in the pressure term. 

The initial condition for $h$ is chosen as 
\begin{align}
  h(x,0) = a+b\cos(2\pi k_0 x)+c\,x(x-1).   \label{eq:general_h_ic}
\end{align}
The wavenumber $k_0$ will typically be $k_0=1$ but will be systematically varied in later sections.
The quadratic term ($c\neq 0$) is used only in Case II, where we allow a nonzero slope at the ends. The initial condition for $p$ is calculated exactly via Eq.~(\ref{eq:mod2}).  One must solve for $u(x,0)$ in order to have a consistent initial condition for the numerical solvers that we use.  We return to this point in Section~\ref{sec:Solving} below.

\begin{table*}
\caption{\label{tab:bc_summary}Summary of boundary conditions on a moving domain with an initial condition $h(x,0)=a+b\cos(2\pi k_0 x)+c\, x(x-1)$, and $0<\nu<1$. The quadratic term in the initial condition is only used in Case II.}
\begin{ruledtabular}
\begin{tabular}{cllccc}
Case    & Fixed end, $x=0$       & Moving end, $x=1+v_0t$ & $a$   & $b$   & $c$   \\
\hline
I      & $h_x=0, \;\;\;p_x=0  $    & $h_x=0   $  & 0.9 & 0.1 & 0            \\
II    & $h_x=-c, \,p_x=0 $      & $h_x=c$  & 0.9 & 0.1 & 0.1   \\
III    & $h_x=0, \;\;\;p_x=0 $     & $(1-\nu)(h-1)+\nu h_x=0$ & 0.9 & 0.1 & 0 \\
IV      &$(1-\nu)(h-1)-\nu h_x=0, \;p_x=0$&$ h_x=0 $ & 0.9 & 0.1 & 0 
\end{tabular}
\end{ruledtabular}
\end{table*}

\section{Numerical solution}
\label{sec:Solving}

To solve the models numerically, we first map from a moving domain $0<x<s(t)$ with $s(t)=1+v_0t$, to a fixed domain $0<\xi<1$ using $\xi=x/s(t)$. On the fixed domain, the unknowns become $H(\xi,t)=h(x,t)$ and $U(\xi,t)=u(x,t)$.  We then apply the mapping to both the weak and moderate elasticity models.  

\subsection{Weak Elasticity}

After mapping Eqs.~(\ref{eq:1}), (\ref{eq:2}), (\ref{eq:bc1}) and (\ref{eq:bc2}) to the fixed domain, one obtains
\begin{align}
  H_t -  v0 (\xi/s) H_\xi + (1/s)(U H)_\xi  = 0, \\
  (4/s^2)(U_\xi H)_\xi  + (\gamma/2s^3) H H_{\xi \xi \xi}  = 0,\\
  H_\xi(0,t)  = 0, \ \ H_\xi(1,t)  = 0, \\
  U(0,t)  = 0, \ \ U(1,t)  = v_0,\\
  H(\xi,0)= a + b \cos  (2 \pi k_0 x). 
\end{align}
Note that for this model there are only two BCs for $h$ and two BCs for $u$;\cite{cummings2014extensional} we do not impose the BC on $p$.

\subsection{Moderate Elasticity}

For the moderate elasticity case, the problem defined in Eqs.~(\ref{eq:mod1}), (\ref{eq:mod2}), and (\ref{eq:mod3}), along with the boundary conditions for Case III, becomes 
\begin{align}
    H_t - v_0 \frac{\xi}{s}H_\xi + \frac{1}{s}(UH)_\xi =0,\\
    \left(U_\xi H\right)_\xi - \left(H^2P_\xi\right)_\xi =0,\\
    P+\frac{\gamma}{2 s^2}H_{\xi\xi}=0,\\
    H_\xi(0,t)=0,\; P_\xi(0,t)=0,\; (1-\nu) s H(1,t) + \nu H_\xi(1,t)=0,\\
    H(\xi,0)=a+b\cos(2\pi k_0 \xi)+c\xi(\xi-1).
\end{align}
Boundary condition Case I is recovered by setting $\nu=1$ in Case III here, while Cases II and IV are transformed similarly.

\subsection{Numerical methods}

We describe the implementation for the moderate elasticity case here in detail; the weak elasticity case is treated similarly.
After mapping to a fixed domain, we apply a version of the method of lines; we implement two approaches to validate our results.  The spatial derivatives are approximated via collocation with either finite difference or Chebyshev spectral discretization. When utilizing finite difference methods, we use a uniform grid. Second-order centered formulas are used inside the domain, and the appropriate second-order non-centered formulas are used to approximate the derivatives at the left and right ends of the sheet. The result is a system of differential algebraic equations (DAEs) at the grid points that we solve forward in time in \textsc{Matlab} (MathWorks, Natick, MA, USA) using \texttt{ode15s}. In general, the number of grid points is $N=512$. As a check, we use the trapezoidal method to calculate the fluid volume, and observe that it is conserved to the order of our imposed tolerances of $10^{-4}$.

Alternatively, we use Chebyshev spectral discretization in space, which also results in a DAE system solved in the same way. \cite{trefethen2000spectral} Typically, the number of grid points we used was $N=128$ for this method. 

For either discretization method, the initial sheet thickness $h(x,0)$ was first specified, then $p(x,0)$ computed from its definition in Eq.~(\ref{eq:mod3}).
Finally, the discrete version of the axial force balance Eq.~(\ref{eq:mod2}) was solved for $u$ on the grid points using the backslash.  

The results using both methods agree, until the final times when error accumulates at the ends with the finite difference method. However, the spectral method could not complete computations over as wide a range of parameter values (for example, for surface tension) as could the finite difference method.

\section{Results}
\label{sec:Results}
We begin by showing solutions for thickness and velocity for the simple case of a flat sheet.
We then present solutions for thickness, velocity, and pressure obtained for the various boundary conditions outlined in Table~\ref{tab:bc_summary} in the case of moderate elasticity, and we compare them with the corresponding results in the case of weak elasticity (where the condition $p_x(0,t)=0$ is not used). We note how the location of the sheet's minimum thickness changes depending on the boundary conditions imposed. Next, we vary both the surface tension and speed of the moving end, and demonstrate the effect for both moderate and weak elasticity. We investigate dynamics resulting from increasing the number of sinusoidal waves in the initial condition, and show examples of how the wave profile changes through time depending on the surface tension value, and the amplitude and period of the imposed initial waves.  Finally, we discuss mechanism responsible for those dynamics.

\subsection{Neumann conditions on $h$\label{ss:neumann}}
\begin{figure}[htbp]
\centering
\includegraphics[width=.32\textwidth]{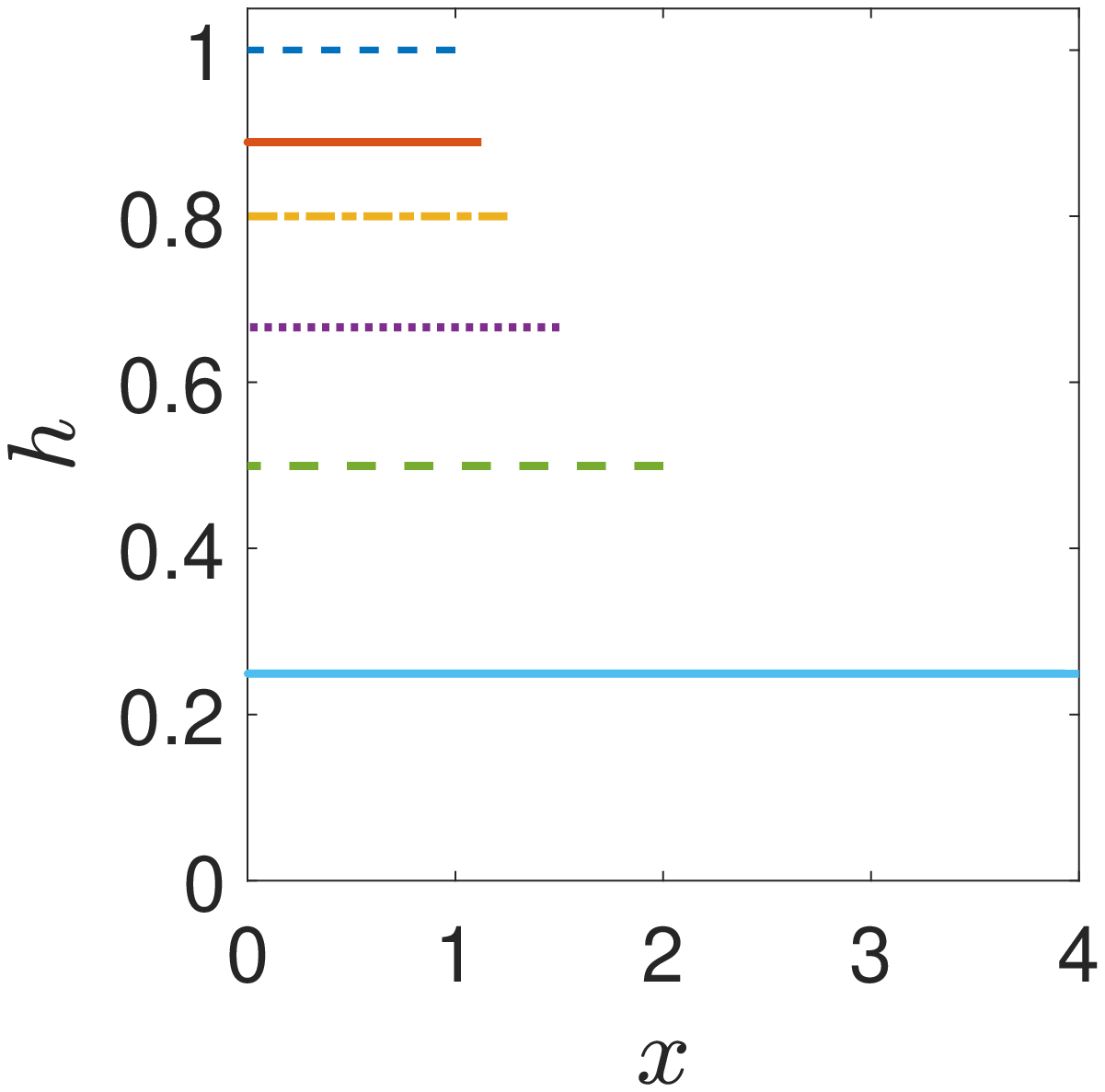}
\includegraphics[width=.32\textwidth]{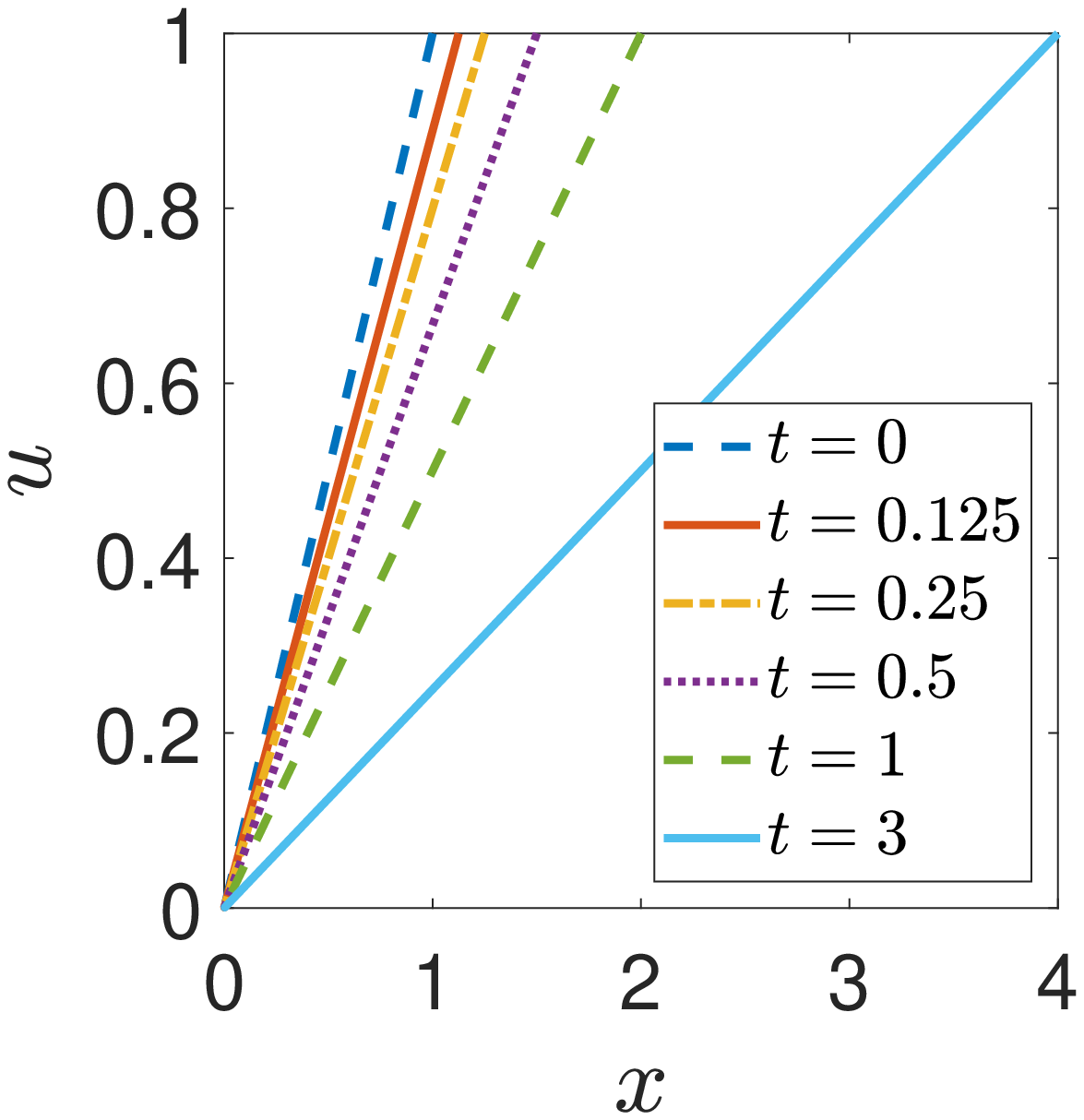}
\caption{Profiles of sheet thickness, $h$, and fluid velocity, $u$, for an initially flat sheet with homogeneous Neumann boundary conditions and $v_0=1$ (both weak and moderate elasticity cases have the same evolution for these variables).}
\label{fig:full_flatic_gam1}
\end{figure}

We begin by showing solutions for an initially flat sheet. We take $h(x,0)=1$, with BC Case I given by $h_x(0,t)=h_x(1,t)=0$, $u(x,0)=v_0x$, $u(0,t)=0$, and $u(1,t)=v_0=1$.  In this scenario, the sheet remains spatially uniform for all time, and the PDEs governing $h$ and $u$ for both weak and moderate elasticity are the same, as the terms containing surface tension are lost. Solutions for $h$ and $u$ are shown in Fig.~\ref{fig:full_flatic_gam1} on the moving domain. The thickness decreases uniformly, and the velocity increases linearly across the sheet. 
Note that $p_x=0$ trivially for all $x$ and $t$ for both moderate and weak elasticity models. In the case of moderate elasticity, the pressure is zero at each time level.  In the case of weak elasticity, from Eq.~(\ref{eq:p_weak}), $p=-2u_x$ and so $p$ is constant in $x$ but decreasing in time.

\begin{figure}[htbp]
\centering
\includegraphics[width=.32\textwidth]{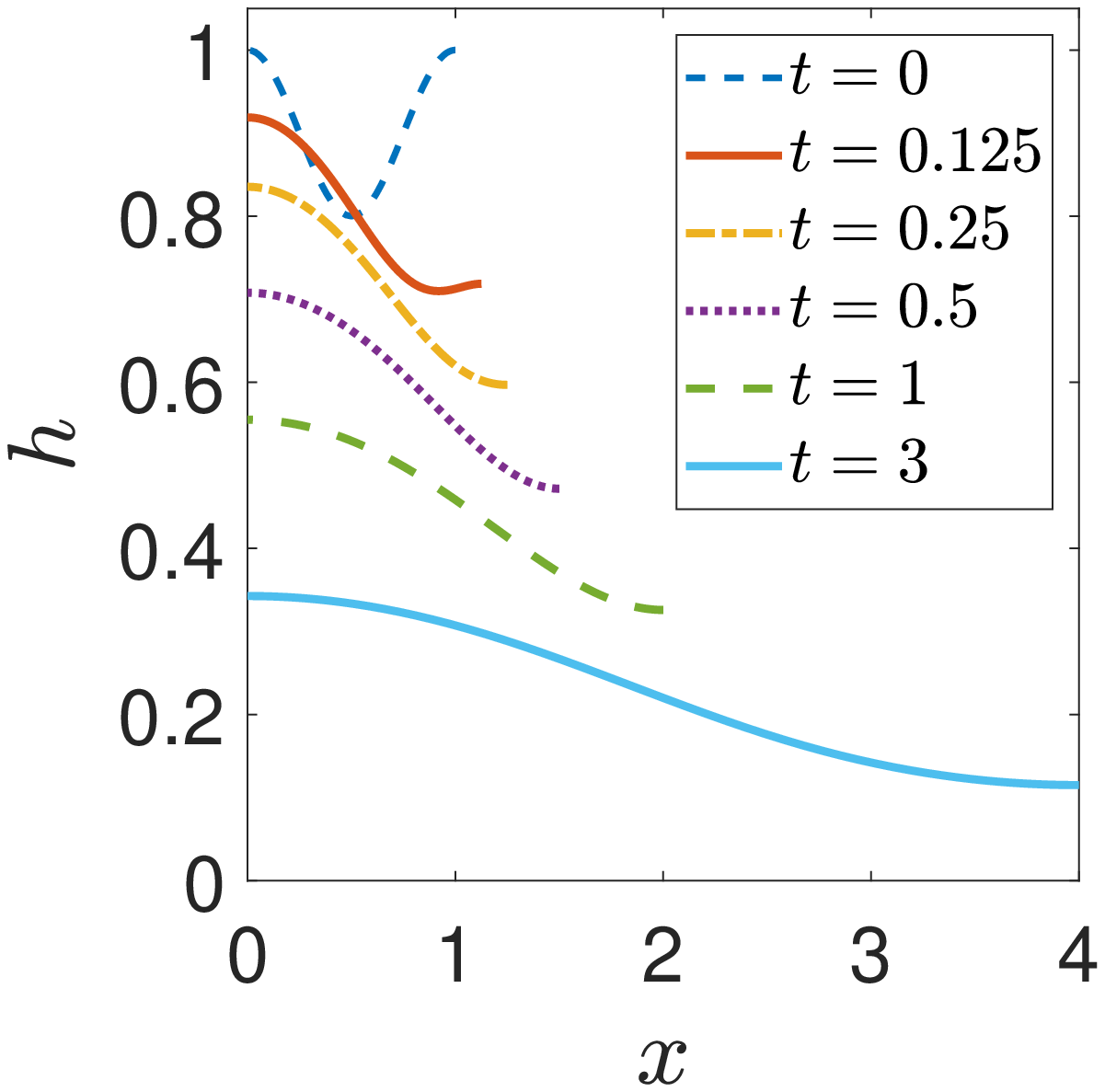}
\includegraphics[width=.32\textwidth]{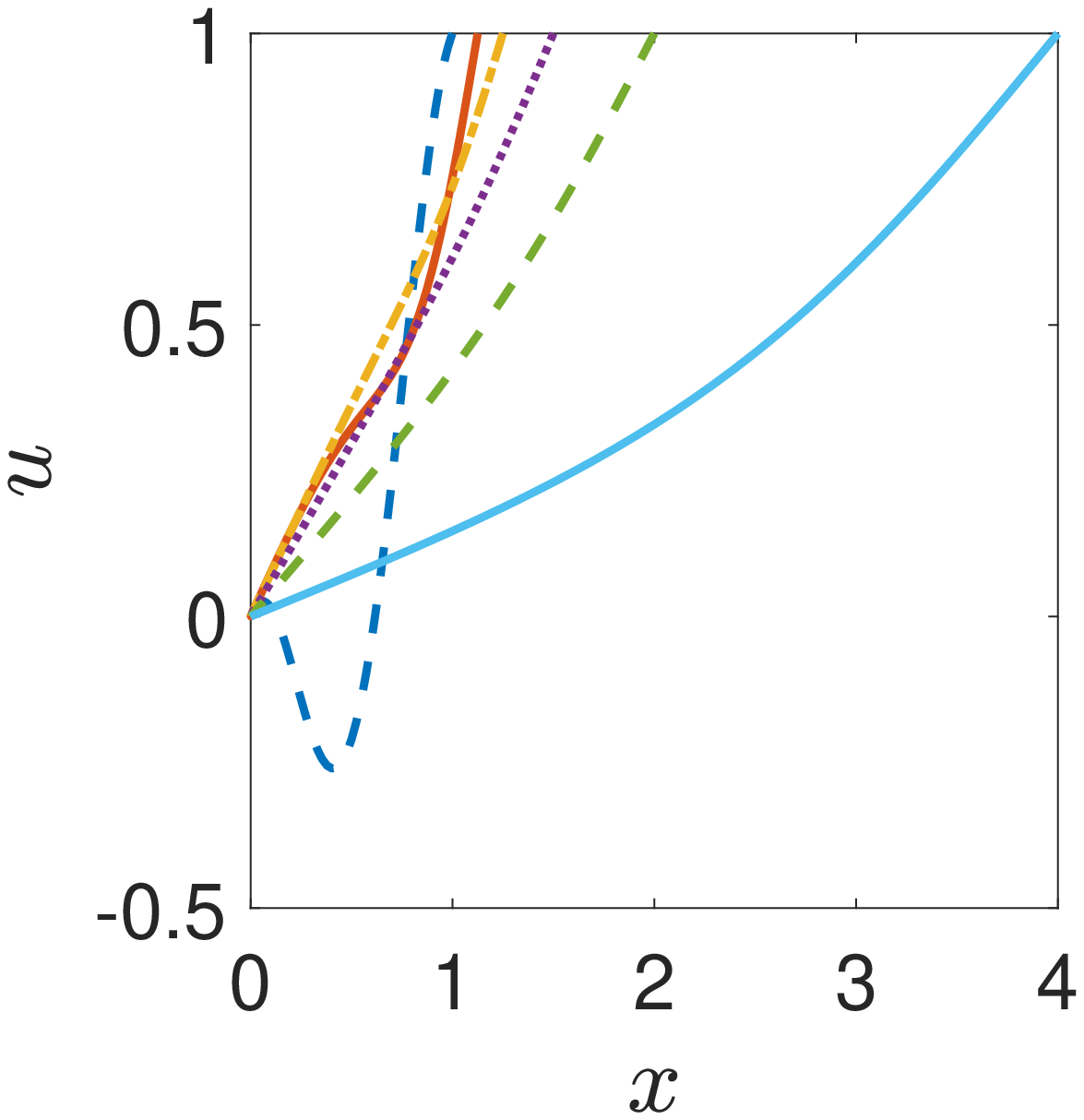}
\includegraphics[width=.32\textwidth]{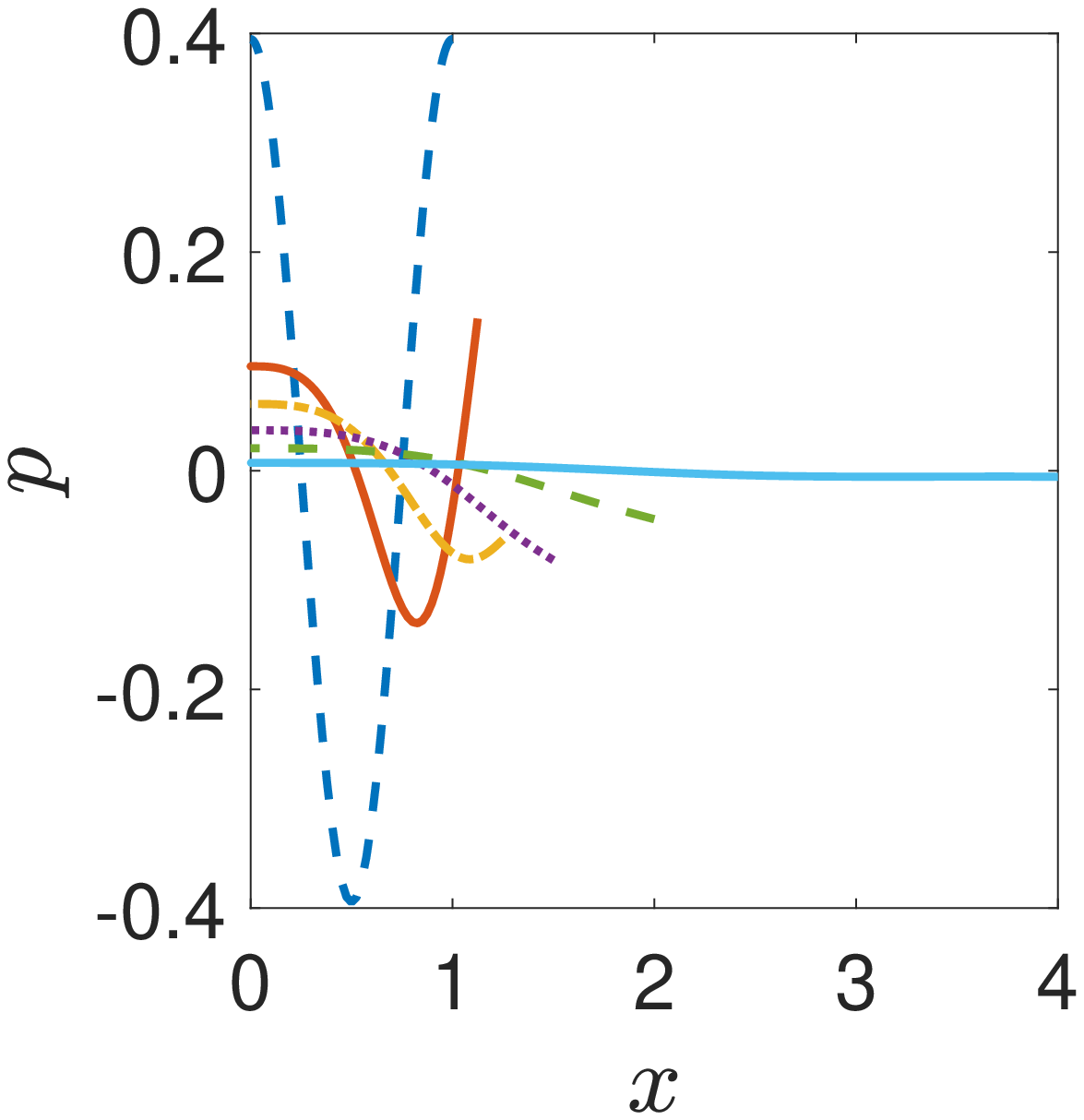}
\caption{Profiles of sheet thickness, $h$, fluid velocity, $u$, and pressure, $p$, for moderate elasticity with a sinusoidal IC and BC Case I when $\gamma=0.1$ and $v_0$=1.}
\label{fig:full_gam1}
\end{figure}

Next, we consider the moderate elasticity solutions for the sheet thickness, axial velocity, and pressure when $\gamma=0.1$ for a sinusoidal IC with $a=0.9$, $b=0.1$ and $c=0$. Results are shown in Fig.~\ref{fig:full_gam1}. Initially, the axial velocity is negative for much of the sheet, meaning that the fluid in these areas is moving to the left. This changes the profile of the sheet thickness very quickly, and extensional flow leads to thinning of the sheet at the right end. The fluid away from the right end is left behind, and by $t=0.25$, there is no longer a local maximum in the thickness at the right end. From that time until $t=3$, the sheet thickness has lost approximately half a wave from the initial one full period.

\begin{figure}[htbp]
\centering
\includegraphics[width=.32\textwidth]{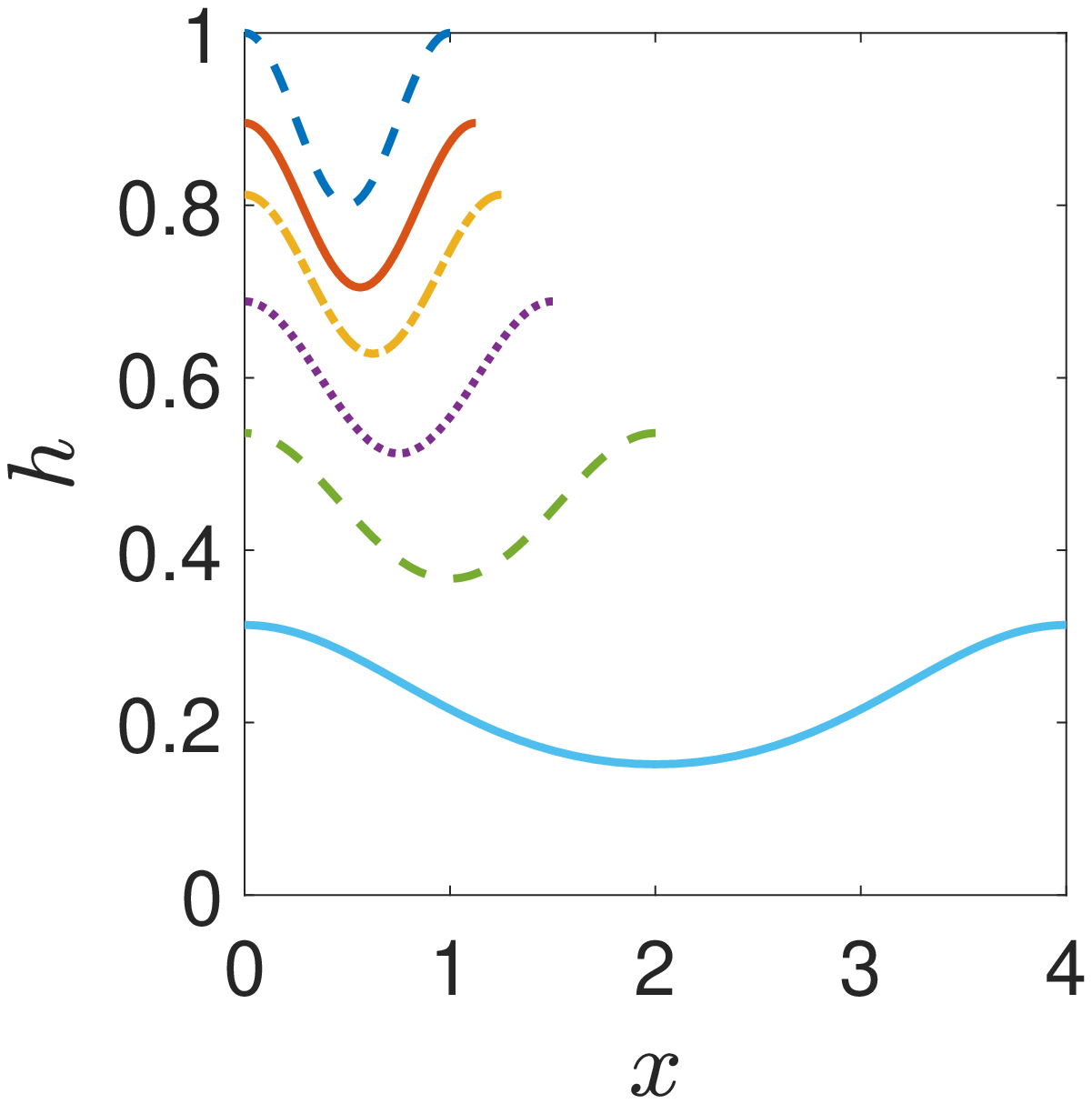}
\includegraphics[width=.32\textwidth]{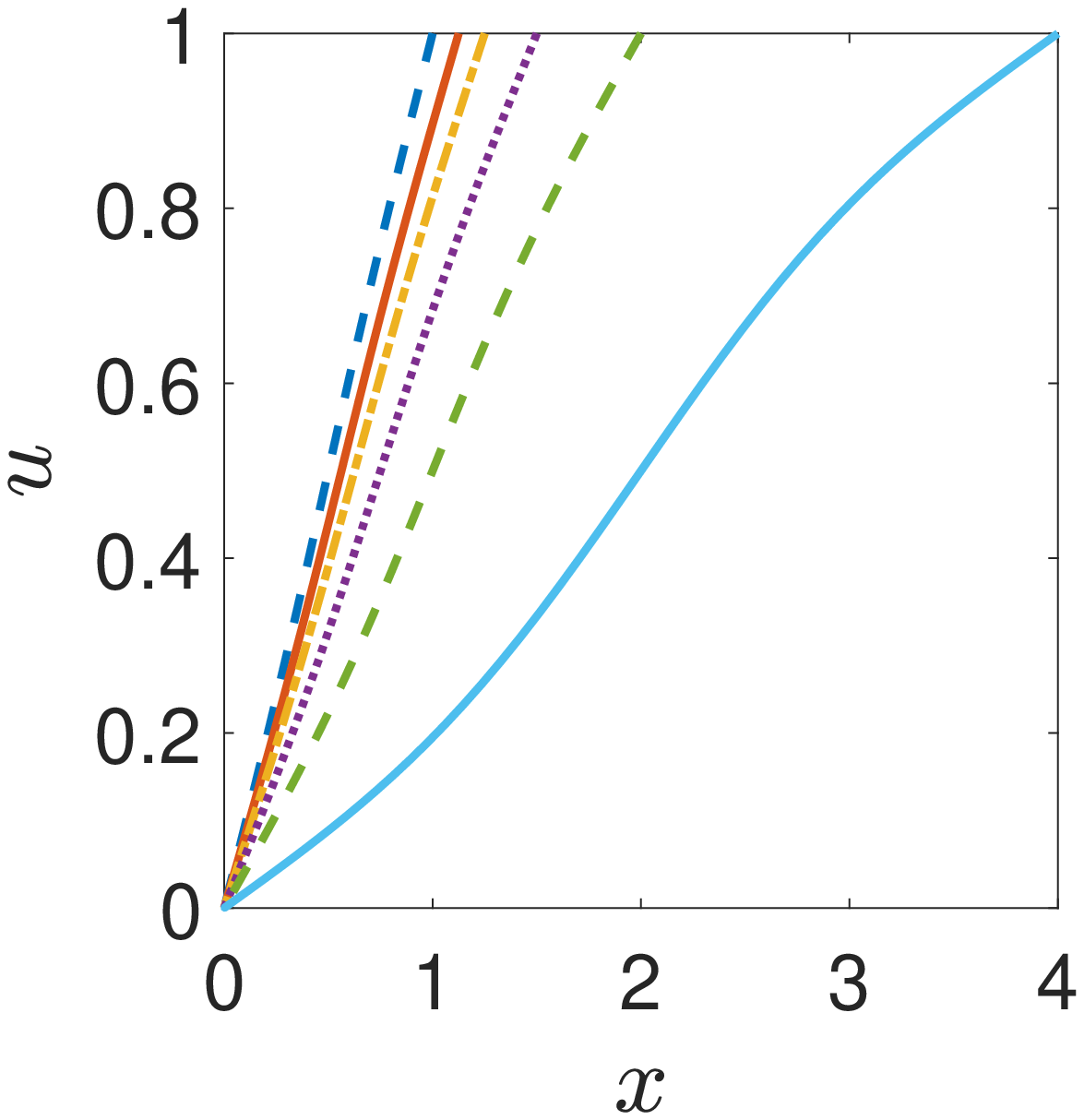}
\includegraphics[width=.32\textwidth]{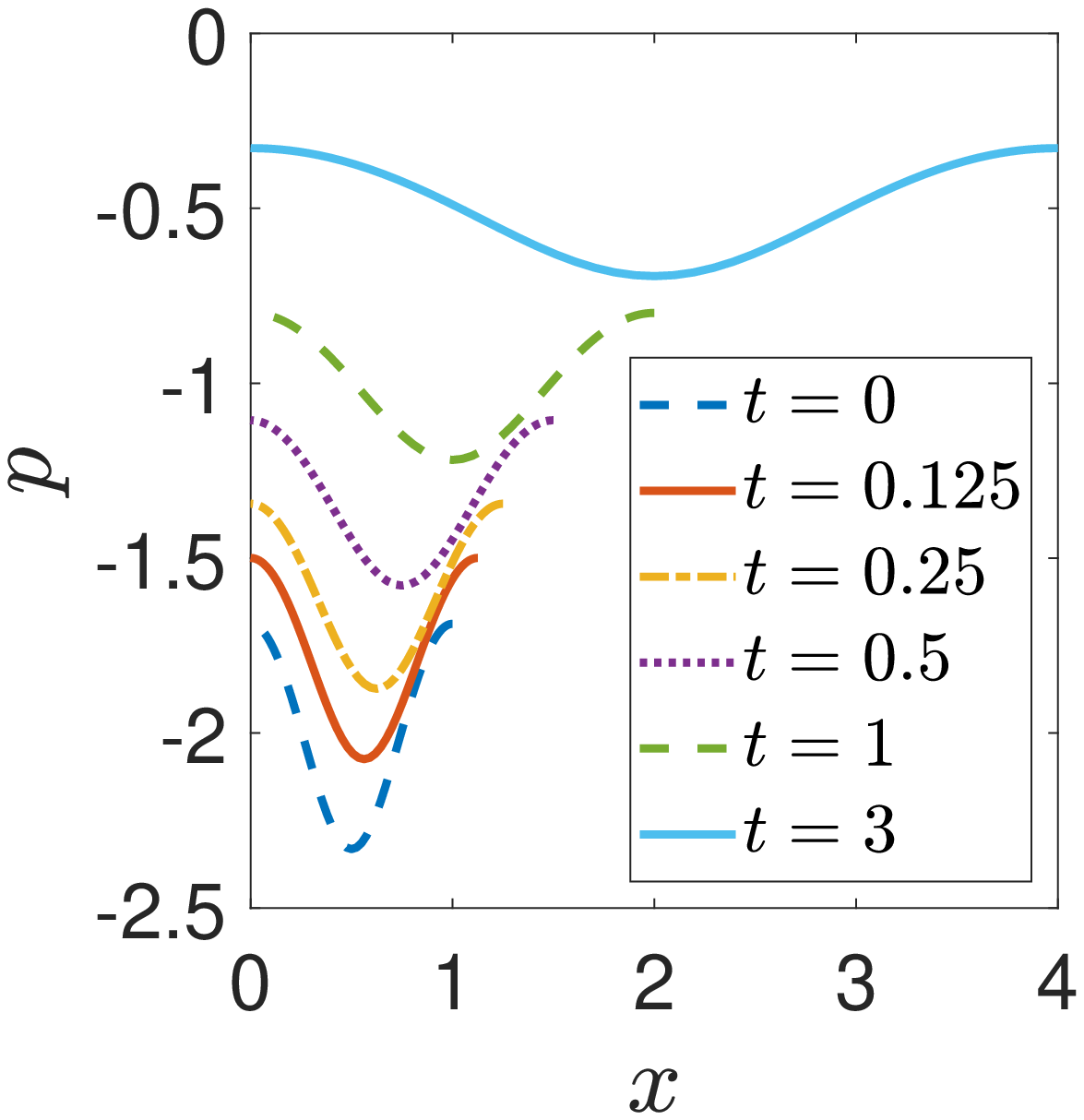}
\caption{Profiles of sheet thickness, $h$, fluid velocity, $u$, and pressure, $p$, for weak elasticity when $\gamma=0.1$ and $v_0$=1.}
\label{fig:full_newt}
\end{figure}

The tendency of fluid to gather at the left end while the right end becomes thinner is a characteristic of moderate elasticity that is not seen in the case of weak elasticity. Fig.~\ref{fig:full_newt} shows the analogous solutions for $h$, $u$ and $p$ of a sheet of fluid with weak elasticity. The sheet remains symmetric about its midpoint throughout the computation, and $h_{min}$ occurs in the middle of the sheet. The middle plot of Fig.~\ref{fig:full_newt} shows that as time progresses, the strain rate $u_x$ is largest in the middle of the sheet, and the sheet thins fastest there. The pressure remains negative throughout the sheet, but the pressure and its gradient decrease as time increases. 

\subsection{Robin boundary conditions (moderate elasticity)\label{ss:robin}}
\begin{figure}[htbp]
\centering
\includegraphics[width=.32\textwidth]{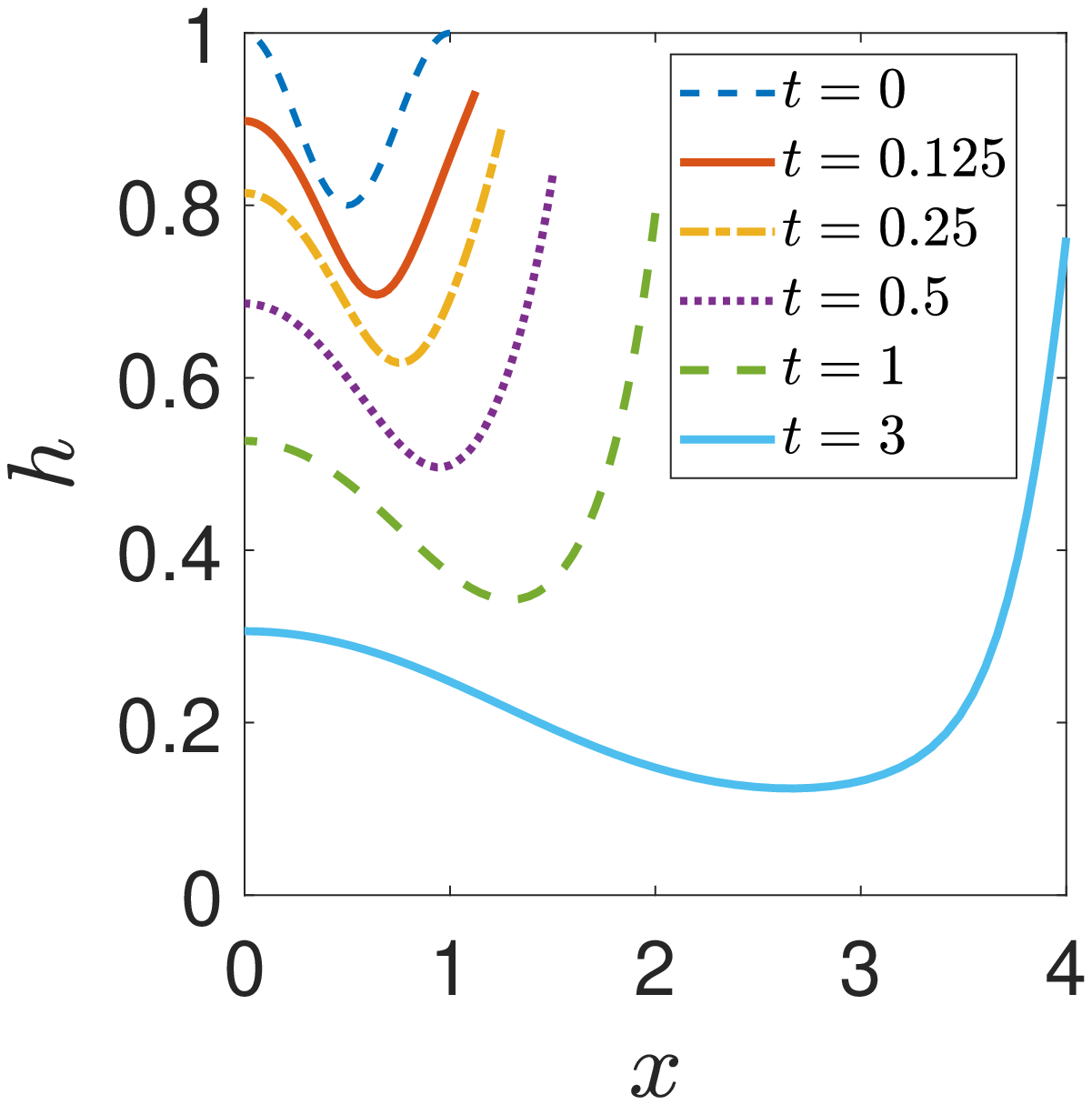}
\includegraphics[width=.32\textwidth]{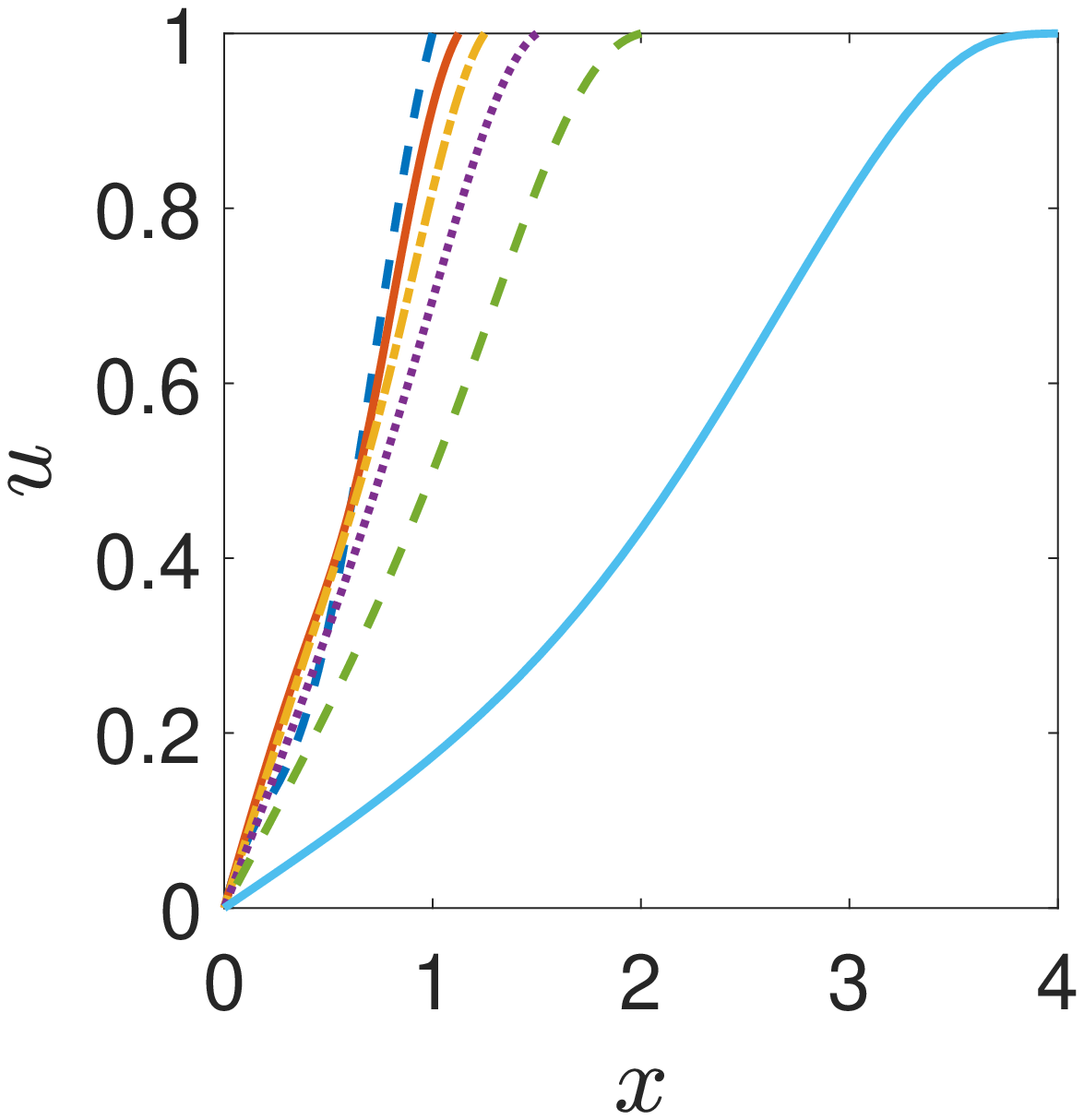}
\includegraphics[width=.32\textwidth]{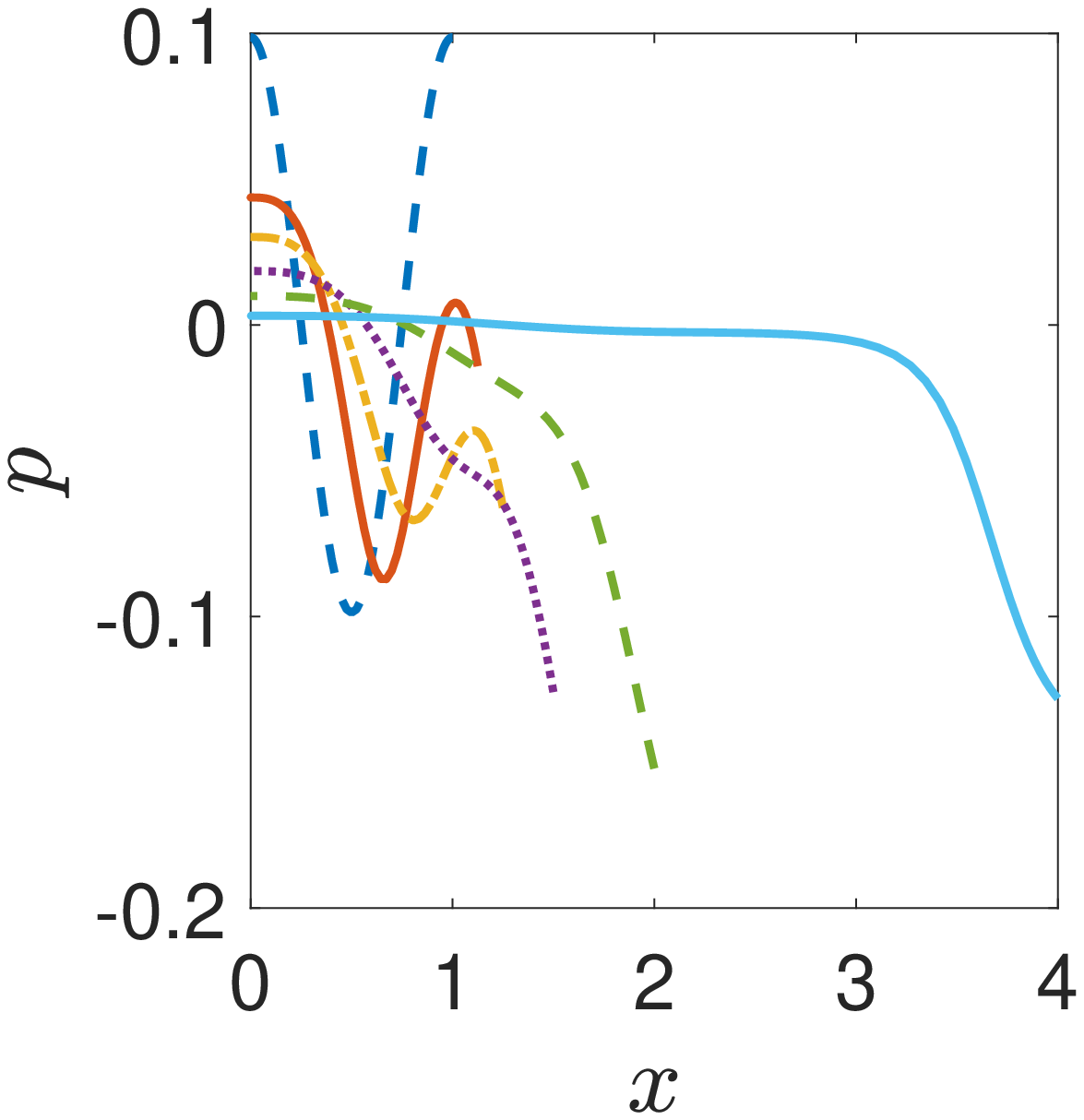}
\caption{Profiles of sheet thickness, $h$, fluid velocity, $u$, and pressure, $p$, for moderate elasticity with BC Case III, a Robin boundary condition at the right end; $\nu=0.1$, $\gamma=0.025$, and $v_0=1$.}
\label{fig:full_robinright}
\end{figure}

Imposing a Robin boundary condition at the moving end (BC Case III) of a sheet with moderate elasticity leads to the formation of a meniscus there, as shown in Fig.~\ref{fig:full_robinright}. The sheet thins primarily in the middle and left (fixed) end of the sheet, with a narrow portion of the fluid at the right traveling at roughly the same speed as the right (moving) end.  The pressure remains positive at the left end due to capillarity, but becomes negative throughout the part of the sheet that forms the meniscus. 

\begin{figure}[htbp]
\centering
\includegraphics[width=.32\textwidth]{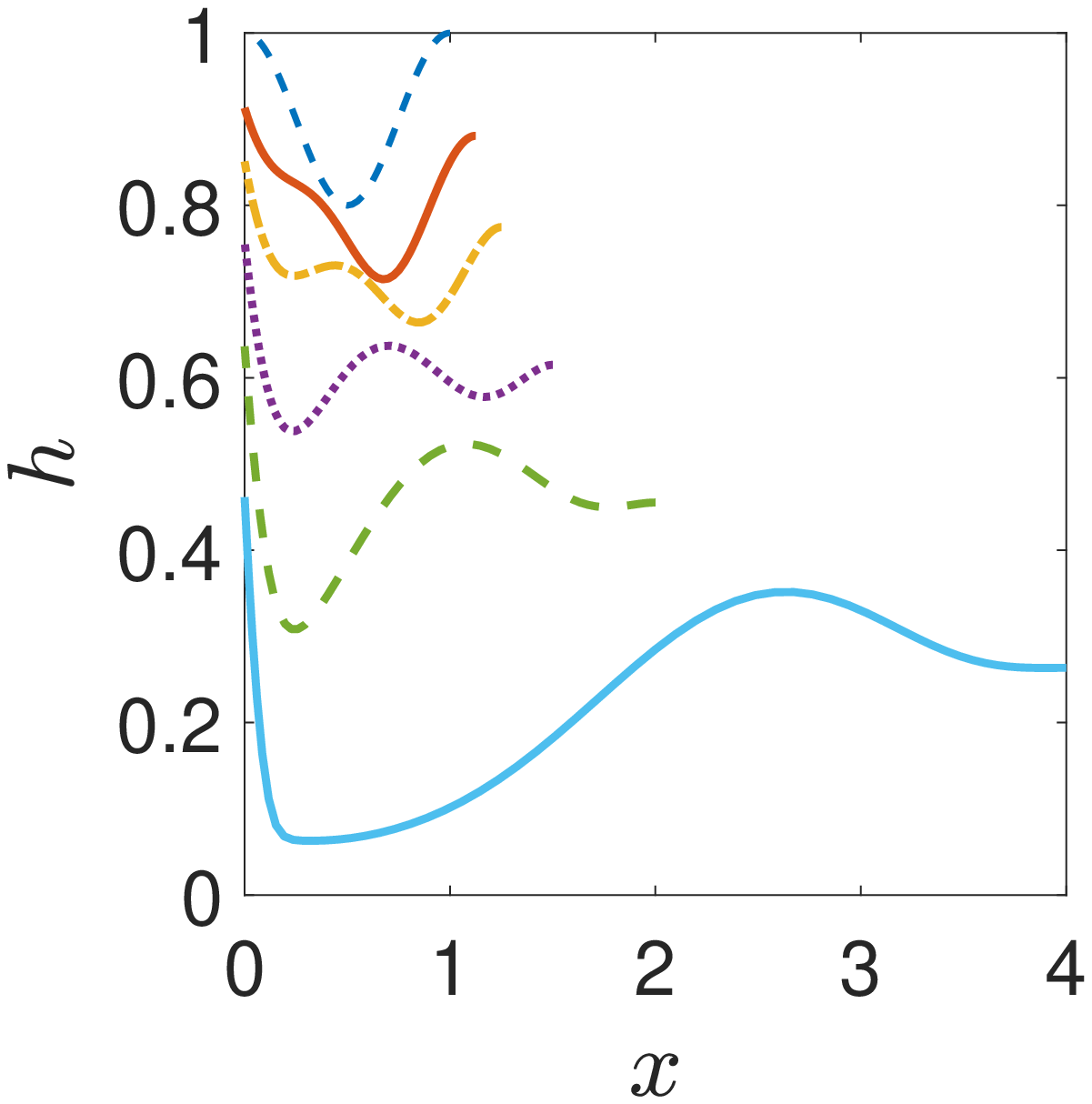}
\includegraphics[width=.32\textwidth]{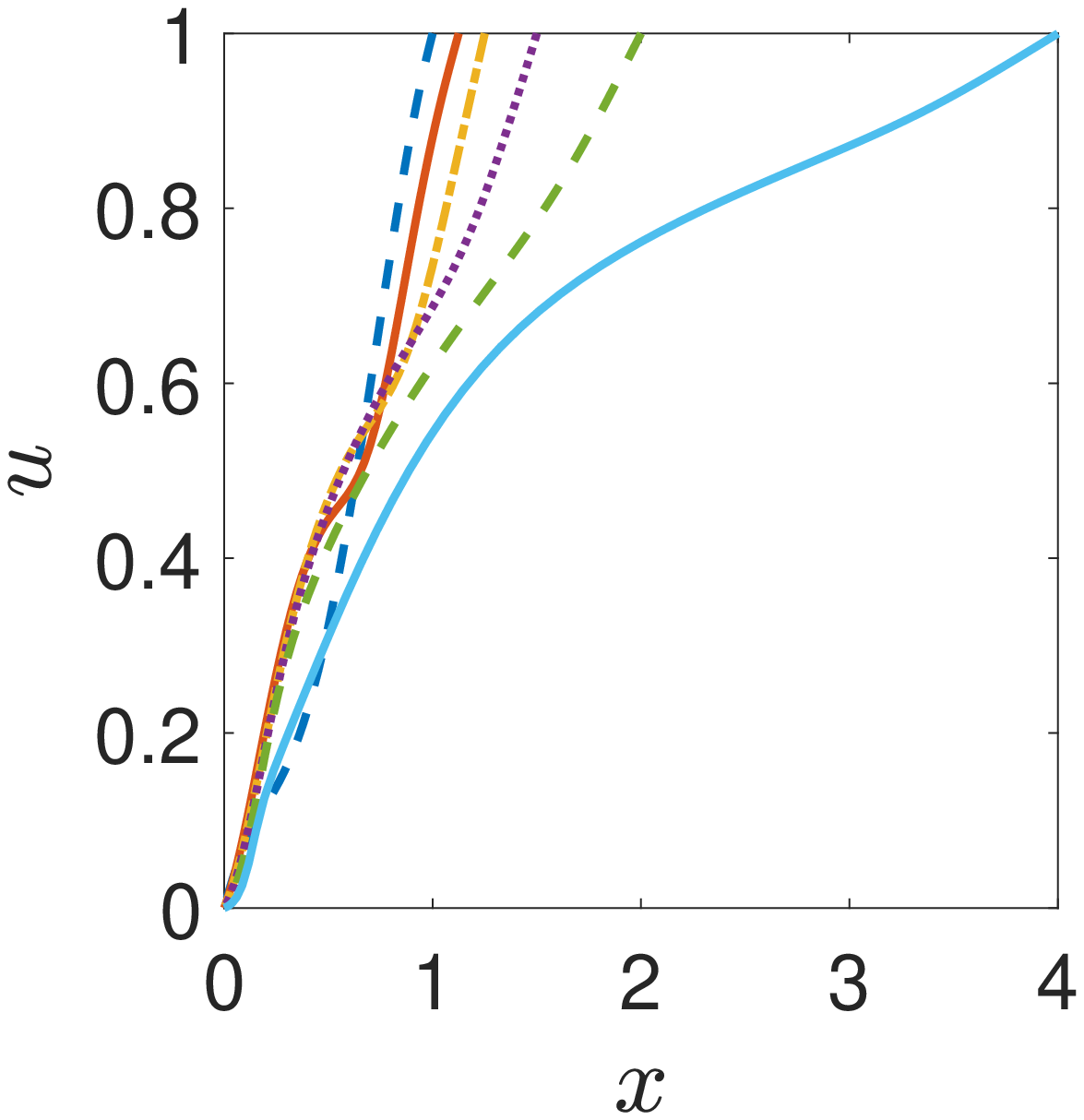}
\includegraphics[width=.32\textwidth]{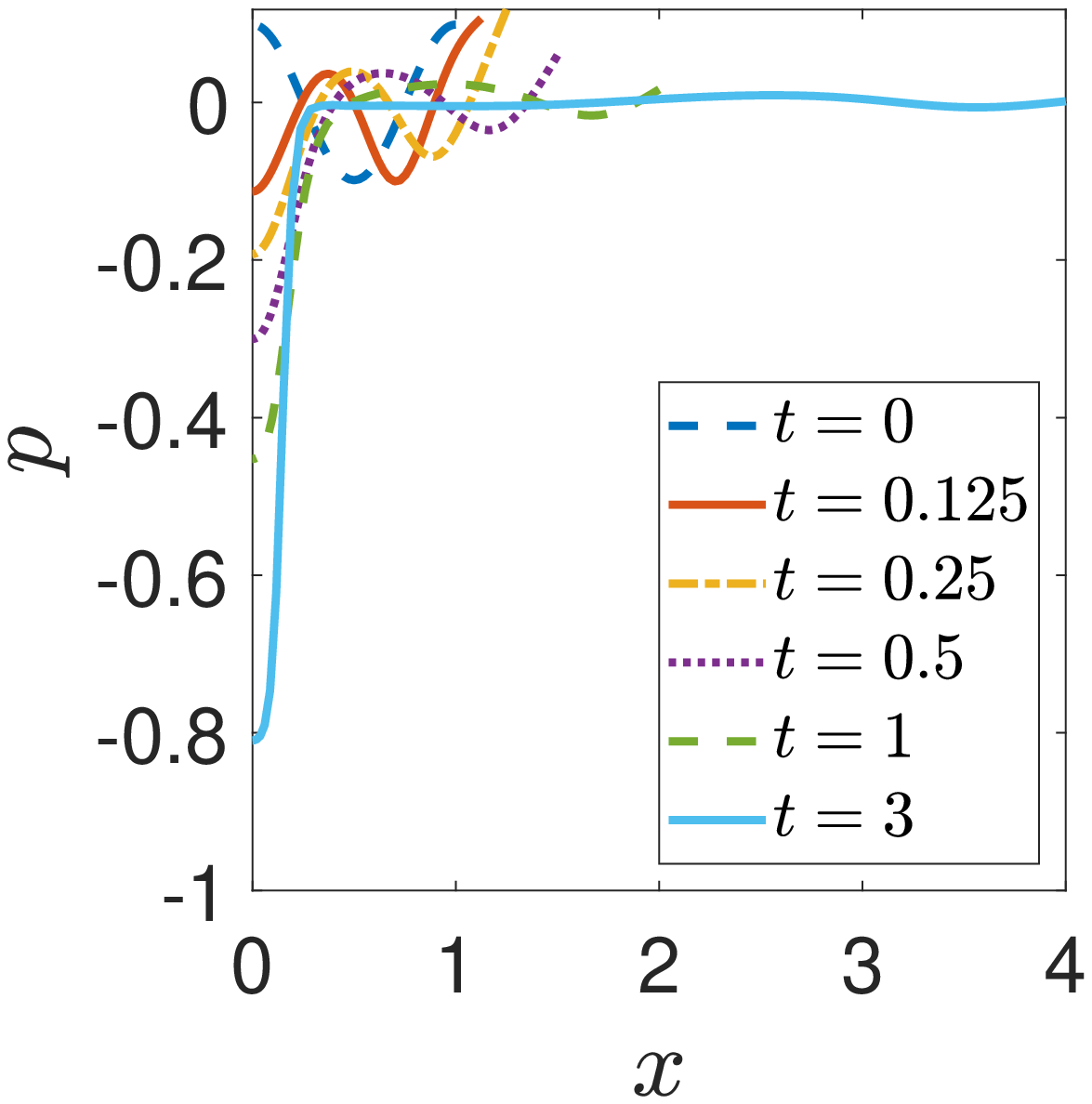}
\caption{Profiles of sheet thickness, $h$, fluid velocity, $u$, and pressure, $p$, for moderate elasticity with BC Case IV, a Robin boundary condition at the left end; $\nu=0.1$, $\gamma=0.025$, and $v_0=1$.}
\label{fig:full_robinleft}
\end{figure}

Fig.~\ref{fig:full_robinleft} shows the results of imposing a Robin condition at the fixed end on the left (BC Case IV). This meniscus is smaller in both height and width than that of Fig.~\ref{fig:full_robinright}, where the Robin condition is imposed at the right. As observed in Fig.~\ref{fig:full_robinright}, thinning corresponds to increased strain rate $u_x$ in the portion of the sheet where it occurs. As time increases the meniscus grows, and the pressure becomes large and negative at $x=0$, while approaching zero in the rest of the film.  

Table \ref{tab:robin_comp} summarizes the differences: in both cases, the maximum sheet thickness $h_{max}$ occurs at the end where the Robin condition is enforced. When the condition is enforced at the left (Case IV), both $h_{max}$ and the range of observed sheet thicknesses ($\Delta h=h_{max}-h_{min}$) are smaller, and at the final time $t=4$, $h_{min}$ is less than half the corresponding value when the Robin condition is imposed at the right (Case III). 

\begin{table}
\caption{\label{tab:robin_comp}Comparison of imposing Robin boundary conditions at either end of the sheet (moderate elasticity solutions of Figs.~\ref{fig:full_robinright} and \ref{fig:full_robinleft}). Here $h_{max}$ is the maximum sheet thickness, $h_{min}$ is the minimum sheet thickness, and $\Delta h$ = $h_{max} - h_{min}$.}
\begin{ruledtabular}
\begin{tabular}{lllllll}
& \multicolumn{3}{c}{$t=0.5$} & \multicolumn{3}{c}{$t=4$}\\
 Robin Location & $h_{max}$   & $h_{min}$  & $\Delta h$  & $h_{max}$   & $h_{min}$& $\Delta h$ \\
\hline
Right end (BC Case III, Fig.~\ref{fig:full_robinright})   & 0.835  & 0.497 & 0.338    & 0.763  & 0.086 & 0.676\\
Left end (BC Case IV, Fig.~\ref{fig:full_robinleft})   & 0.758  & 0.538 & 0.220   & 0.431  & 0.037& 0.394
\end{tabular}
\end{ruledtabular}
\end{table}

\subsection{Location of minimum thickness}
\begin{figure}[htbp]
\centering
\includegraphics[width=.6\textwidth]{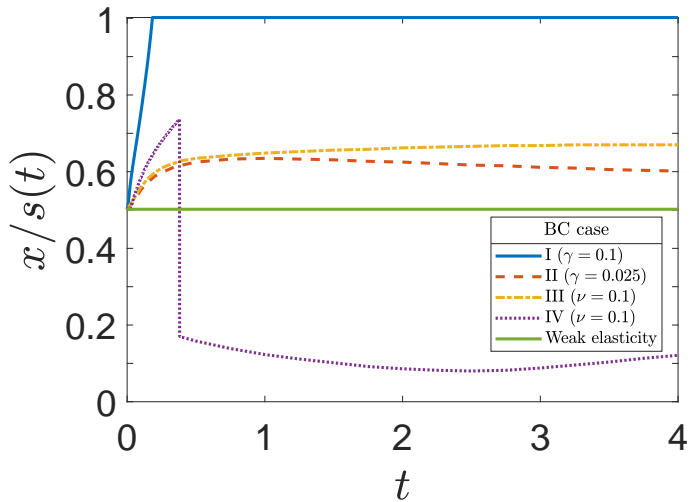}
\caption{Location of minimum sheet thickness shown on a fixed domain through time. In all cases, $h(x,0)=0.9+0.1 \cos (2 \pi x)$, and $\gamma=0.025$ except where otherwise stated.  In Case II, $c=0.1$ (see Table~\ref{tab:bc_summary}). For the weakly elastic case, the minimum remains at $x=0.5$ for all time.}
\label{fig:min_thickness}
\end{figure}

As mentioned before, in the weak elasticity case, the evolution of the sheet is symmetric about the midpoint for the chosen boundary and initial conditions.  The minimum sheet thickness $h_{min}$ begins, and remains, at the midpoint throughout the evolution. For moderate elasticity, however, the situation is more complicated. Fig.~\ref{fig:min_thickness}
summarizes a range of results for different BCs, with the initial condition $h(x,0)=a+b \cos (2 \pi x)$ with $a=0.9$, $b=0.1$ (as used in the results of Secs.~\ref{ss:neumann} and \ref{ss:robin} above), and 
$\gamma = 0.025$ unless otherwise noted. 
Fig.~\ref{fig:min_thickness} demonstrates that, even with a simple initial film shape that is 
initially symmetric about the midpoint, the minimum thickness migrates from the midpoint, and can occur in a variety of locations on the 
sheet that depend on the boundary conditions imposed. If we consider BC Case I (homogeneous Neumann conditions on $h$), with $\gamma=0.1$, then 
the minimum rapidly migrates to the right end of the domain, by about $t=0.25$.  For BC Case I with $\gamma=0.025$ (not shown in Fig.~\ref{fig:min_thickness}), the minimum remains in the right half of the domain near $x=0.5$.  
Allowing a slight slope on the end (Case II, with $c = 0.1$) keeps the minimum slightly more centered than BC Case I for the same $\gamma$. With $\gamma=0.025$ and Case I and II BCs, the minimum starts in the center of the sheet (as dictated by the initial condition), shifts to the right by $t=0.15$ or so, and then slowly begins to approach the center of the sheet again. A Robin boundary condition on the right (Case III) leads to a minimum location that begins similarly to Case I: the minimum shifts to about $\xi=x/s(t)=0.7$, but then stays there. A Robin boundary condition on the left (Case IV) causes the location of the minimum to move around the most. Referring to Fig.~\ref{fig:full_robinleft}, we see that for early times, the sheet has two local minima, with the global minimum closest to the moving end. As the sheet lengthens, that dip flattens, and the global minimum shifts to the bottom of the steep meniscus near the fixed end. For the remaining time, the minimum stays close to the left (fixed) end. This switch in the location of the global minimum is clearly seen in Fig.~\ref{fig:min_thickness}.

\subsection{Varying the surface tension (moderate elasticity)}

\begin{figure}[htbp]
\centering
\includegraphics[width=.49\textwidth]{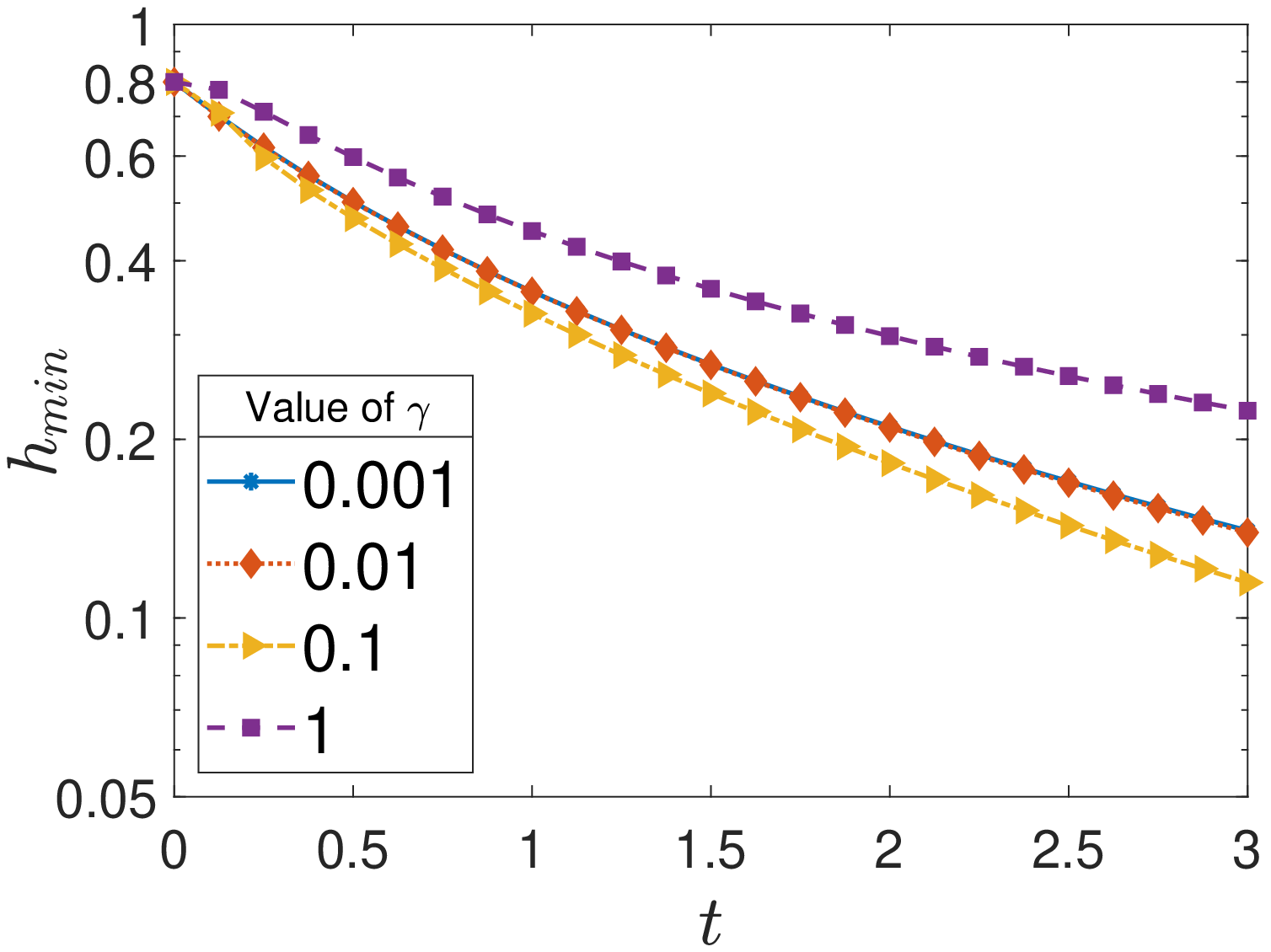}
\includegraphics[width=.49\textwidth]{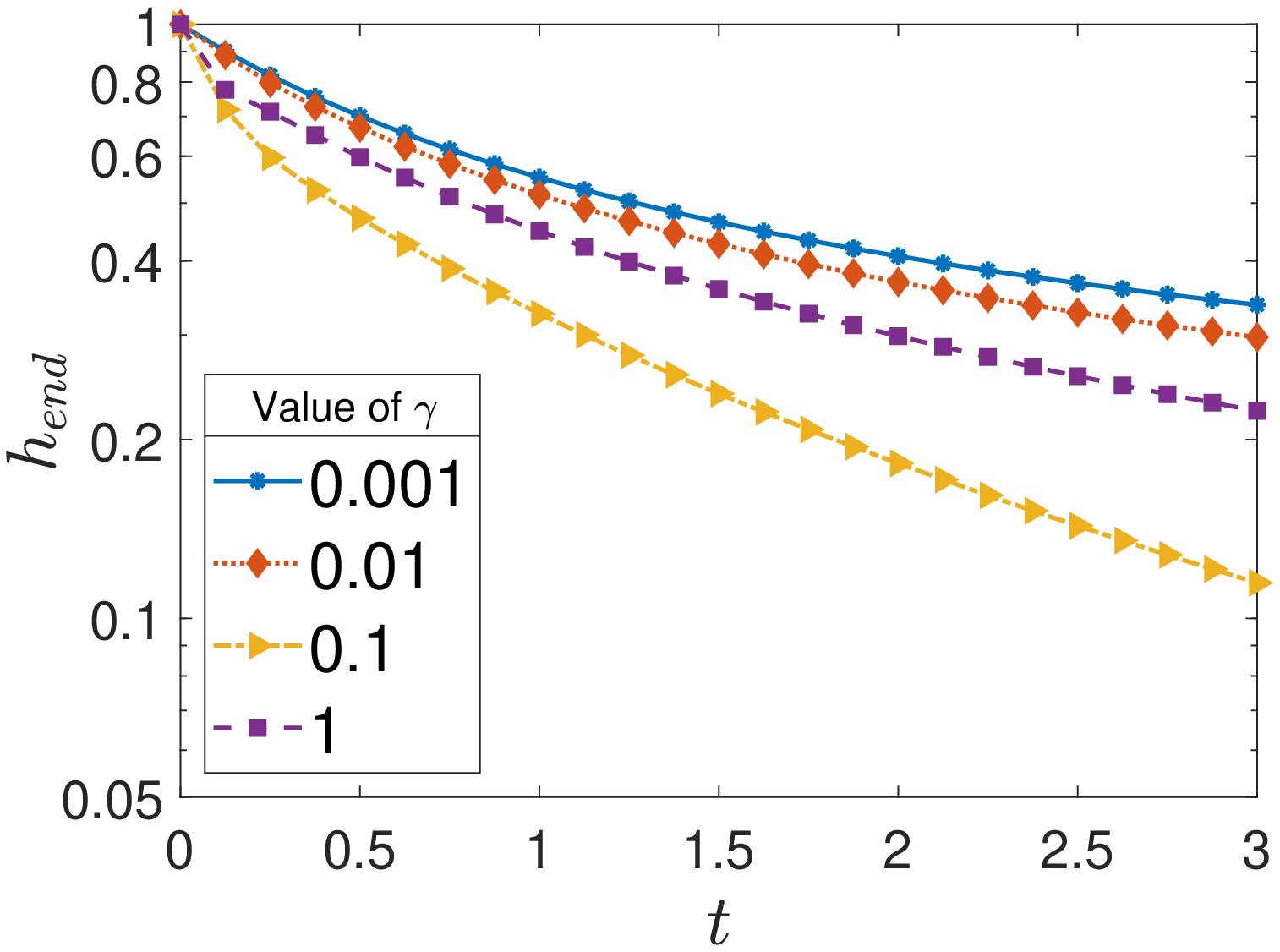}
\caption{\label{fig:one}Evolution of the minimum sheet thickness, $h_{min}$, and thickness $h_{end}$ at the right end of the sheet, as the surface tension varies for moderate elasticity. Results are for BC Case I with $v_0=1$.  In some cases, the minimum occurs at the right end. In the left plot, the curve for $\gamma = 0.001$ (in blue) lies directly under that of $\gamma=0.01$ (in red).} 
\end{figure}

We summarize the effect of the surface tension $\gamma$ on the sheet thickness for the moderate elasticity model in Fig.~\ref{fig:one}, where we compare a range of $\gamma$-values, spanning four orders of magnitude. 
The first plot of Fig.~\ref{fig:one} shows the minimum sheet thickness $h_{min}$ versus time $t$ on a semilog scale. The relationship between $h_{min}$ 
and $\gamma$ is not monotone; the largest values of $h_{min}$ for all values of time occur when surface tension 
is largest ($\gamma = 1$), while the smallest values occur at $\gamma=0.1$. Smaller values of $\gamma$ lead to intermediate minimum thickness values. The second plot of Fig.~\ref{fig:one} shows the film thickness at the right end of the sheet, $h_{end}=h(s(t),t)$, versus $t$, on a semilog scale. The minimum thickness may occur at the right end (see Fig.~\ref{fig:full_gam1}). 

\subsection{Varying the speed of the moving end}
\label{sec:speed}
\begin{figure}[htbp]
\includegraphics[width=.49\textwidth]{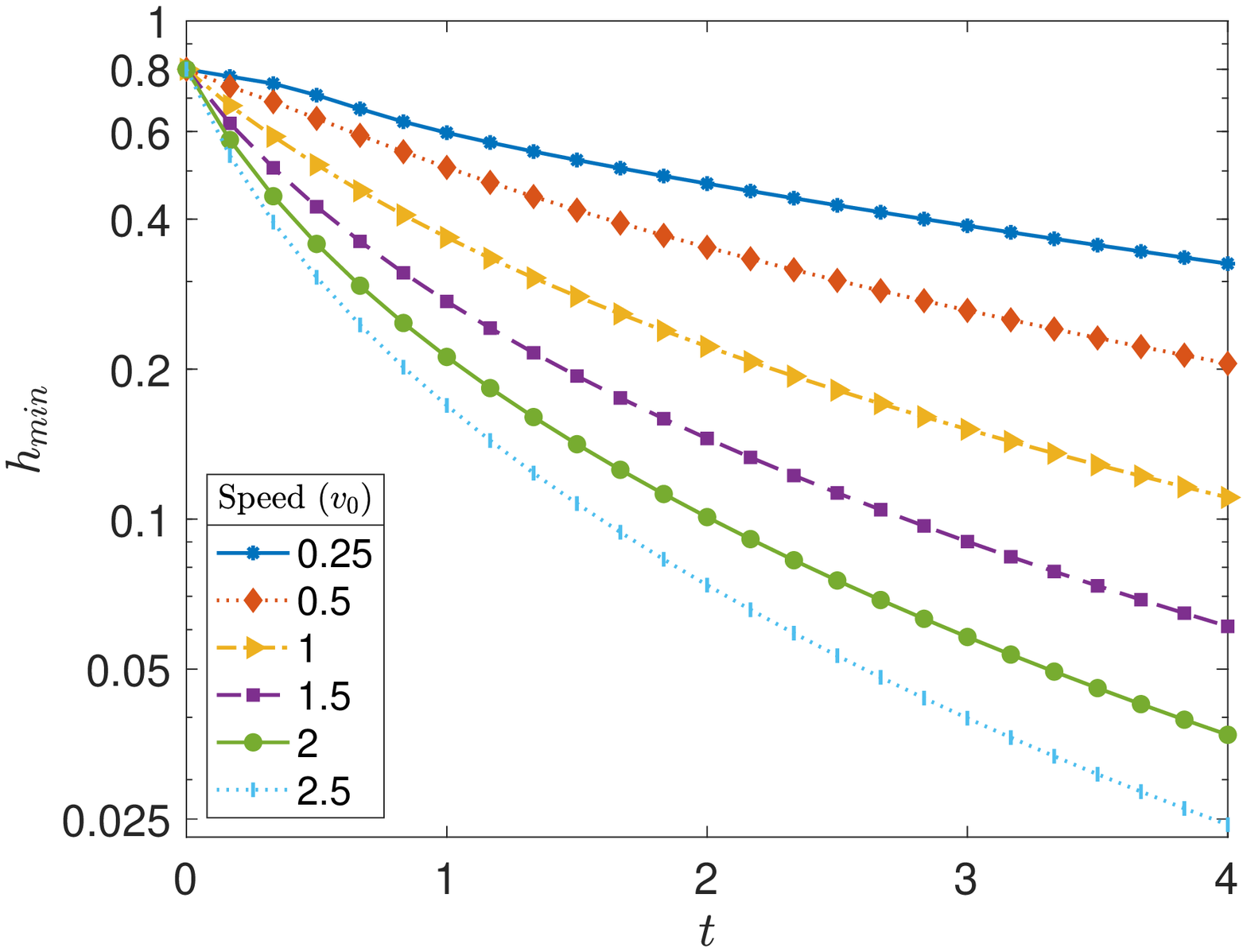}
\includegraphics[width=.49\textwidth]{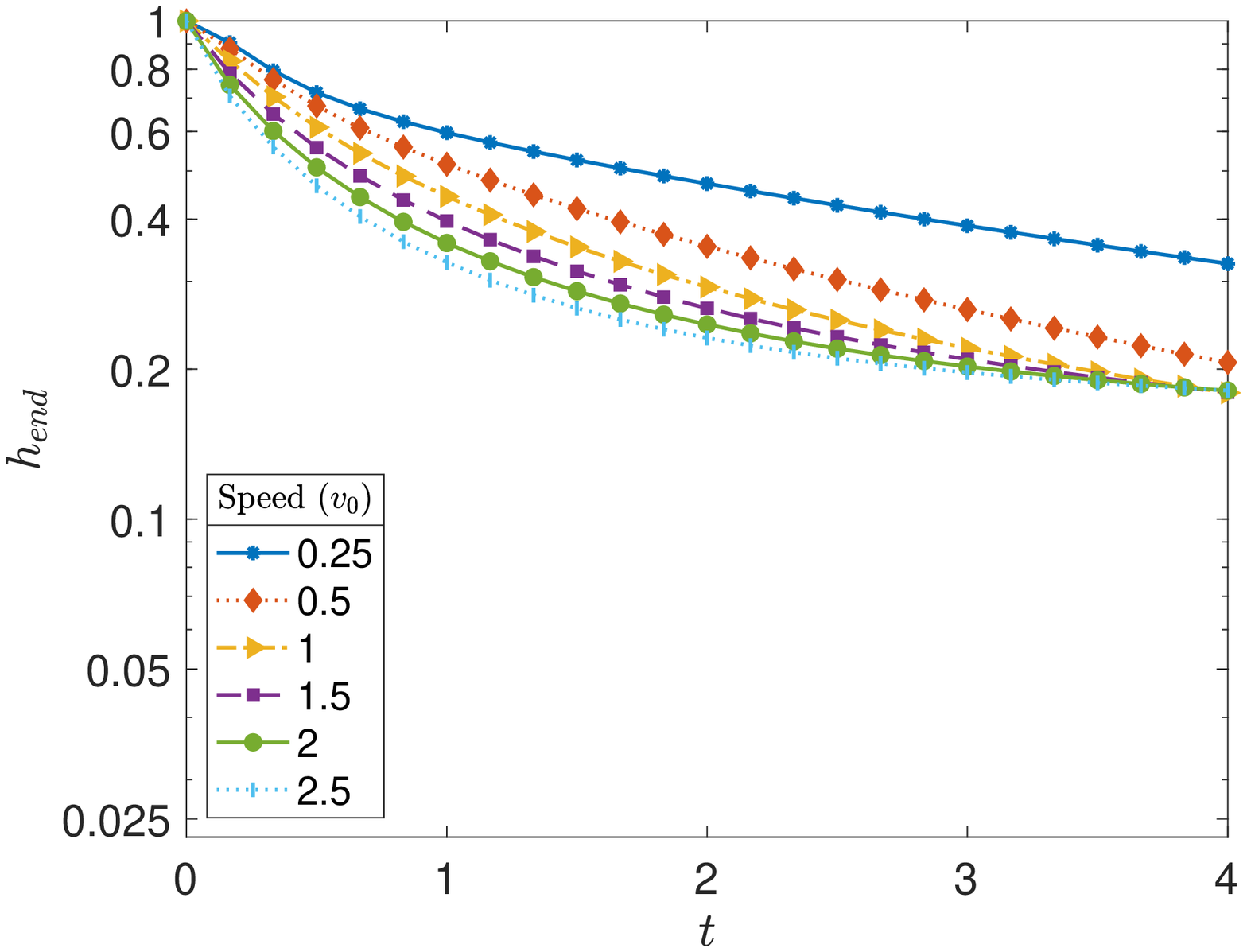}
  \caption{Evolution of the minimum sheet thickness, $h_{min}$, and height of the right end of the sheet, $h_{end}$, for moderate elasticity as the speed varies from 0.25 to 2.5. Results are for BC Case I with $\gamma=0.025$.  In some cases the minimum occurs at the right end.}
	\label{fig:two}
\end{figure}

\begin{figure}[ht]
		\includegraphics[width=.49\textwidth]{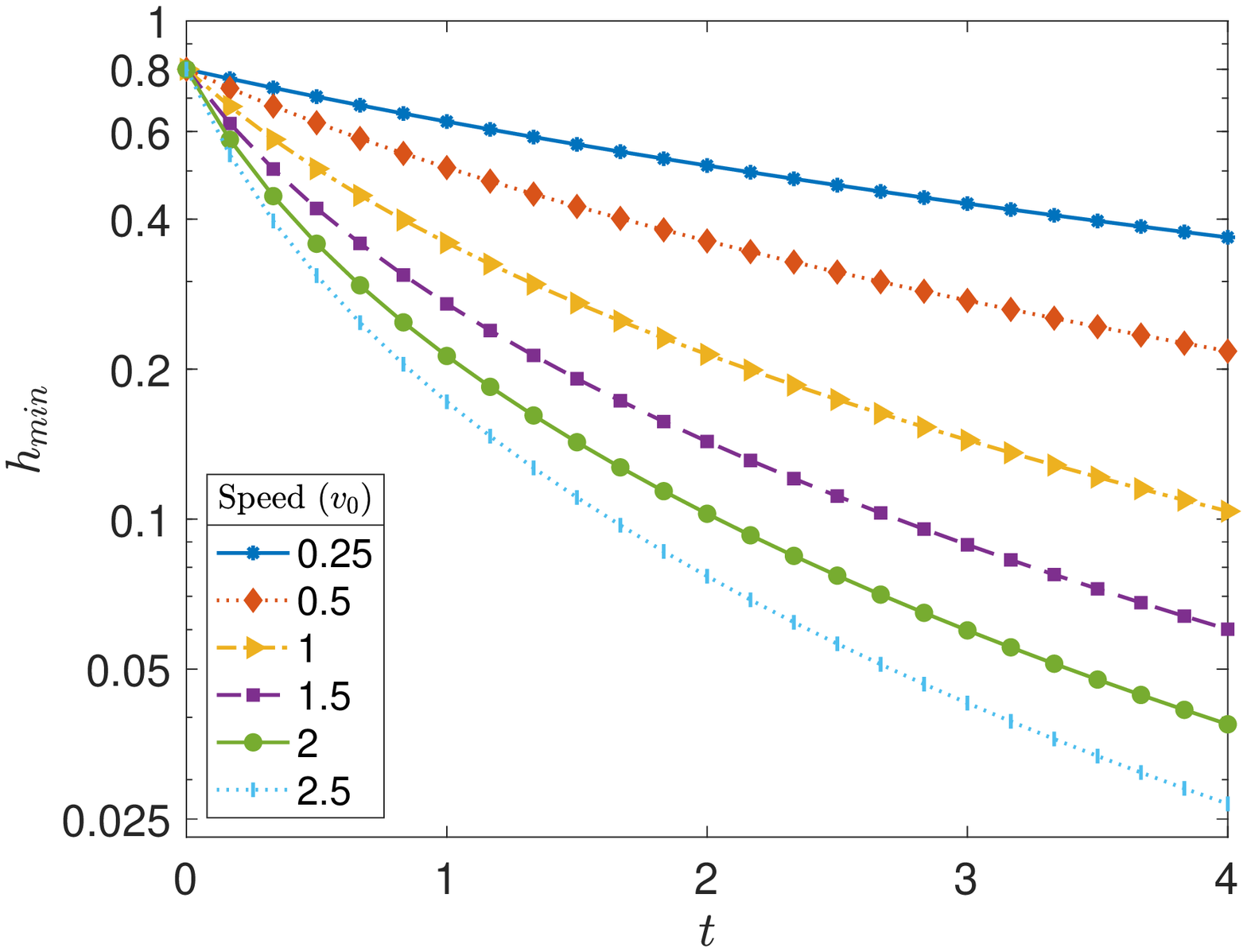}
		\includegraphics[width=.49\textwidth]{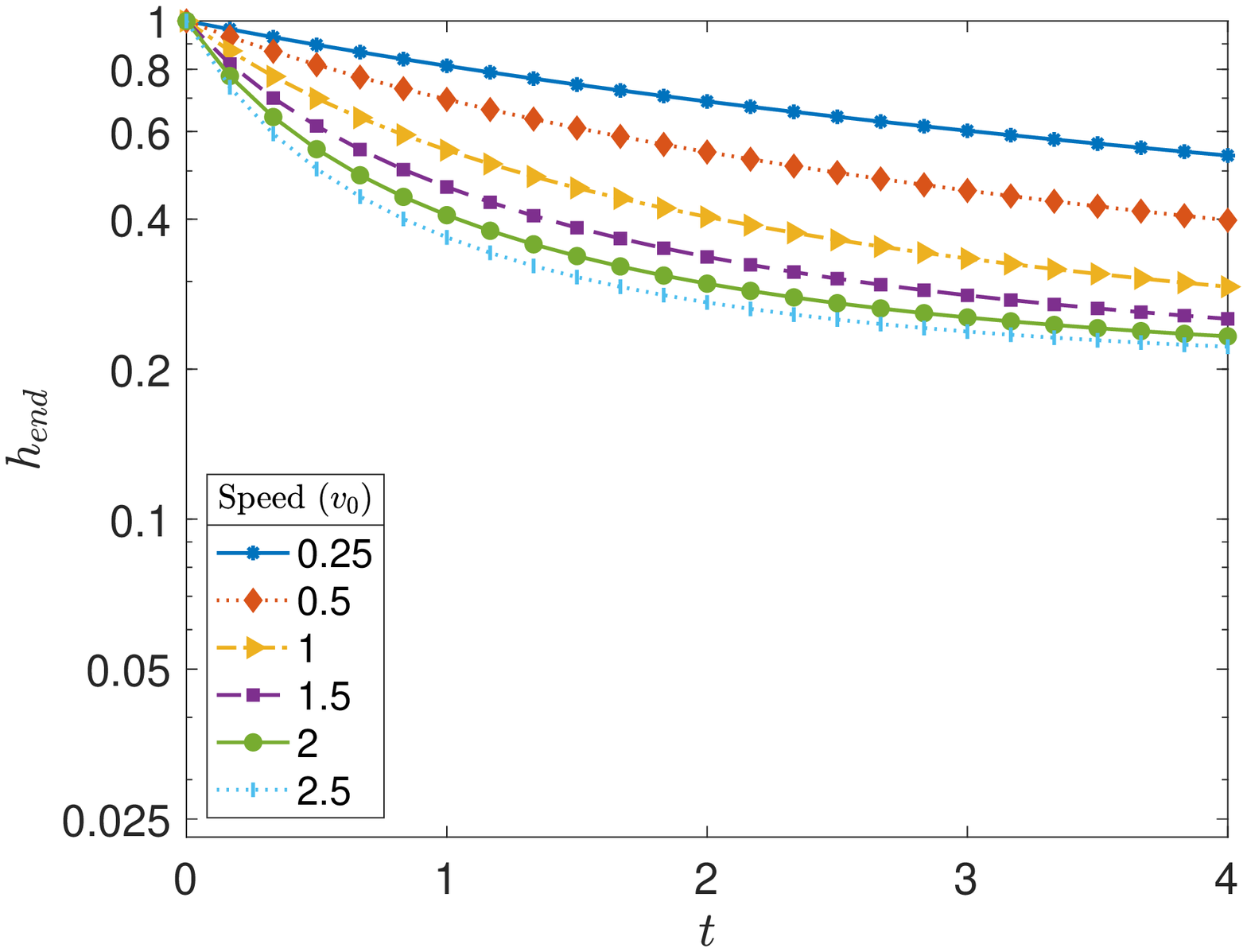}
		\caption{Evolution of the minimum sheet thickness, $h_{min}$, and height of the right end of the sheet, $h_{end}$, for weak elasticity as the speed varies from 0.25 to 2.5. Results are for BC Case I with $\gamma=0.025$.  In some cases the minimum occurs at the right end.}
	\label{fig:three}
\end{figure}

In the previous results, we varied surface tension $\gamma$, while fixing the speed of the moving end at $v_0=1$. Now we vary the speed, for fixed surface tension $\gamma=0.025$. Figs.~\ref{fig:two} and \ref{fig:three} show, for moderate and weak elasticity respectively, how the sheet thickness (as characterized by $h_{min}$ and $h_{end}$) is affected when $v_0$ varies from 0.25 to 2.5. The plots show $h_{min}$ and $h_{end}$ versus time $t$, on a semilog scale. 
Unsurprisingly, the faster the speed of the moving end, the thinner the sheet at its minimum, for all time points, and for both moderate and weak elasticity models. Comparing the minimum thickness in Figs.~\ref{fig:two} and \ref{fig:three}, the trend over time is remarkably similar, although $h_{min}$ is slightly lower for the case of moderate elasticity.
We note that as time progresses the thickness of the sheet at the moving end may merge or cross at around $h_{end} \approx 0.2$, even when varying the speed. 
 This contrasts with the weak elasticity case shown in Fig.~\ref{fig:three}: comparing the $h_{end}$ plots in Figs. \ref{fig:two} and \ref{fig:three}, we see that for weak elasticity, the moving end of the sheet continues to decrease in thickness as the speed increases for all points in time. This is another way in which the model with moderate elasticity differs from that with weak elasticity. 

\begin{figure}[htbp]
\centering
\includegraphics[width=.32\textwidth]{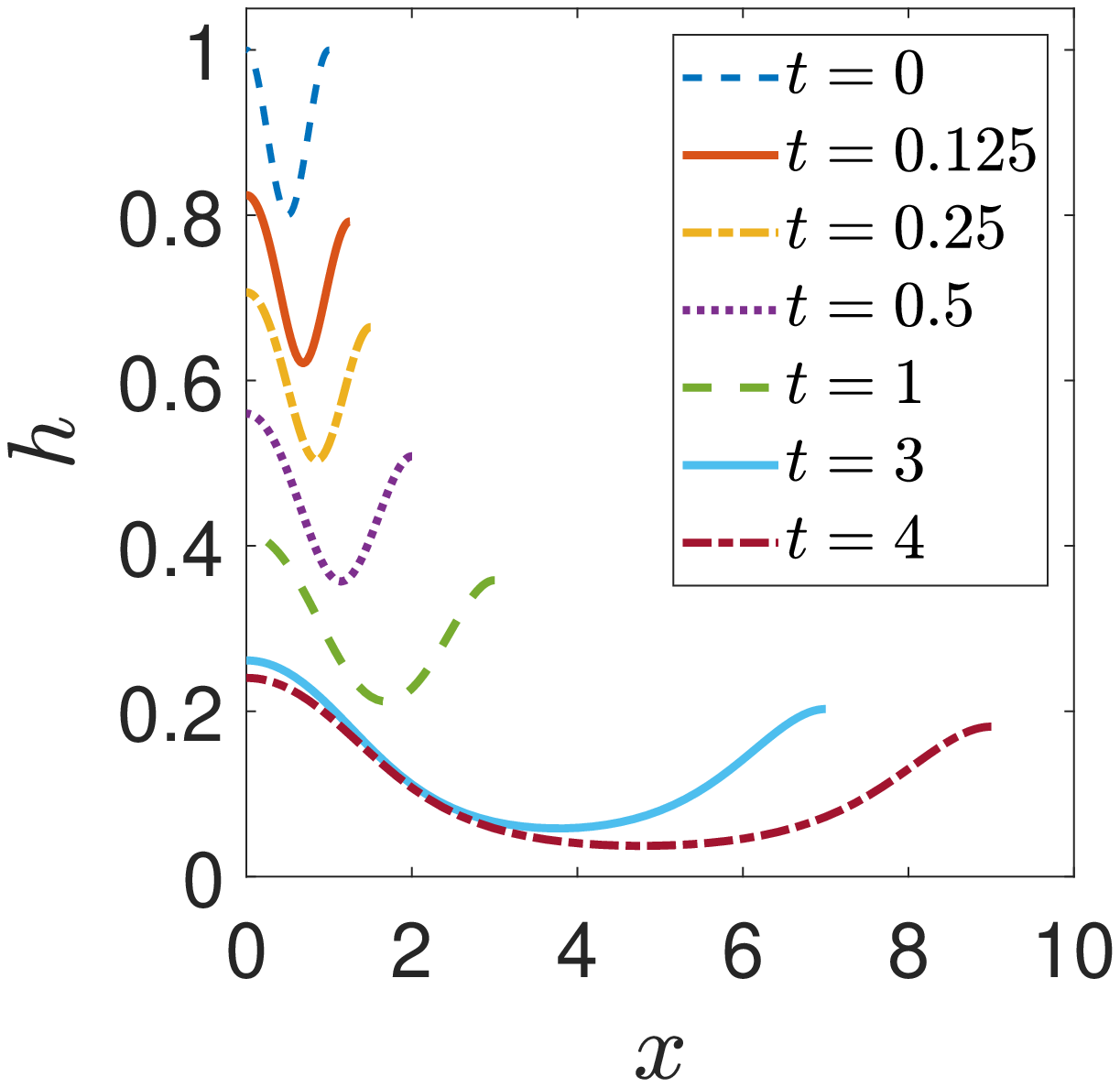}
\includegraphics[width=.32\textwidth]{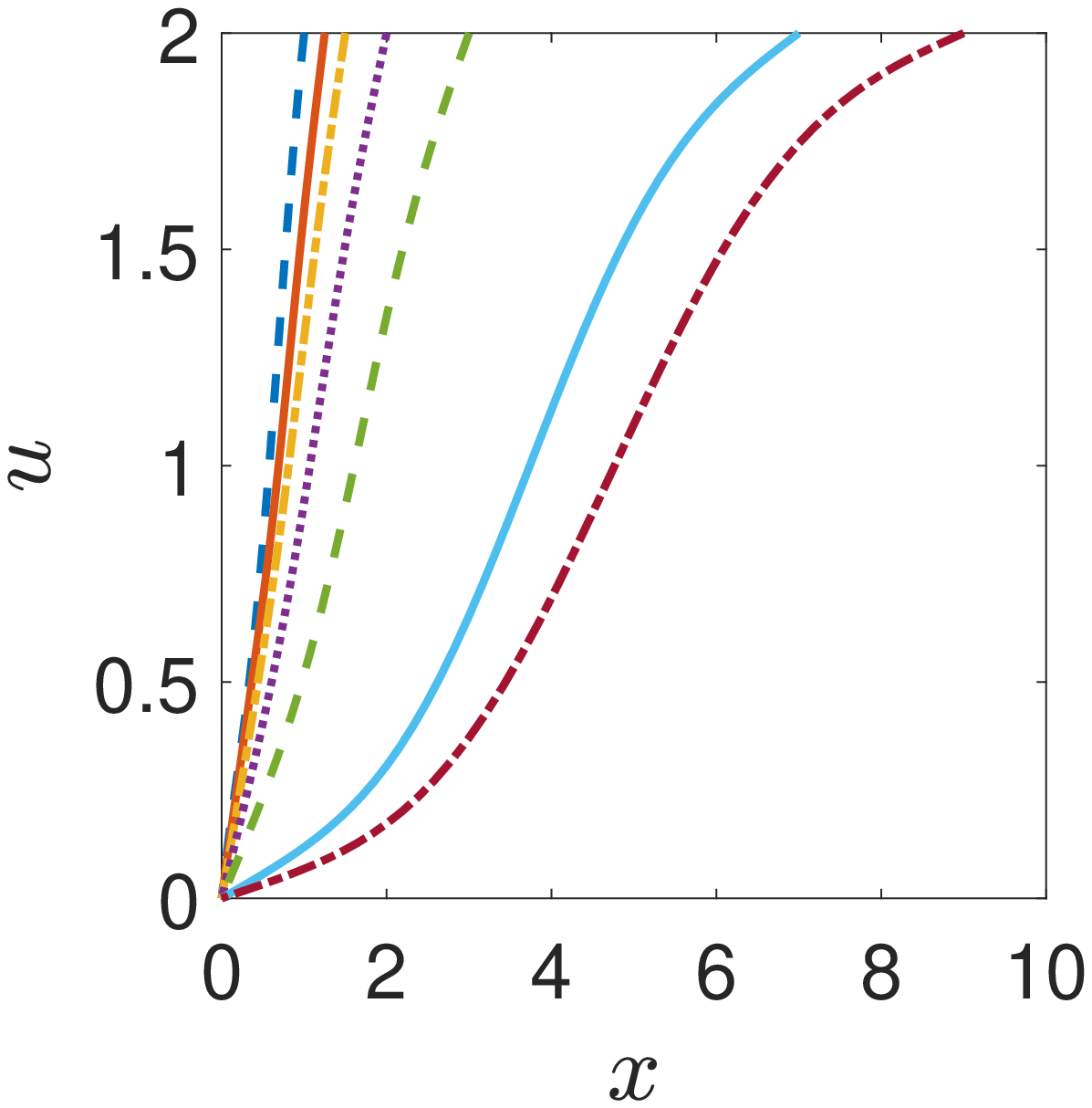}
\includegraphics[width=.32\textwidth]{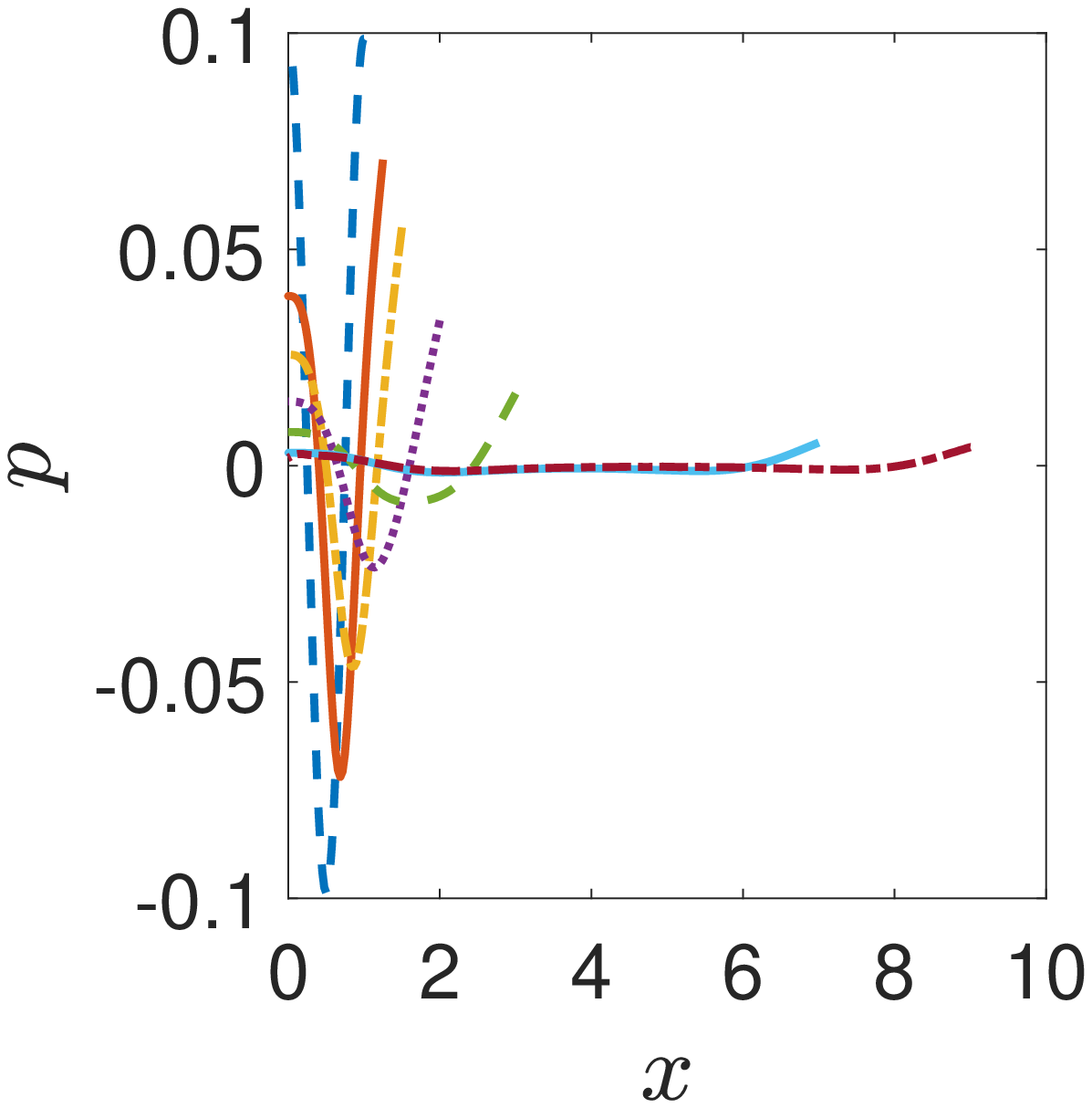}
\caption{Solutions for sheet thickness, $h$, fluid velocity, $u$, and pressure, $p$, for moderate elasticity. Results are for BC Case I with $\gamma=0.025$, and $v_0=2$.}
\label{fig:full_speed2}
\end{figure}

The moderate elasticity solution for the sheet thickness, axial velocity, and pressure corresponding to Fig.~\ref{fig:two} with $v_0=2$ are shown in Fig.~\ref{fig:full_speed2}. While the initial sheet profile is retained, qualitatively, under stretching, the right end is slightly thinner than the left.  The slower the speed of the moving end, the more of the original wave is lost as time progresses. 

\begin{figure}[htbp]
\centering
\includegraphics[width=.5\textwidth]{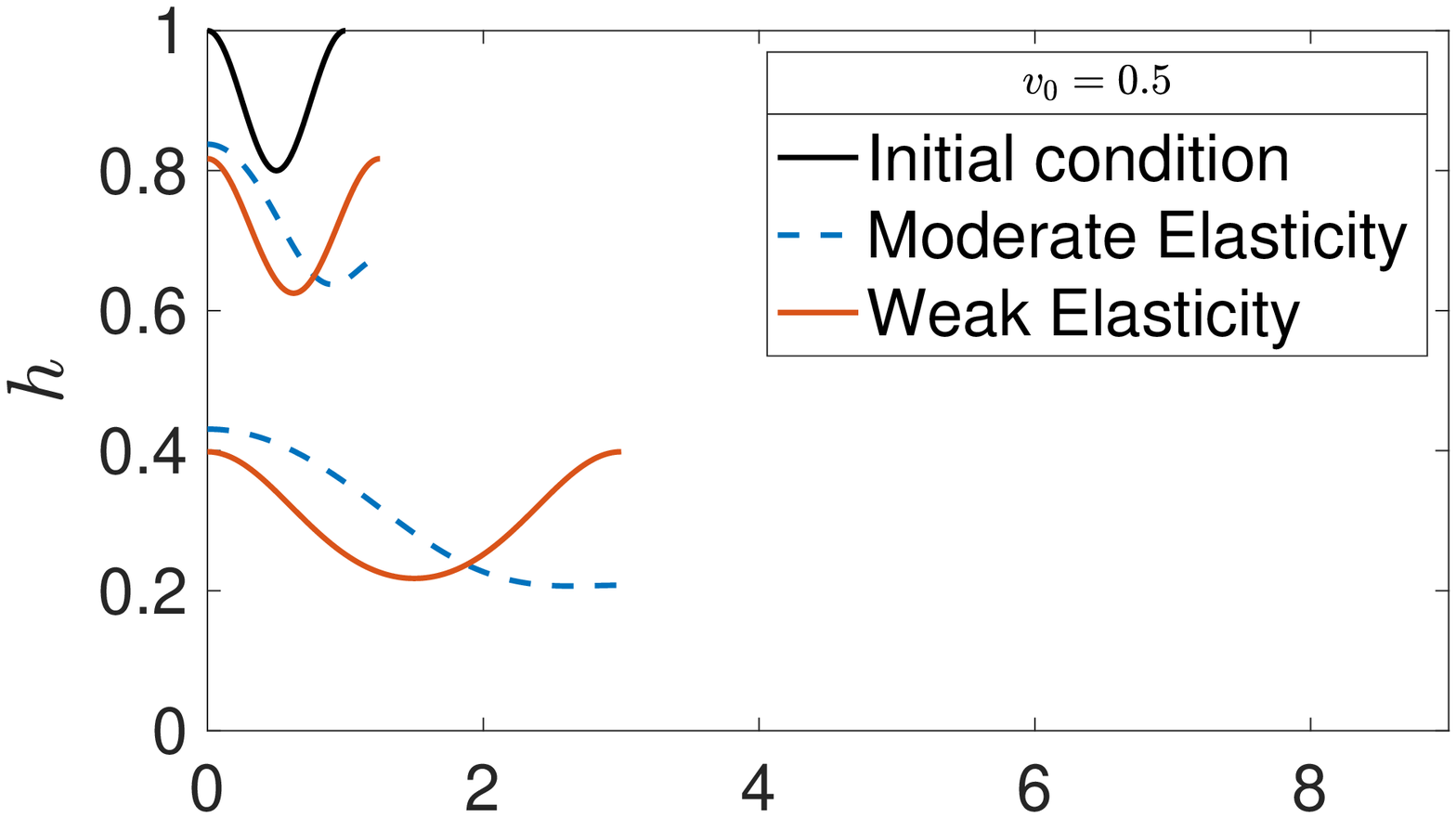}\\
\includegraphics[width=.5\textwidth]{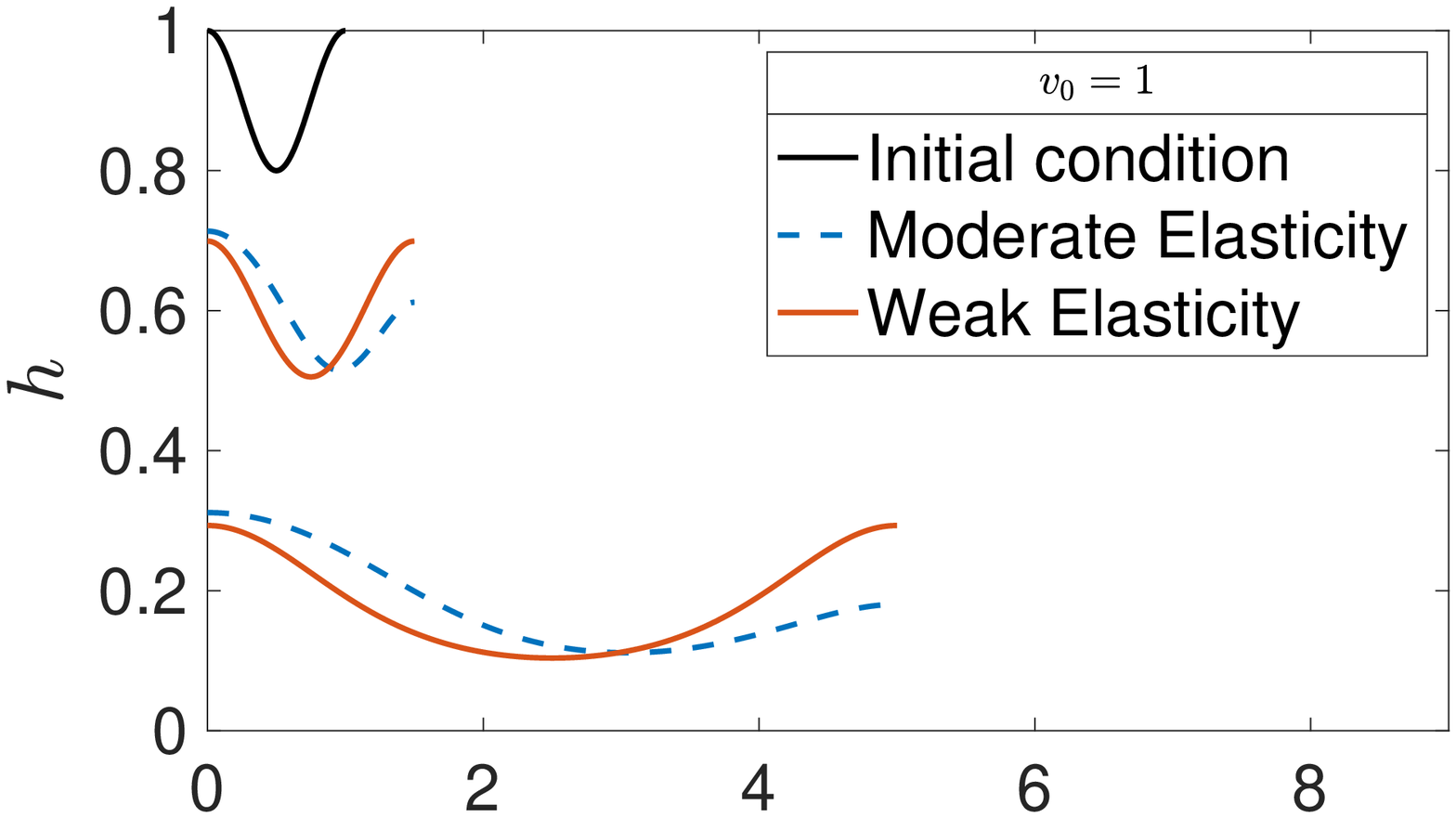}\\
\includegraphics[width=.5\textwidth]{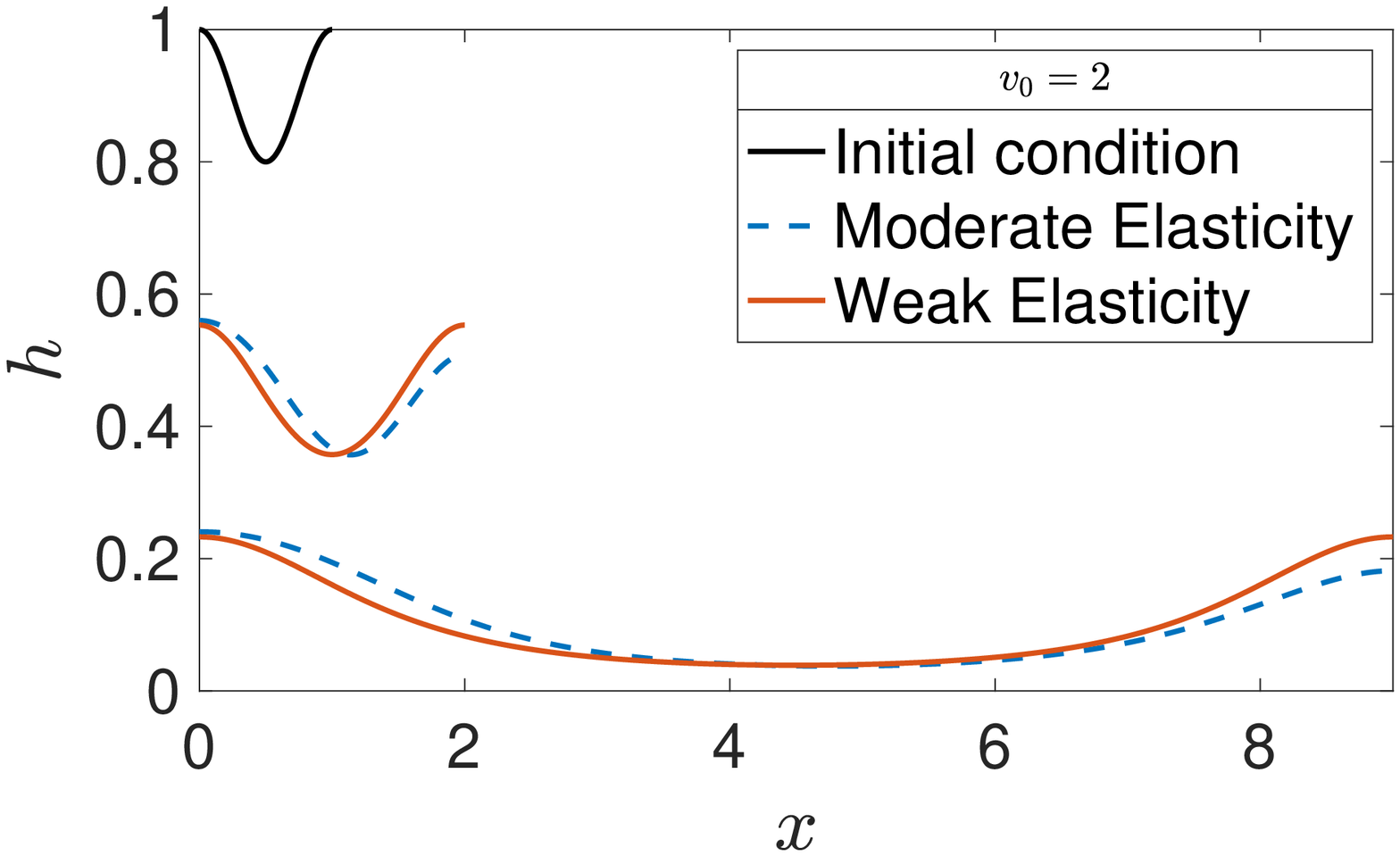}
\caption{\label{fig:speed_comp}Sheet thicknesses for moderate and weak elasticity models for increasing values of extension speed $v_0$, shown at times $t=0.5$ and $t=4$. The solid black curve shows the initial sheet profile for both models at $t=0$. Results are shown for BC Case I, with $\gamma = 0.025$.}
\end{figure}

Fig.~\ref{fig:speed_comp} compares sheet evolution in time, for the moderate and weak elasticity cases, for three different values of the sheet extension speed $v_0$ at times $t=0.5$ and $t=4$. At all speeds, the sheet with weak elasticity remains symmetric about its midpoint, and retains the wavenumber of the initial condition while being stretched over the increasing domain. This is not the case for moderate elasticity solutions. For the slowest extension speed $v_0=0.5$, the moving end of the sheet thins significantly, such that roughly half of the initial wave is lost by $t=0.5$, leading to very large differences between the weak and moderate elasticity predictions. When $v_0=1$, more of the original shape is retained, but the moving end still thins significantly relative to the left end; the prediction is again substantially different from the weak elasticity case. The differences between the two models are least pronounced for the fastest extension speed $v_0=2$. At both time points shown, the moderate elasticity model yields a sheet that is only slightly thicker over the left half than the right. The sheet thickness at the left end remains  very similar for the two models, but the moving end of the sheet with moderate elasticity is thinner. 

\subsection{Increasing wavenumber in ICs}
Imaging of the tear film has on occasion shown stripes or ridges in the lipid layer. \cite{braun2015dynamics} To investigate whether our model can sustain multiple waves during extensional flow, we experiment with increasing the wavenumber $k_0$ in the initial condition Eq.~(\ref{eq:general_h_ic}).
For all of the following results, we use the case of moderate elasticity with BC Case I and $v_0=1$. 

\begin{figure}[htbp]
\centering
\includegraphics[width=.5\textwidth]{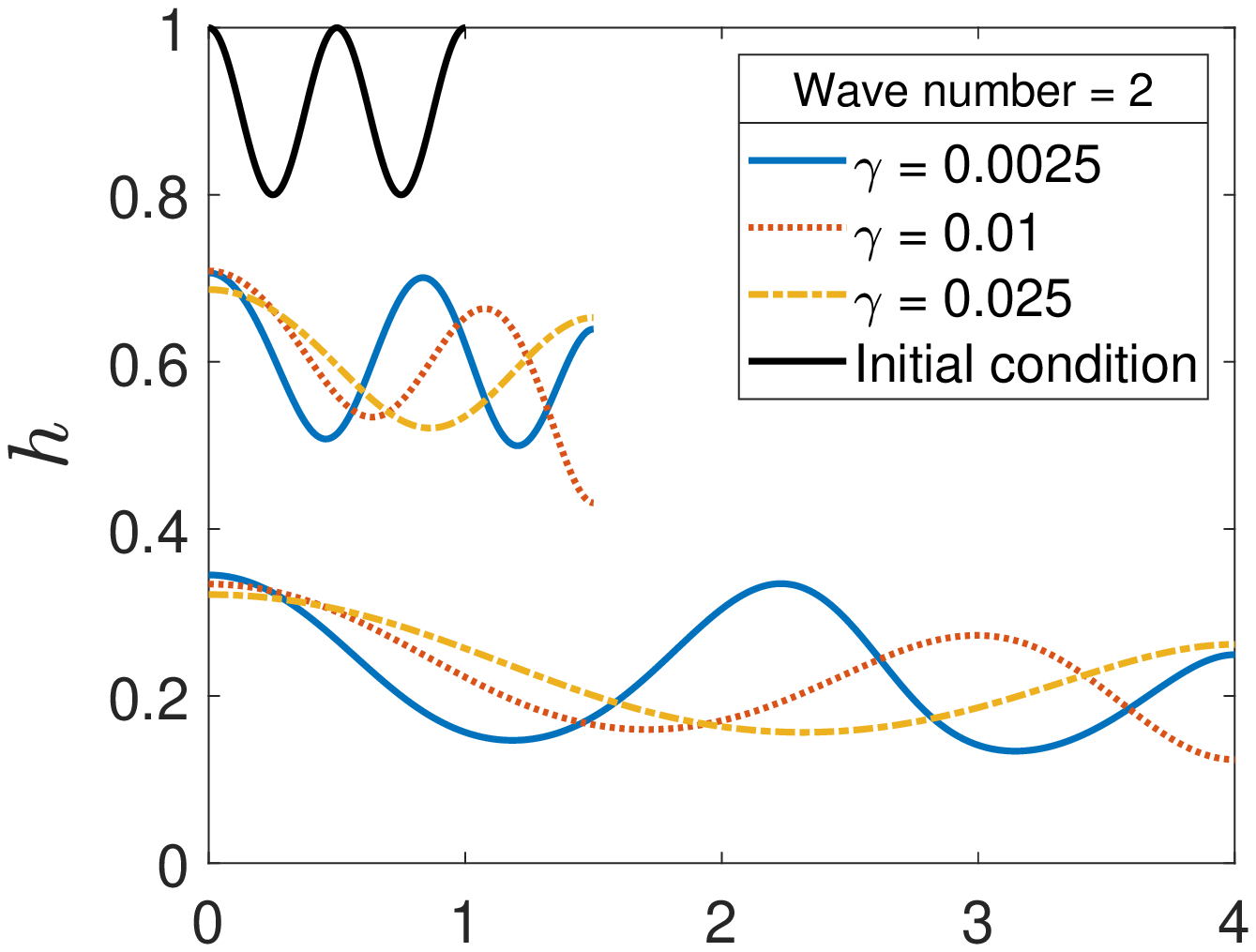}
\includegraphics[width=.5\textwidth]{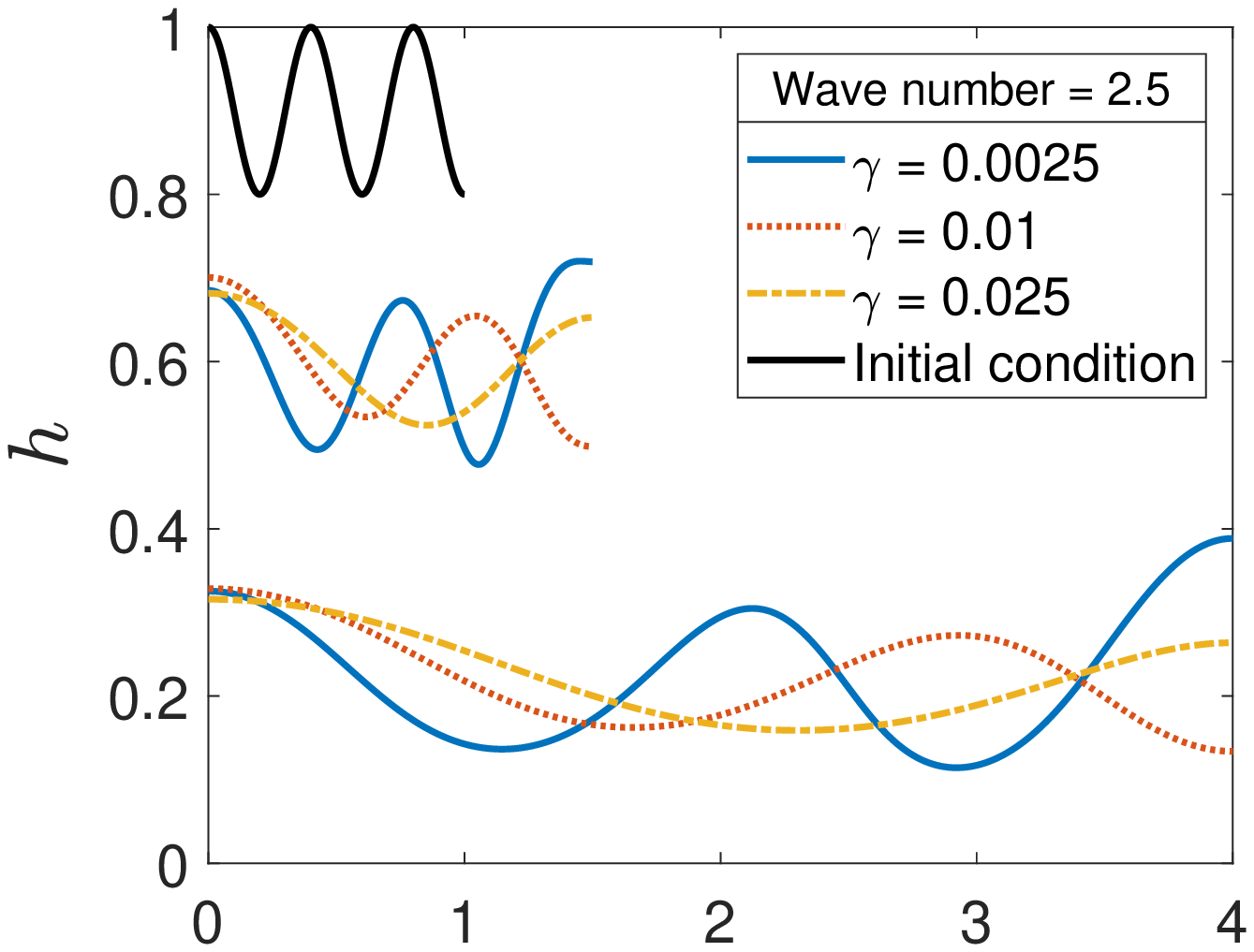}
\includegraphics[width=.5\textwidth]{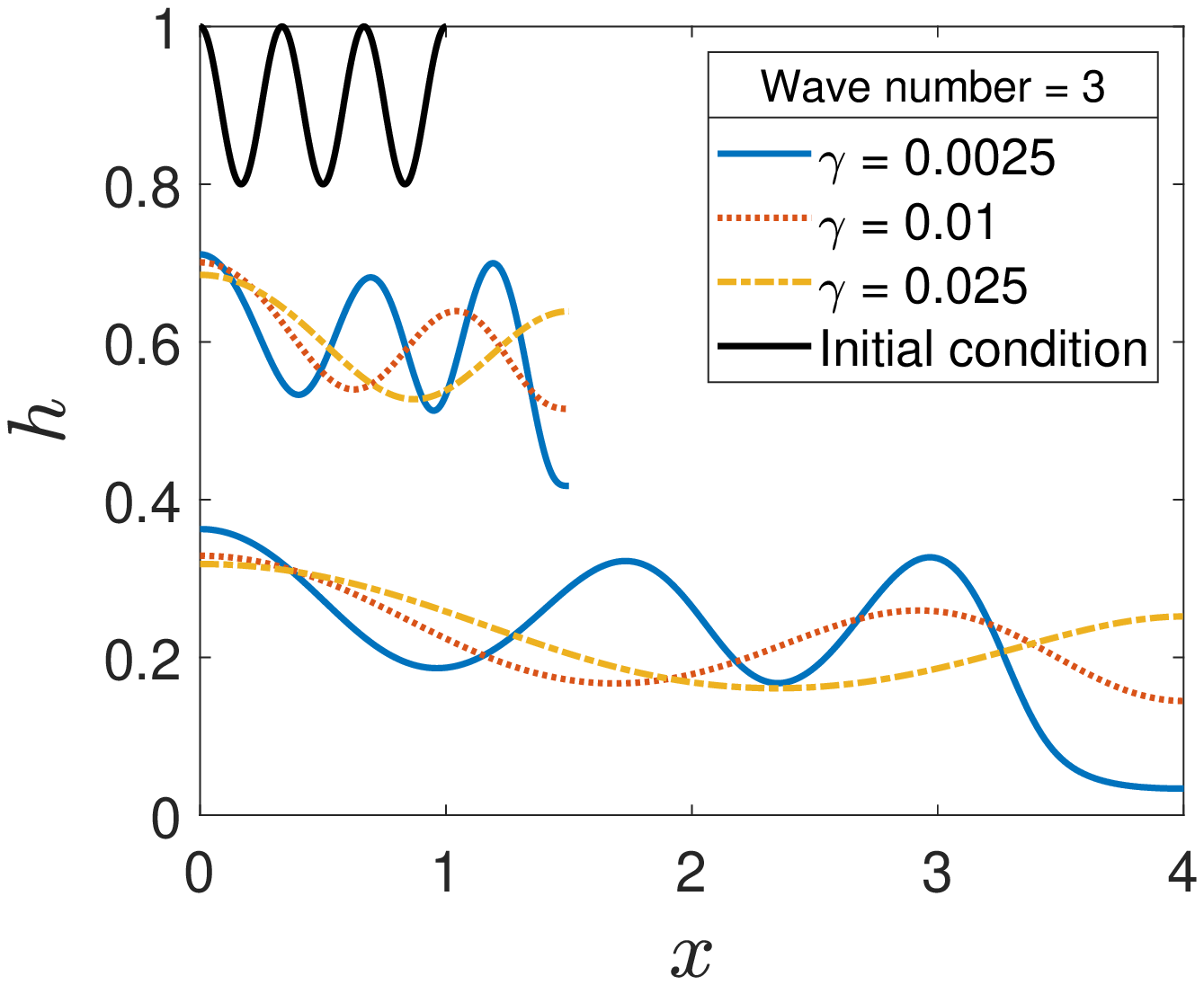}
\caption{\label{fig:ic_profiles}Profiles of sheet thickness, $h$, at $t=0$ (top curve in each plot), $t=0.5$ (middle curves), and $t=3$ (bottom curves) when the initial condition Eq.~(\ref{eq:general_h_ic}) has wavenumber $k_0=2$, 2.5, and 3 ($b=0.1, a=0.9$ in all cases). The higher the surface tension, the more waves are lost over time. Note that the shape of the sheet appears to be largely determined by $t=0.5$; subsequent evolution results in the extension of the sheet shape, but not the loss of more waves. }
\end{figure}

Fig.~\ref{fig:ic_profiles} shows the sheet solution profiles at $t=0.5$ and $t=3$ for three different values of the initial wavenumber $k_0$. For each IC, the sheet thickness is shown for three different values of the surface tension, $\gamma=0.0025,\ 0.01$, and $0.025$. The lower the surface tension, the more of the original waves are retained as time progresses. We note that the reduction of wavenumber appears to be complete by time $t=0.5$; after that, the resulting shape primarily stretches as the sheet lengthens (this point is discussed further below). In particular, in the first example with wavenumber $k_0=2$, both waves are retained for the smallest value of $\gamma$, while for $\gamma=0.01$ and $\gamma=0.025$, half a wave and a full wave (respectively) are lost from the initial shape by the final time. Similar differences are also apparent at higher wavenumbers: for $k_0=2.5$ the smallest surface tension simulation ($\gamma=0.0025$) loses just half a wave by the final time, while $\gamma=0.01$ loses a full wave and $\gamma=0.025$ loses 1.5 waves; and for $k_0=3$ the simulation for $\gamma=0.0025$ again loses just half a wave, while $\gamma=0.01$ loses 1.5 waves and $\gamma=0.025$ loses 2 full waves. 

We further investigate simulations for $\gamma=0.0025$, since this value leads to persistent waves in the sheet. For this value of $\gamma$ we vary the wave amplitude $b$ in the initial condition Eq.~(\ref{eq:general_h_ic}) and observe the change of wavenumber over time as the sheet is stretched (specifically, the number of complete waves that are lost); the results are summarized in Table~\ref{tab:ampvwave}. The top row of this table corresponds to the $\gamma=0.0025$ simulations of Fig.~\ref{fig:ic_profiles}. We see that the value of the initial wavenumber $k_0$ is more influential than the initial wave amplitude $b$. 

\begin{table}
\caption{\label{tab:ampvwave} Table entries show number of waves lost from initial condition Eq.~(\ref{eq:general_h_ic}) at $t=4$ as amplitude $b$ and wavenumber $k_0$ are varied. Here, $a=1-b$ in Eq.~(\ref{eq:general_h_ic}) and $\gamma=0.0025$.}
\begin{ruledtabular}
\begin{tabular}{lccccccc}
 & \multicolumn{7}{c}{wavenumber $k_0$} \\
 Amplitude $b$    & 1  & 1.5  & 2  & 2.5    & 3    & 3.5 & 4 \\
\hline
0.2   & 0  & 0  & 0  & 1/2  & 1/2  & -  & - \\
0.1   & 0  & 0  & 0  & 1/2  & 1/2  & 1  & 1 \\
0.05  & 0  & 0  & 0  & 1/2  & 1/2  & 1  & 1 \\
0.025 & 0  & 0  & 0  & 1/2  & 1/2  & 1  & 1 
\end{tabular}
\end{ruledtabular}
\end{table}

\begin{figure}[htbp]
\centering
\includegraphics[width=.45\textwidth]{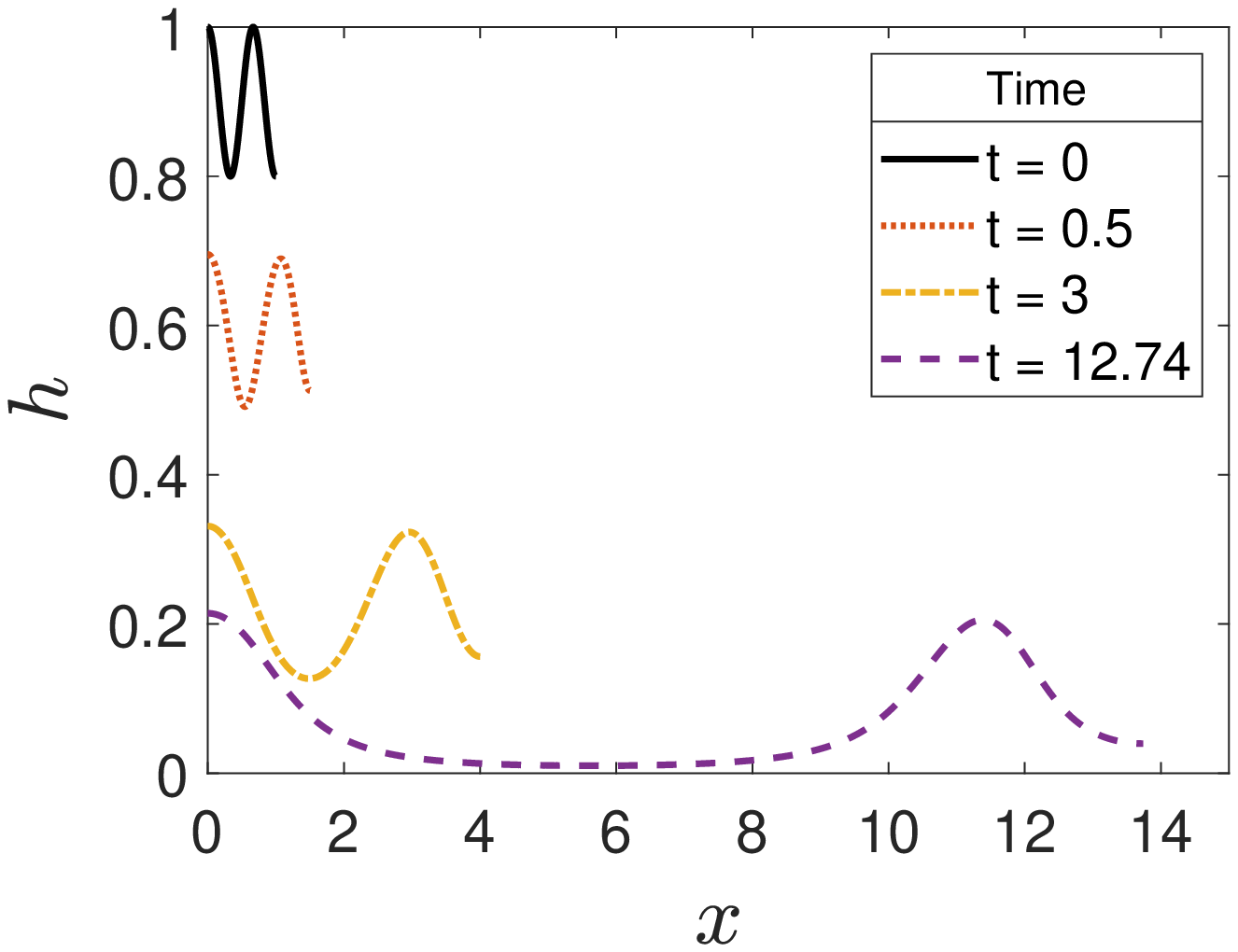}
\includegraphics[width=.45\textwidth]{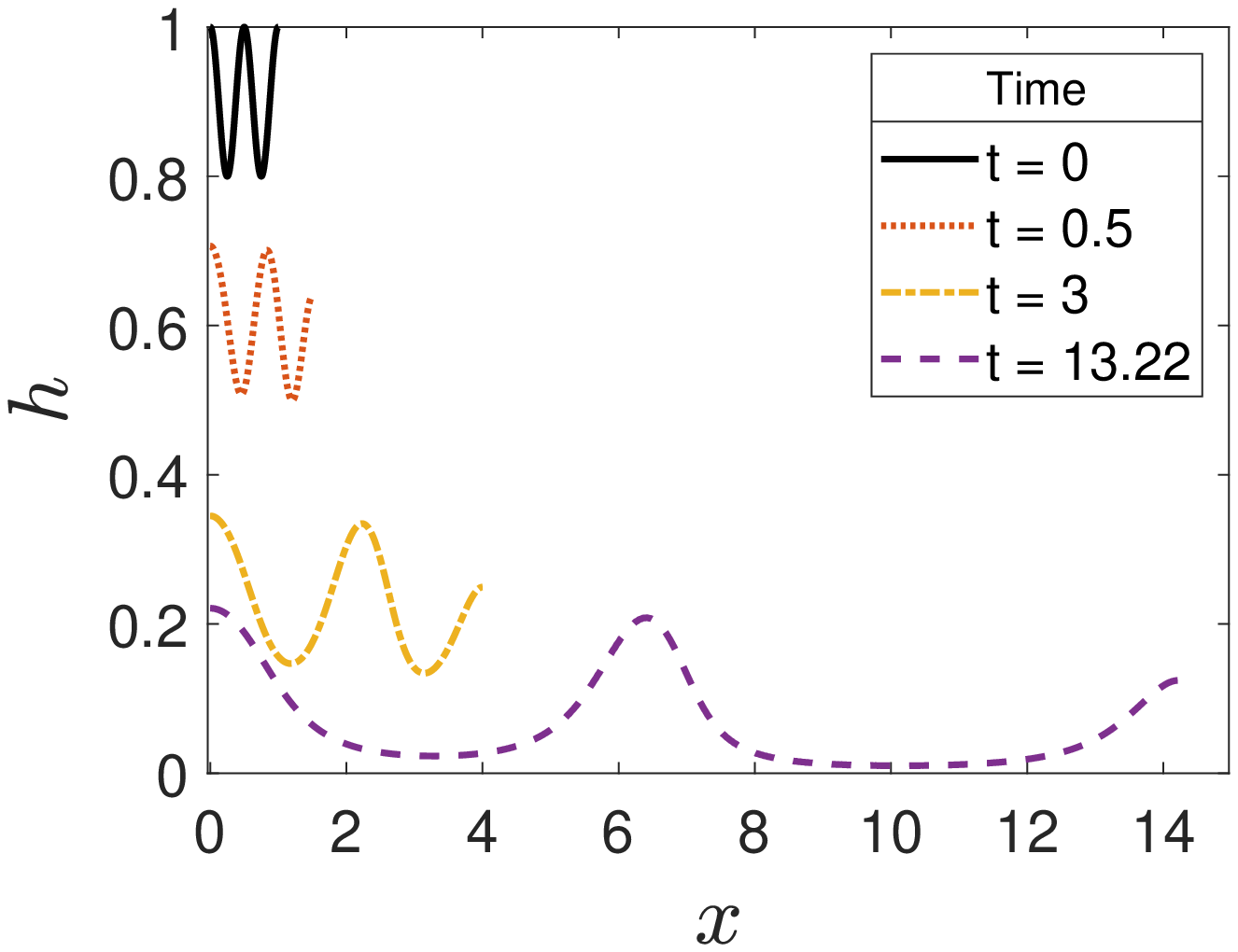}
\includegraphics[width=.45\textwidth]{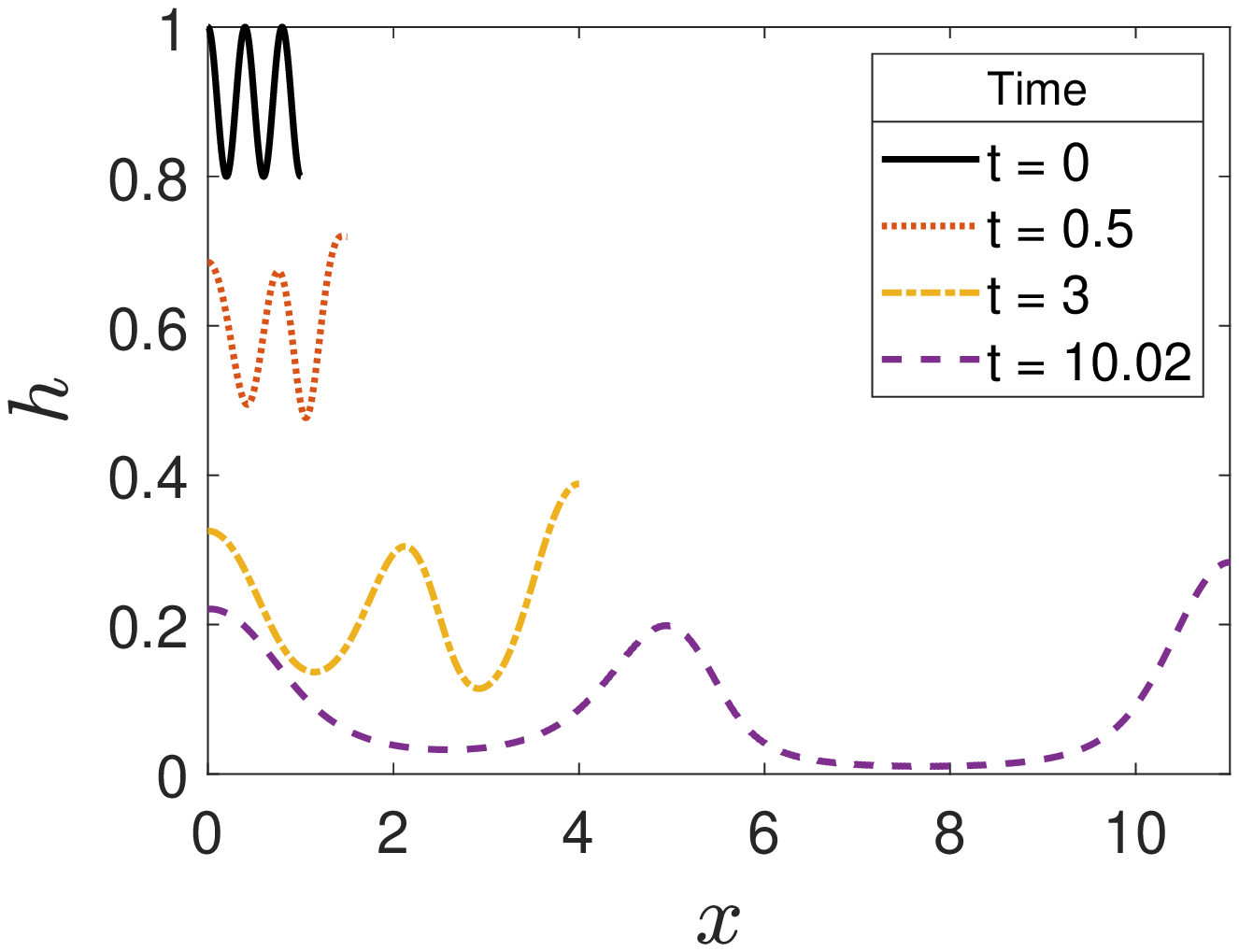}
\includegraphics[width=.45\textwidth]{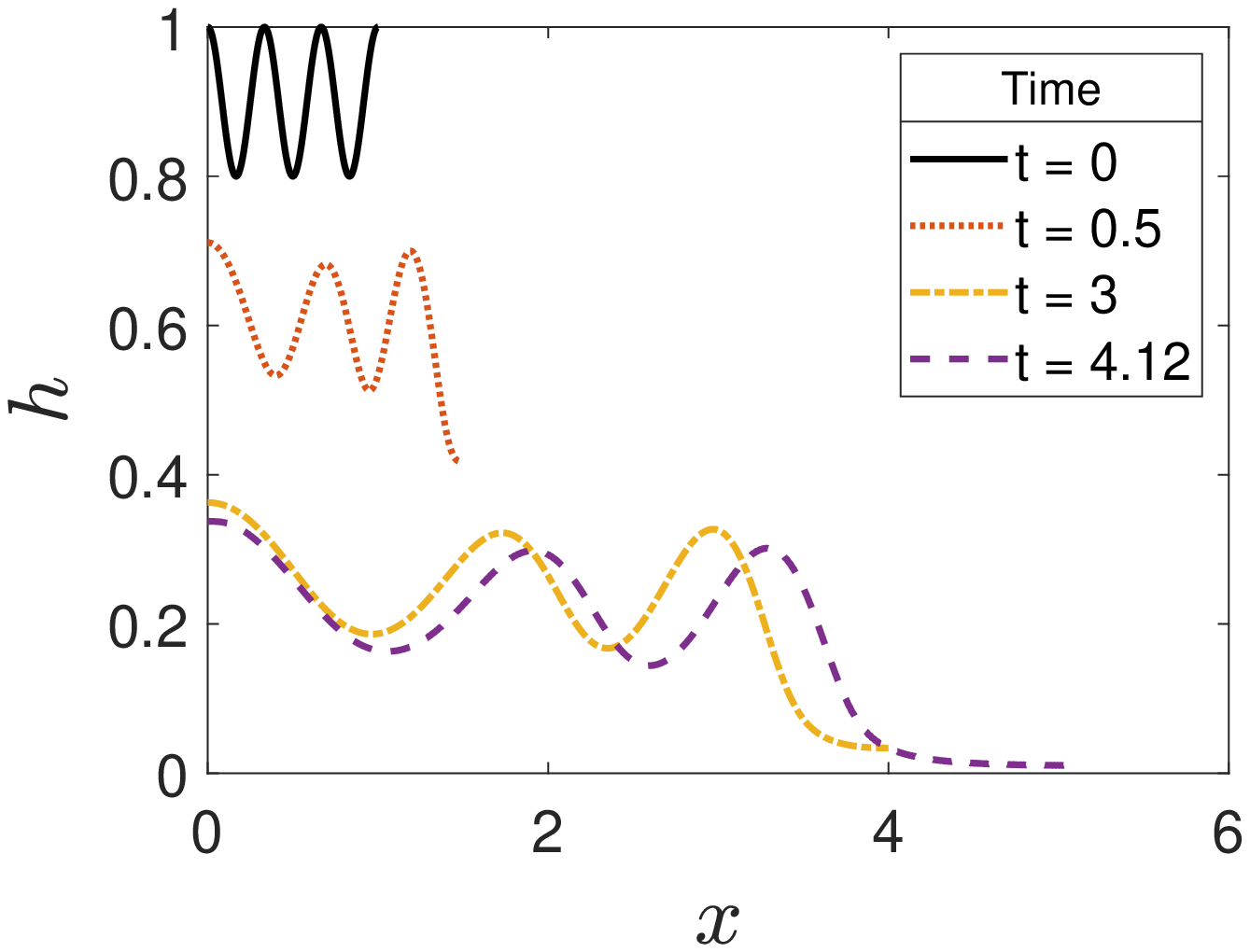}
\caption{Sheet thickness, $h$, at $t=0$ (top curve in each plot), $t=0.5$ (second curve), and $t=3$ (third curve) and the time to reach the threshold thickness of $h<0.01$ (bottom curve) for initial condition Eq.~(\ref{eq:general_h_ic}) with $a=0.9, b=0.1$ and wavenumbers $k_0= 1.5,\,2, \, 2.5$, and 3 with $v_0=1$.}
\label{fig:final_profile}
\end{figure}

We also test our earlier assertion, that the reduction in wavenumber appears to be determined at an early stage of the stretching, by running simulations to larger times.
We used the event detection option in \textsc{Matlab} and let the sheet stretch until $h_{min}<0.01$ (assumed to represent sheet breakup in the model). The results are summarized in Table~\ref{tab:event_detect}, which records the IC used in the simulation, the time to breakup, the number of waves lost from the IC during evolution, and whether the final extremum of sheet thickness at the moving end is a maximum or minimum. In each case, the sheet reached this minimum thickness threshold before any noticeable change in shape from that noted at $t=0.5$. When the moving end of the sheet is (or evolves to) a local minimum, the sheet ``breaks'' faster than when the moving end is a local maximum (sheet contains an integer number of full waves). 
For example, the two ICs with $k_0=2.5$ and $k_0=3$ both lose half a wave under stretching.  The curve resulting from $k_0=2.5$  develops a local maximum at the right end, and can stretch for more than twice the time for $k_0=3$, which develops a local minimum there. Fig.~\ref{fig:final_profile} shows the sheet profiles at the time that the thickness reaches the threshold of $h<0.01$ for four initial conditions. Interestingly, although the sheet profiles that have a minimum at the moving end always appear to break first, the breakup does not always appear at the moving end, but may happen at an interior minimum.

\begin{table*}
\caption{\label{tab:event_detect}Comparison of the time to reach $h<0.01$, which represents sheet breakup, for various initial conditions when $\gamma=0.0025$. Here $a=0.9$ and $b=0.1$. }
\begin{ruledtabular}
\begin{tabular}{lccc}
  Initial condition                 & Time to $h<0.01$ & Waves lost & Final extremum at right\\ \hline
$a+b \cos(3\pi x)$   & 12.9935   &None&     minimum\\
$a+b \cos(4\pi x)$   & 13.4664   &None& maximum    \\
$a+b \cos(5\pi x)$   & 10.2892   &1/2& maximum     \\
$a+b \cos(6\pi x)$  & 4.2099     &1/2& minimum    \\
$a+b/2 \cos(5\pi x)$ & 18.2691   &1/2& maximum     \\
$a+b/2 \cos(6\pi x)$ & 9.5259  &1/2&minimum
\end{tabular}
\end{ruledtabular}
\end{table*}

\subsection{Mechanisms}

\begin{figure}[htbp]
\centering
\includegraphics[width=.32\textwidth]{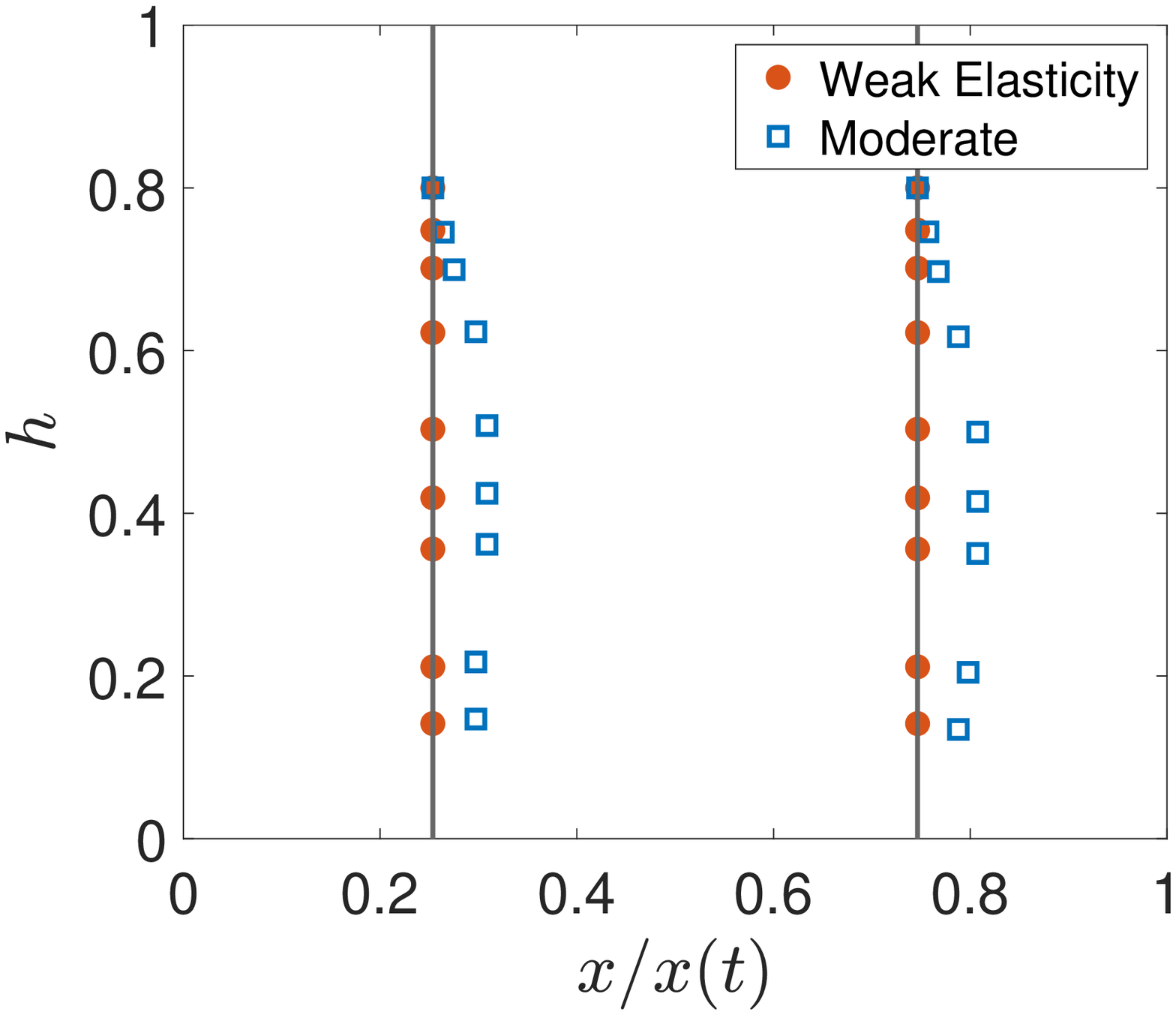}
\includegraphics[width=.32\textwidth]{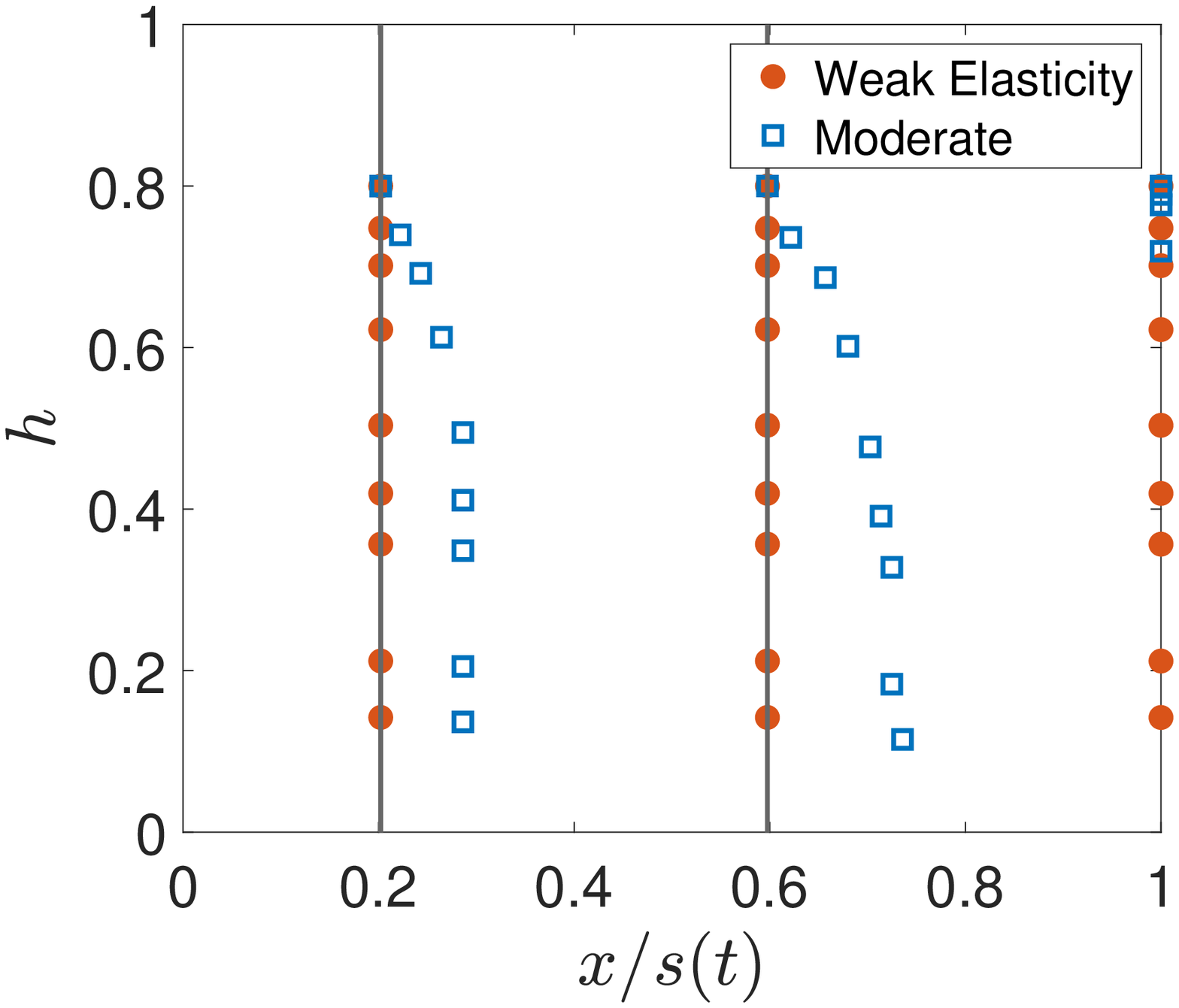}
\includegraphics[width=.32\textwidth]{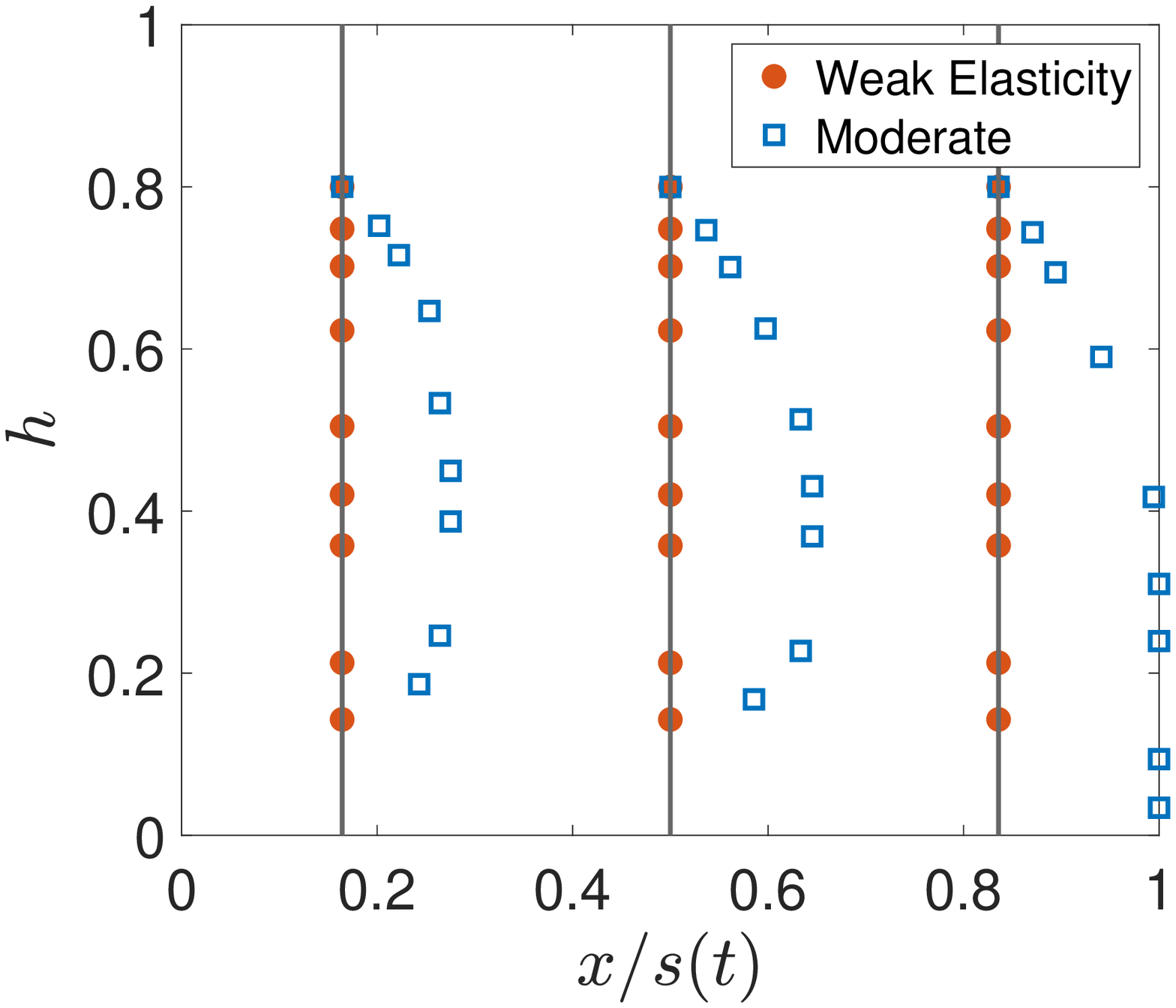}
\caption{\label{fig:minloc_comp}Location of local minima in thickness, for weak and moderate elasticity models, at $t = 0,\,0.0625,\,0.125,\, 0.25 ,\,0.5 ,\,0.75,\, 1,\, 2 ,\,3$ on a fixed domain with $\gamma=0.0025, \,v_0=1$. Each plot corresponds to a different wavenumber: $k_0 =2,\, 2.5,\, 3$ in Eq.~(\ref{eq:general_h_ic}) (with $a=0.9,b=0.1$). The sheets thin in time, so lower points correspond to later times. The straight black lines emphasize that locations of thickness minima are stationary for weak elasticity on a fixed domain. For moderate elasticity with $k_0=3$, the local minimum on the right travels to the moving end and becomes the global minimum.} 
\end{figure}

For weak elasticity, the oscillations contained in the initial condition are retained in the sheet throughout time, and are stretched as the sheet lengthens. The sheet retains any symmetry in the initial condition, and the locations of minimum and maximum thickness are unchanged through time when plotted in terms of the coordinate $\xi=x/s(t)$; see Fig.~\ref{fig:minloc_comp}. If we compare the individual terms of the PDE, as shown in Fig.~\ref{fig:allterms_newt4} (where only the right half of the domain is shown), we see that it is primarily the extensional terms from $(hu_x)_x$ that balance; the role of surface tension is minor. The velocity profile is nearly linear with small fluctuations in the slope. Pressure remains negative through the entire sheet, as extension is dominating capillarity, and decreases in magnitude as the sheet lengthens. 

\begin{figure}[htbp]
\centering
\includegraphics[width=.32\textwidth]{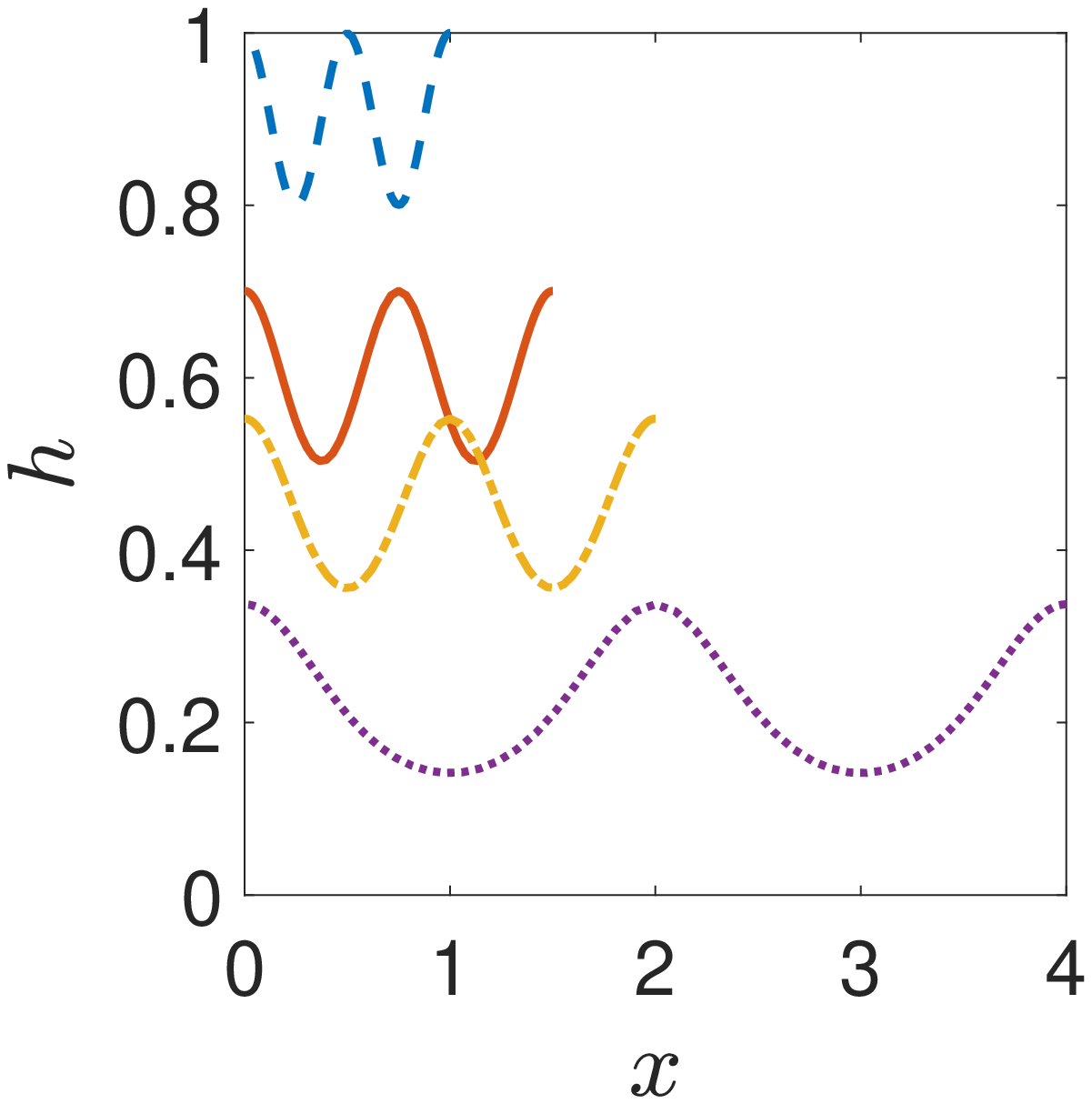}
\includegraphics[width=.32\textwidth]{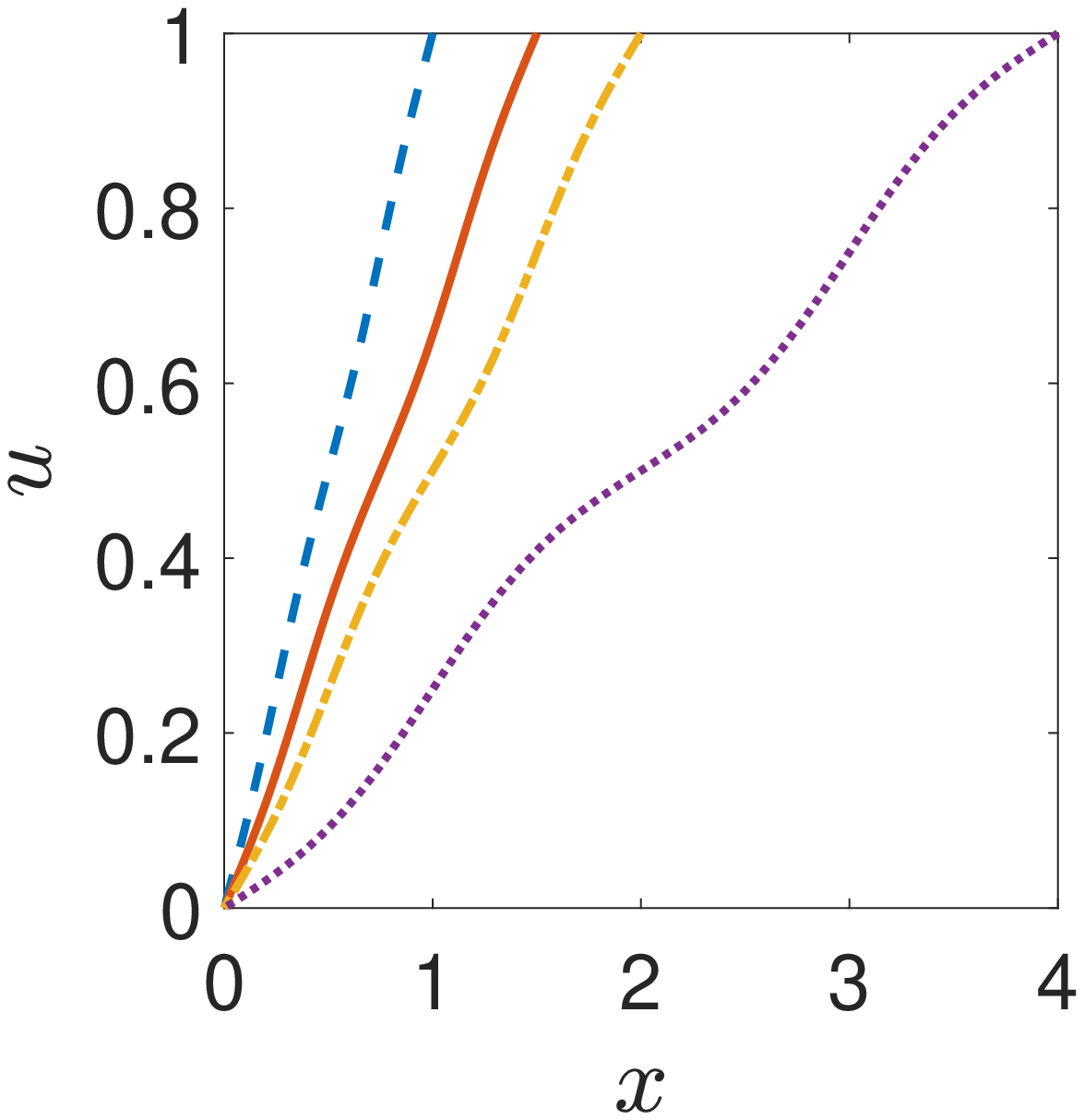}
\includegraphics[width=.32\textwidth]{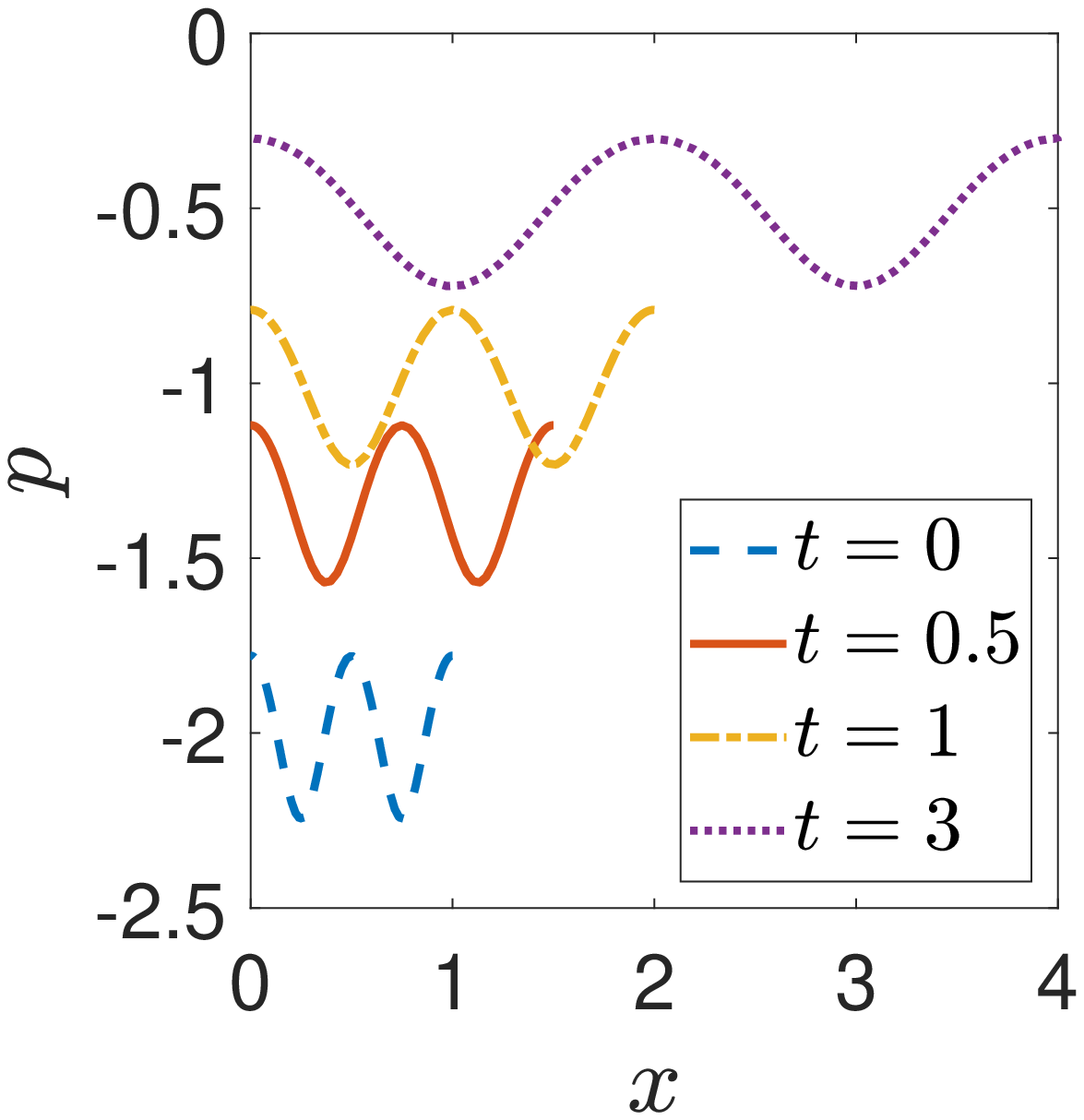}
\caption{Profiles of sheet thickness, $h$, fluid velocity, $u$, and pressure, $p$, for the case of weak elasticity when $k_0=2$. Results are for BC Case I with $\gamma=0.0025$ and $v_0=1$.}
\label{fig:fullprofile_newt4}
\end{figure}

\begin{figure}[htbp]
\centering
\includegraphics[width=.24\textwidth]{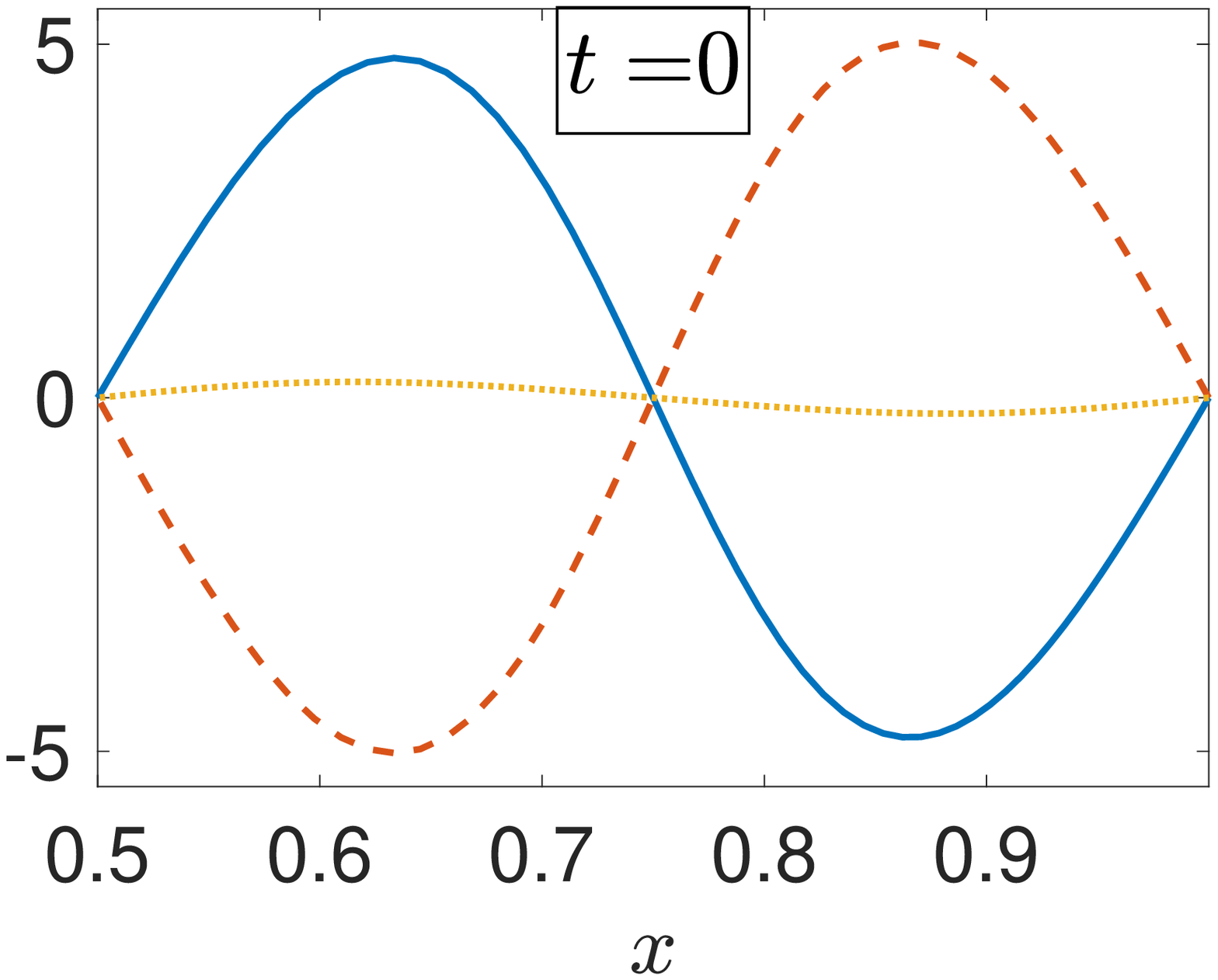}
\includegraphics[width=.24\textwidth]{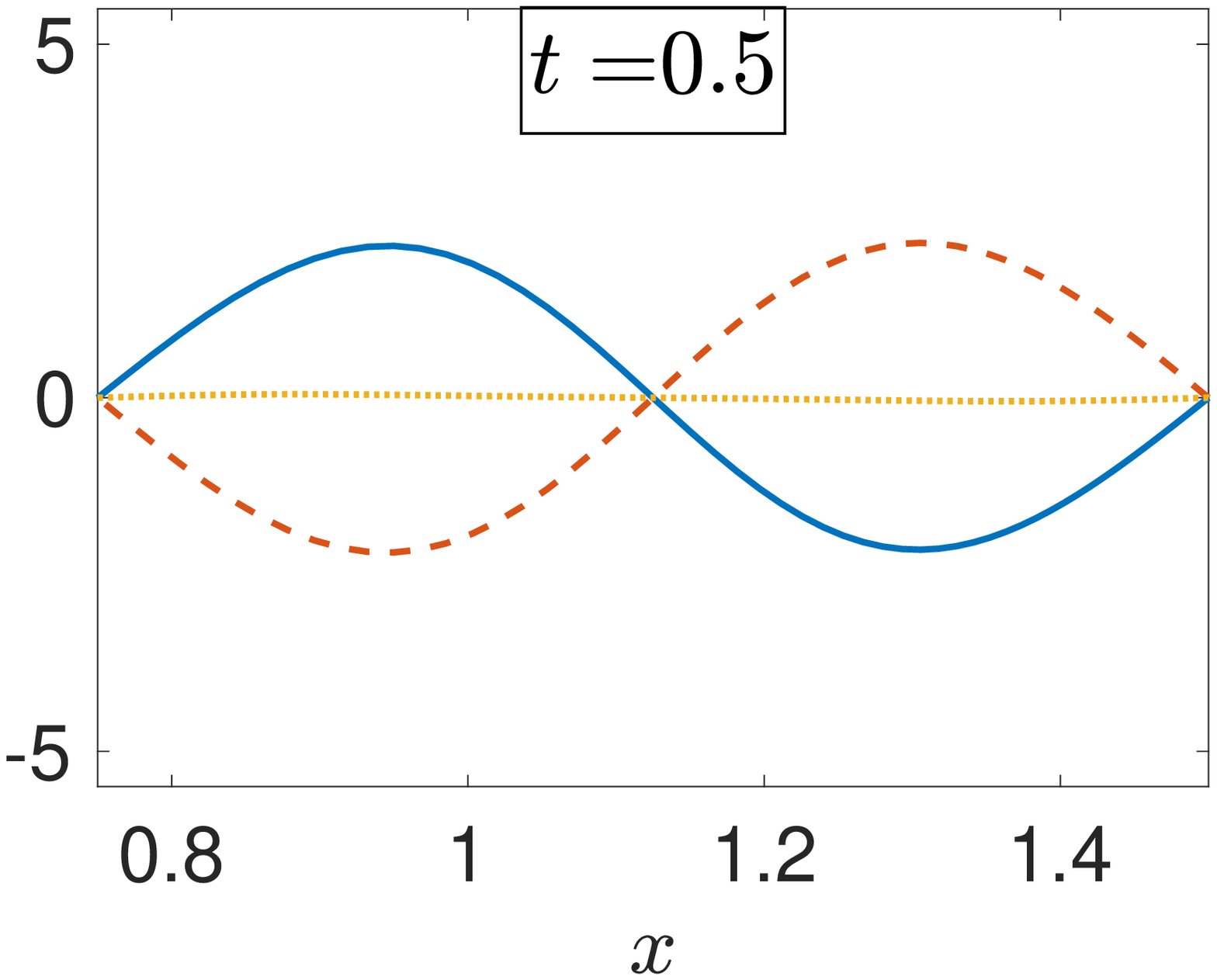}
\includegraphics[width=.24\textwidth]{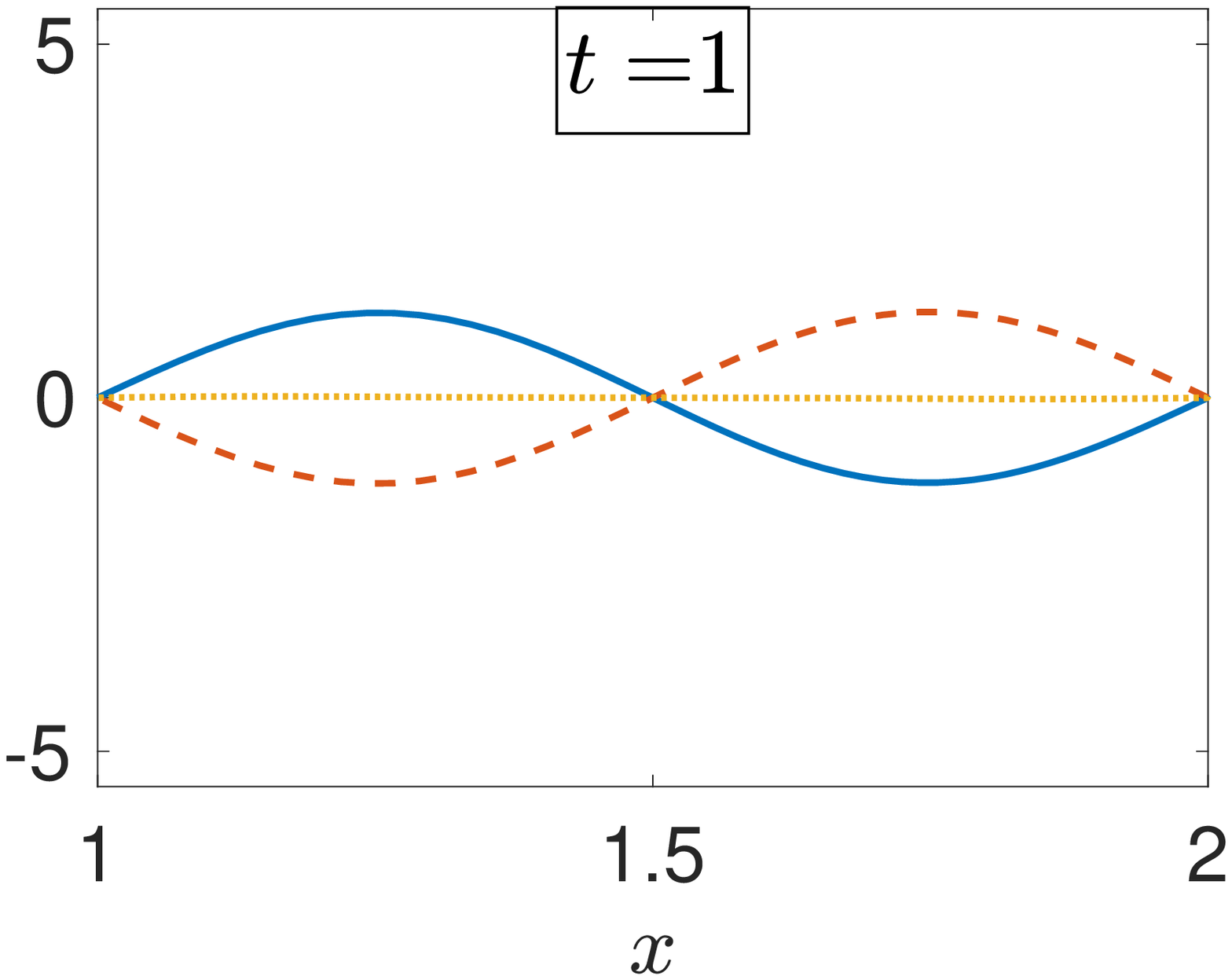}
\includegraphics[width=.24\textwidth]{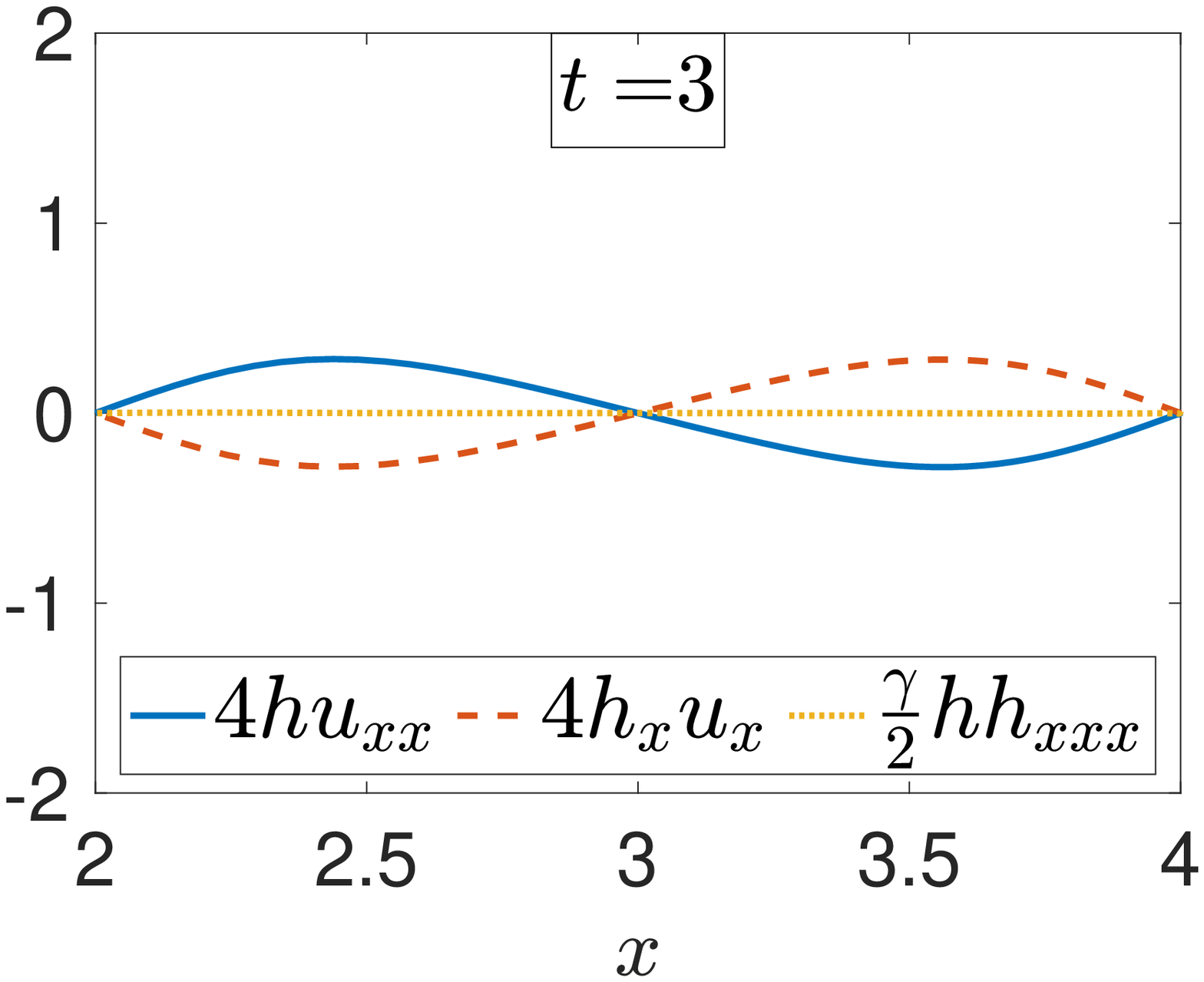}
\caption{Individual terms in the axial force balance Eq.~(\ref{eq:2}) for the case of weak elasticity shown in Fig.~\ref{fig:fullprofile_newt4} ($k_0=2$). Note that only the right half of the domain is shown, and the scale of the vertical axis changes in the final plot. }
\label{fig:allterms_newt4}
\end{figure}

\begin{figure}[htbp]
\centering
\includegraphics[width=.6\textwidth]{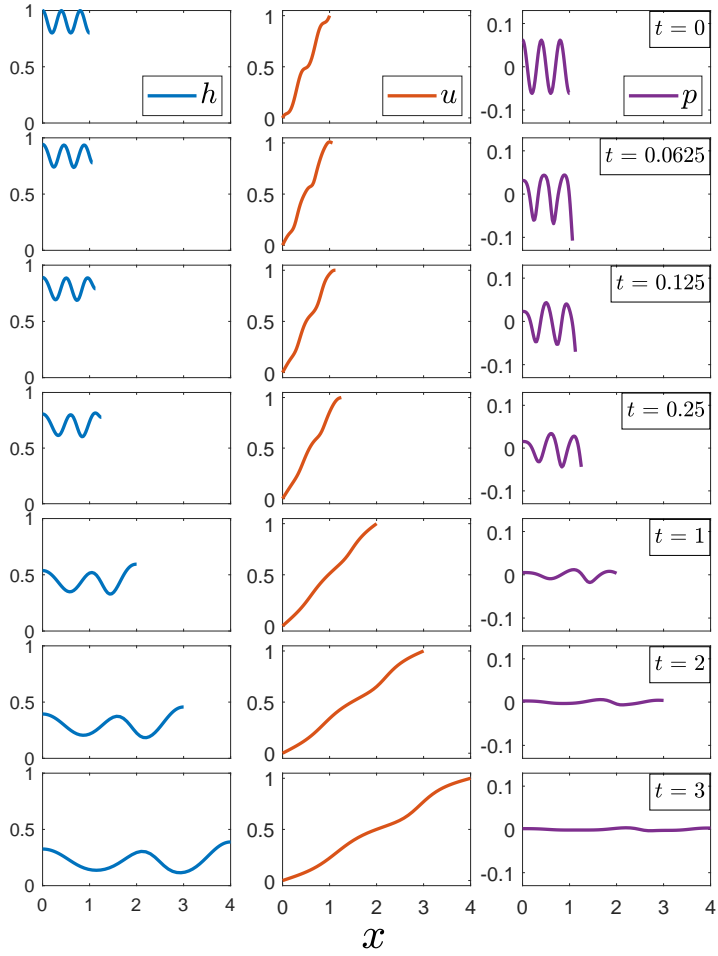}
\caption{Sheet thickness, $h$, fluid velocity, $u$, and pressure, $p$, for the case of moderate elasticity when the initial condition has two and a half waves. Results are for BC Case I with $\gamma=0.0025$ and $v_0=1$. Each row represents a time level; each column shows the respective dependent variable.}
\label{fig:subplot_profile_5}
\end{figure}

\begin{figure}[htbp]
\centering
\includegraphics[width=.32\textwidth]{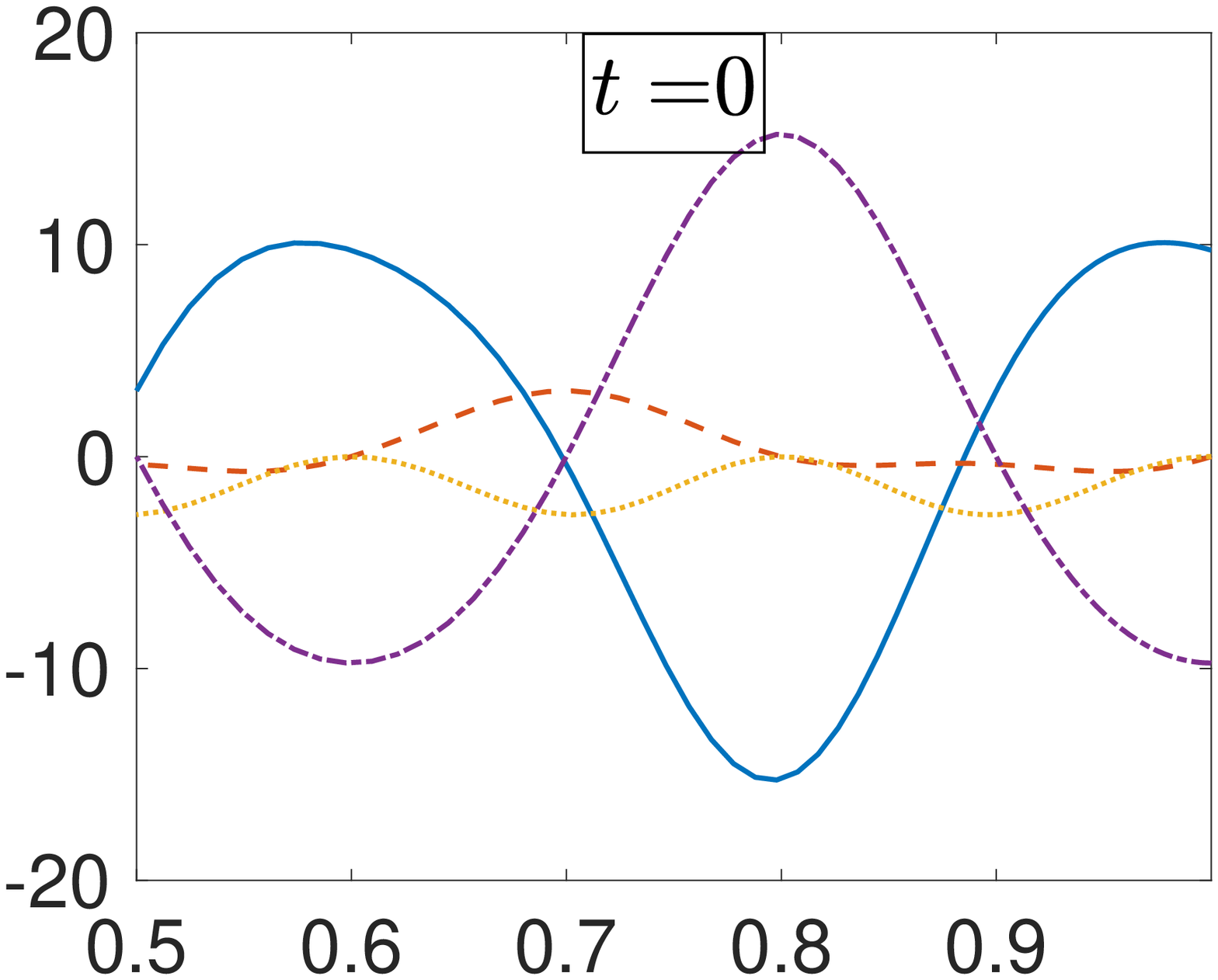}
\includegraphics[width=.32\textwidth]{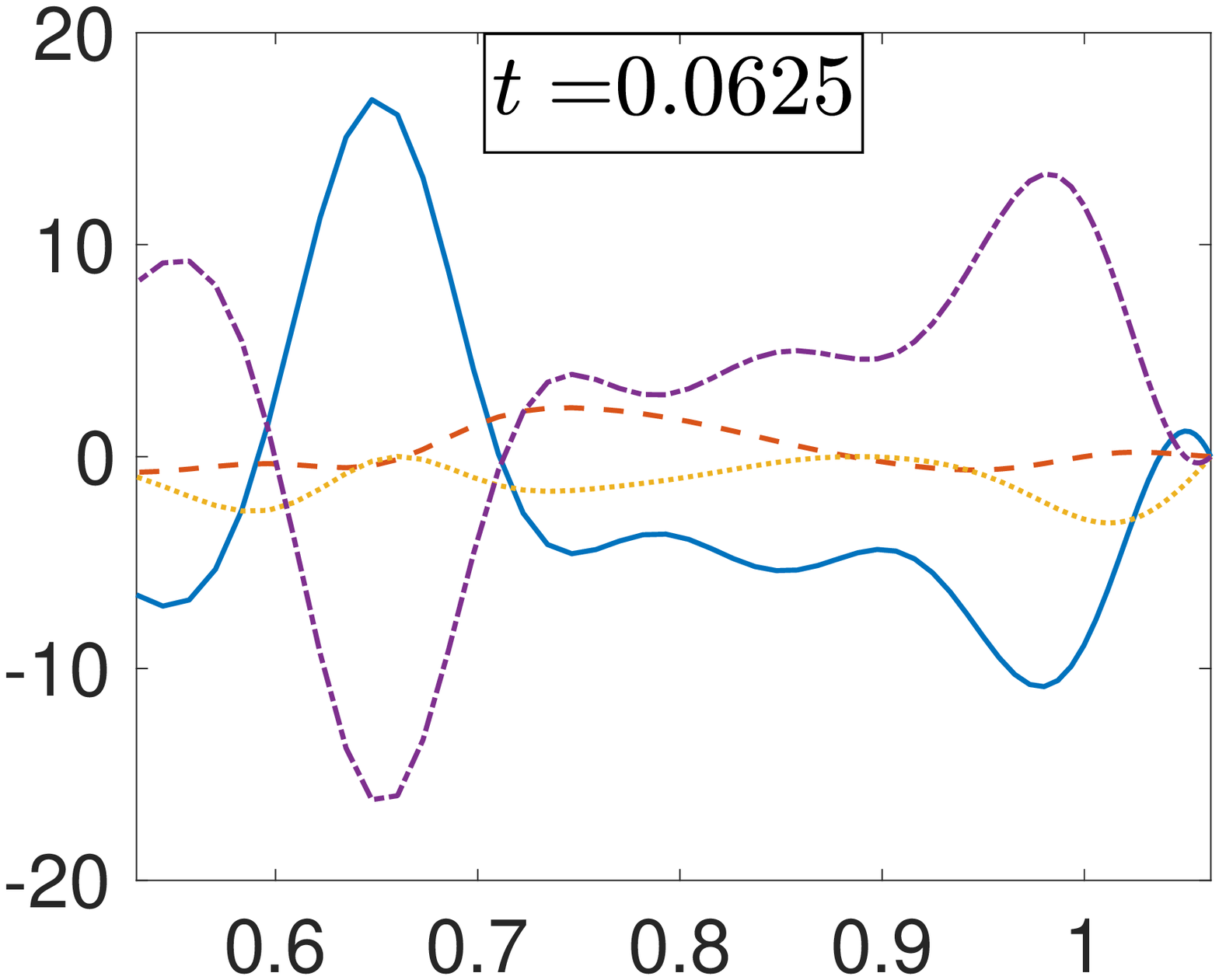}
\includegraphics[width=.32\textwidth]{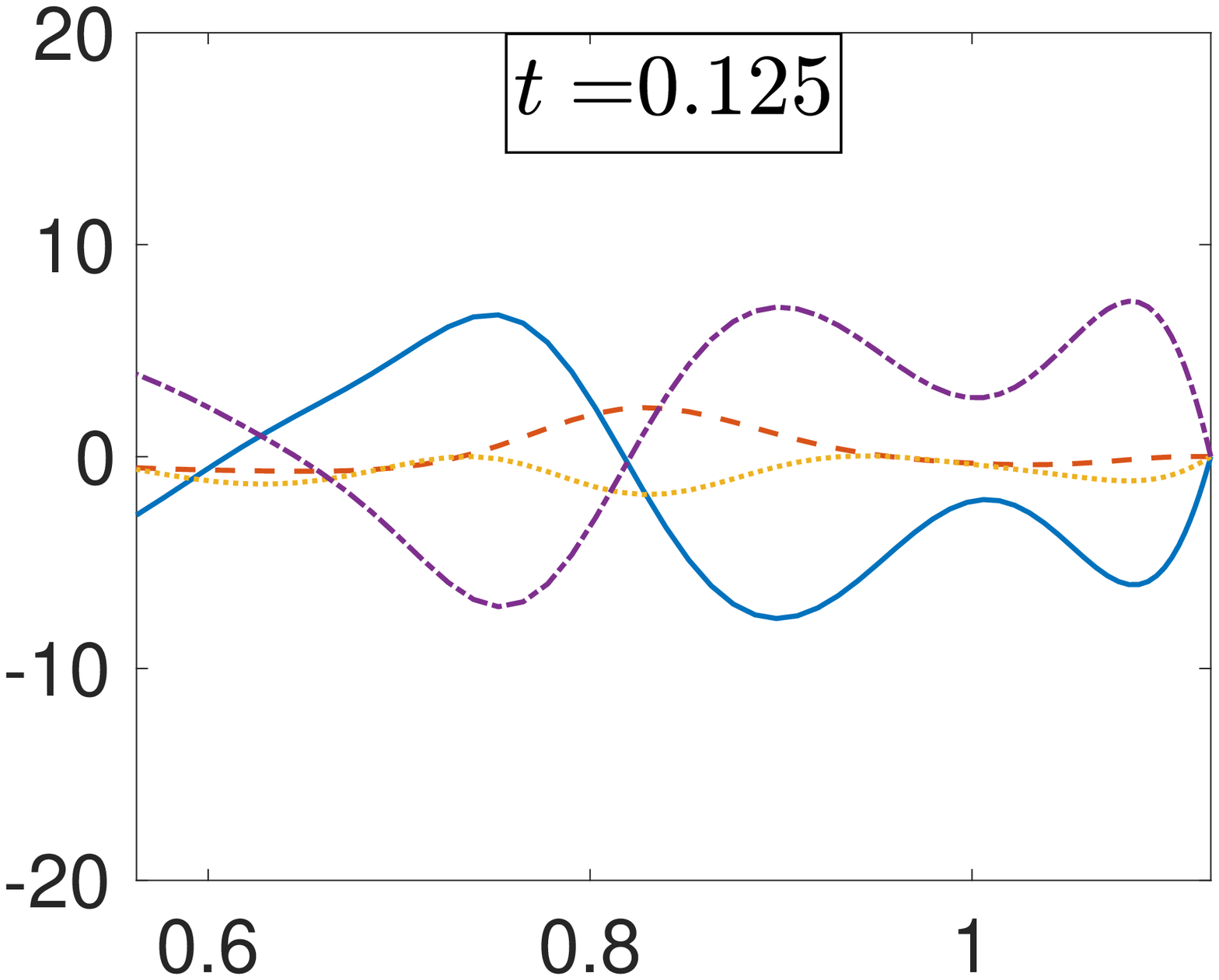}
\includegraphics[width=.32\textwidth]{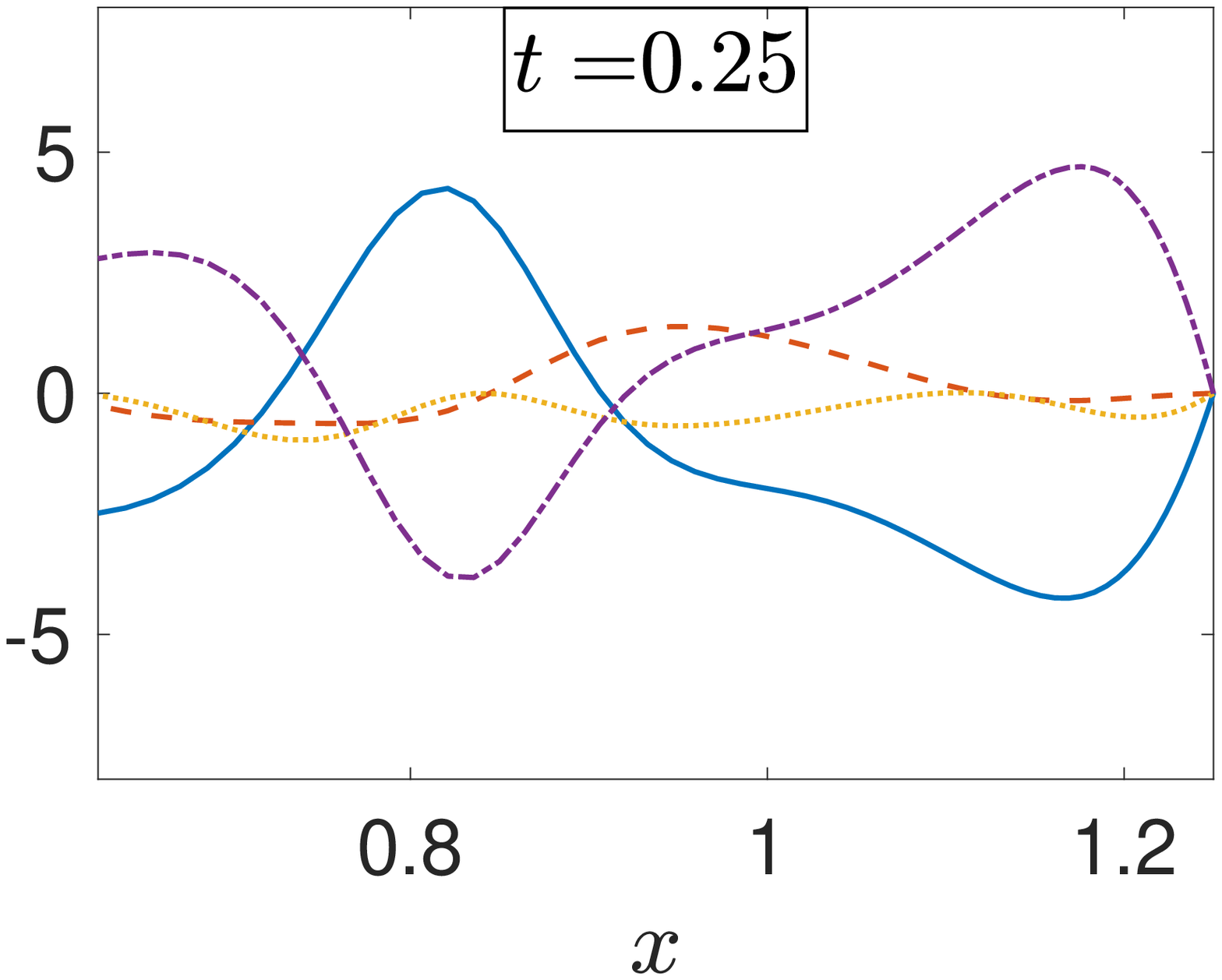}
\includegraphics[width=.32\textwidth]{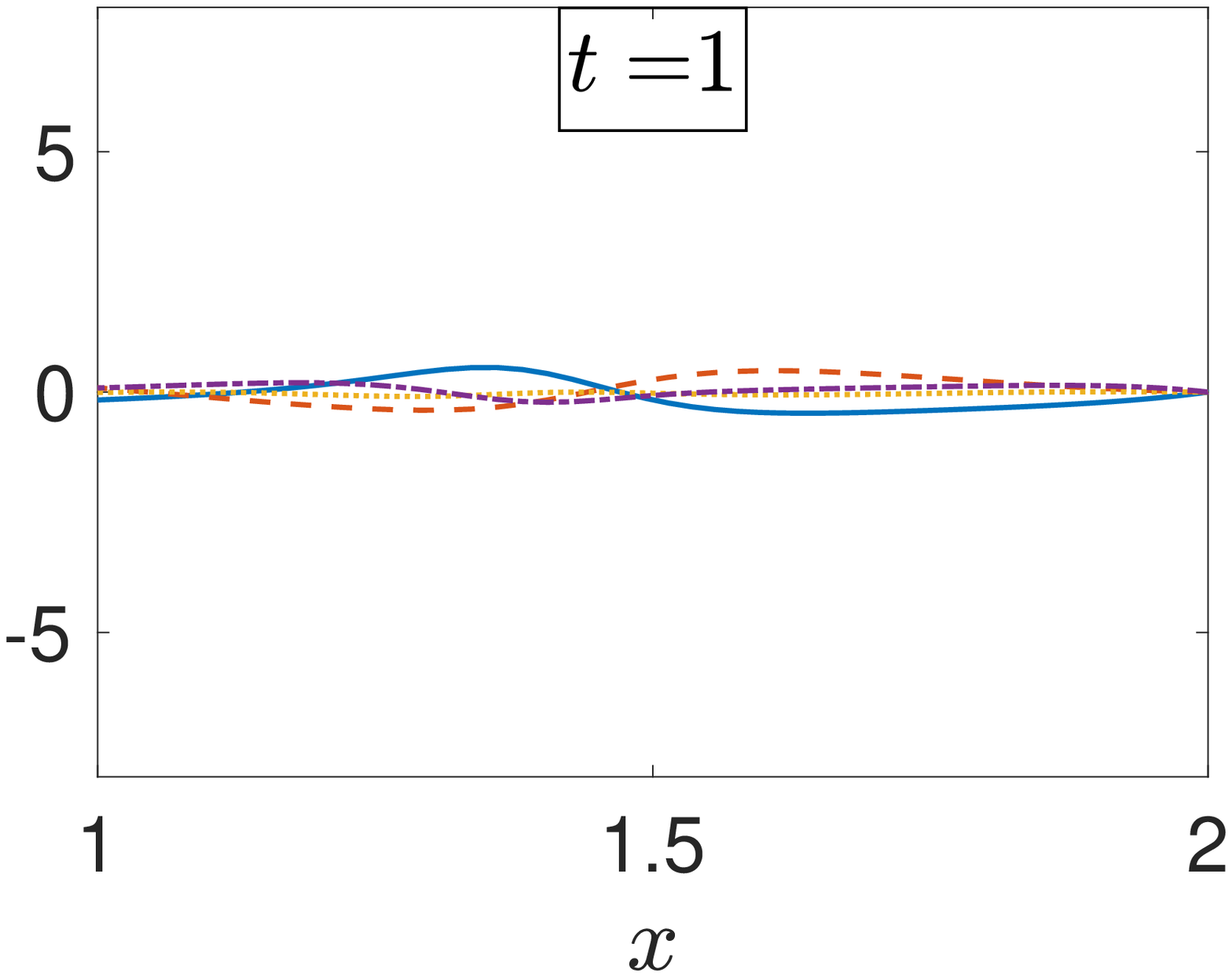}
\includegraphics[width=.32\textwidth]{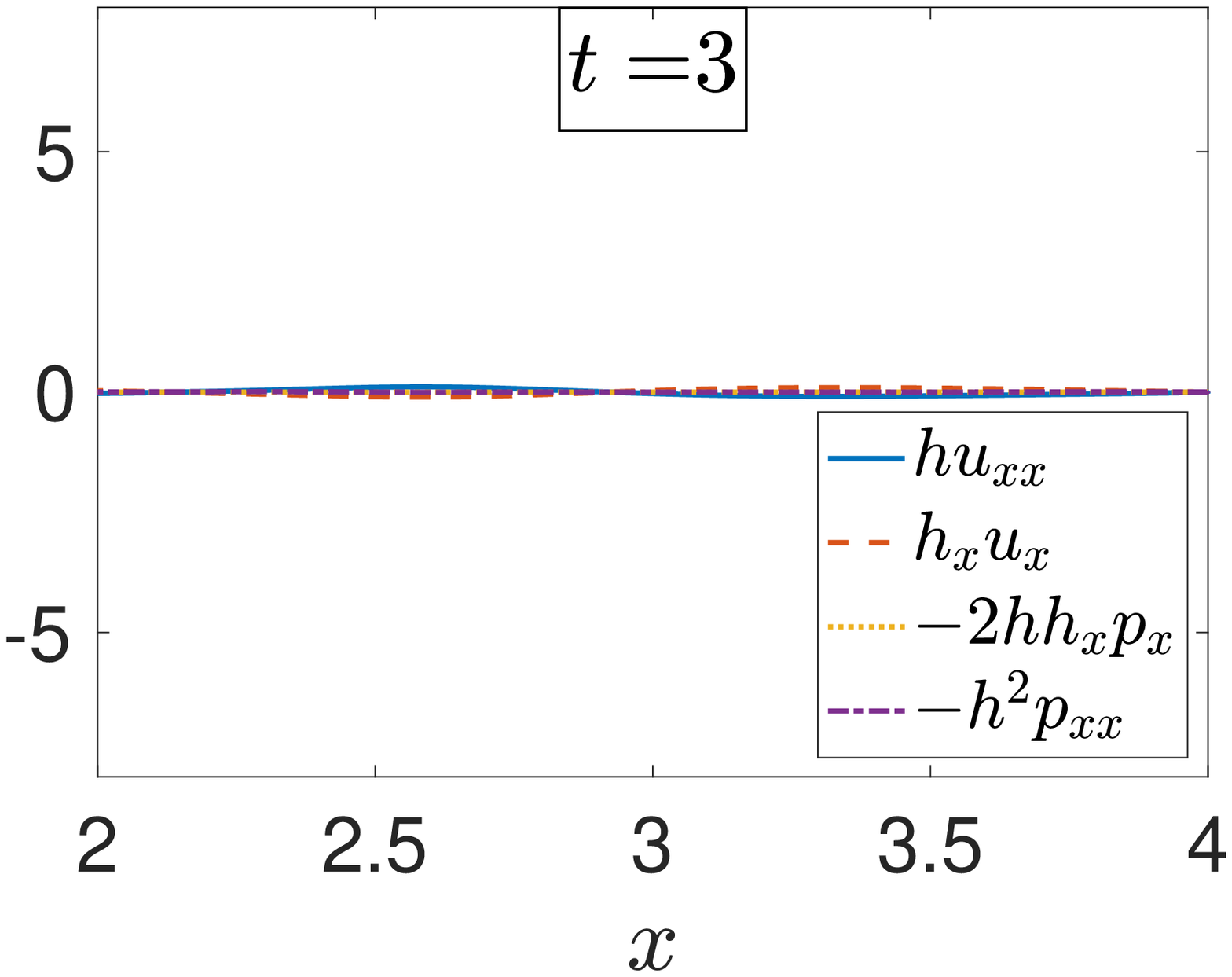}
\caption{Individual terms in the axial force balance Eq.~(\ref{eq:mod2}) for the case of moderate elasticity whose full profiles are shown in the previous Fig.~\ref{fig:subplot_profile_5} (only half the domain is shown here). Note that the scale of the vertical axis changes on the second row. }
\label{fig:allterms_5}
\end{figure}

\begin{figure}[htbp]
\centering
\includegraphics[width=.6\textwidth]{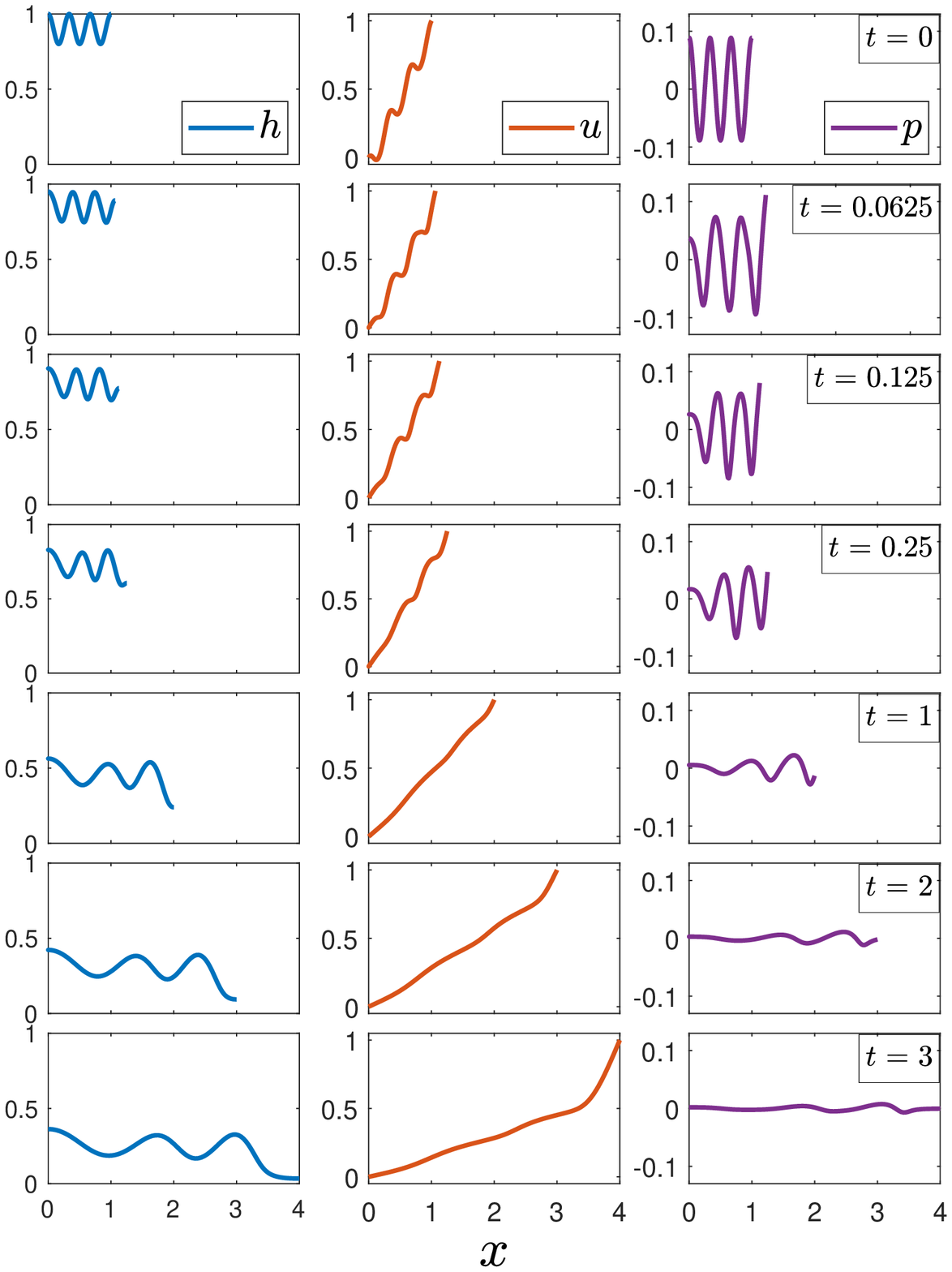}
\caption{\label{fig:subplot_profile_6}Profiles of sheet thickness, $h$, fluid velocity, $u$, and pressure, $p$, for the case of moderate elasticity when the initial condition has three waves and $\gamma=0.0025$. Each row represents a time level; each column shows the dynamics of the respective dependent variable.}
\end{figure}

However, for moderate elasticity, solutions are more complicated.  We compare the sheet thickness, velocity, and pressure when the initial condition contains either two and a half ($k_0=2.5$) or three ($k_0=3$) waves; the solutions are shown in Figs.~\ref{fig:subplot_profile_5} through \ref{fig:subplot_profile_6}. When $k_0=2.5$ (Fig.~\ref{fig:subplot_profile_5}), the moving end begins as a thickness minimum. Local low  pressure draws fluid toward the moving end, and this local minimum becomes a global maximum by time $t=0.5$. The minimum thickness occurs in the interior, in the trough closest to the moving end. The early rapid movement of fluid toward the moving end is shown in the velocity profile at $t=0.0625$, where the velocity briefly increases above the pulling velocity ($v_0=1$) near the moving end. Fluctuations in the velocity profile smooth after this time, and the profile becomes nearly linear. Pressure decays to near zero for $t>1$. Fig.~\ref{fig:allterms_5} shows the role of each term in the PDE. At early times, we see that the terms with the highest derivatives are flipping roles. As pressure diminishes, extensional terms take over.

When $k_0=3$ (Fig.~\ref{fig:subplot_profile_6}), we see the role of pressure has changed. Together, pressure and extension prevent fluid from keeping up with the moving end, and the right end quickly becomes the global minimum. A boundary layer in the velocity profile is seen to form at the right end of the sheet in the middle column of Fig.~\ref{fig:subplot_profile_6}. A maximum in the pressure develops at the right end by $t=0.0625$, and remains a global maximum until about $t=0.25$. The pressure diminishes thereafter. 

In summary, our results for the moderate elasticity model show that, depending on the initial condition, the number of waves in the sheet may be reduced, and there are significant changes in the shape of the sheet as fluid moves due to changes in pressure. Model parameters, in particular the surface tension $\gamma$, can also strongly influence the number of waves retained in the sheet under extension; in this subsection such model parameters were fixed. At early times, pressure either cooperates with or opposes extension at the moving end, which redistributes fluid there and may result in the loss of a maximum or minimum in the sheet thickness there. When a maximum is lost from the moving end, a boundary layer forms in the velocity profile. As time increases, the pressure decreases in magnitude, its influence on the shape of the sheet decreases, and the role of the extension becomes more pronounced. In general, the roles of pressure and extension are more intertwined than in the case of weak elasticity. 

\section{Discussion and conclusion}
\label{sec:Discussion}

We present a new model for describing the extensional 2D flow of nematic liquid crystal sheets with moderate elasticity, and compare results to the analogous weak elasticity model.  For moderate elasticity, the pressure, surface tension and elastic energy were all promoted to larger values compared to the weak elasticity case studied by Cummings et al.\cite{cummings2014extensional} The axial force balance, Eq.~(\ref{eq:axforcebal}), in the new model is of higher spatial order than the model for weak elasticity; in terms of the sheet thickness, the equation is fourth order rather than third in spatial derivatives.  This change necessitates an additional boundary condition.  Consideration of the individual terms in the sheet tension in Eq.~(\ref{eq:T_mod_using_p}) motivated the additional condition that we used, $p_x(0,t)=0$.  Numerical exploration suggested that the single equation Eq.~(\ref{countbc}), describing the sheet profile evolution, may be viewed as being dispersive, and that the additional boundary condition may be considered as specifying the value for an incoming characteristic. 

For initial conditions, we use sinusoidal curves, and we explore a range of initial wavenumbers.  Further work could include formulating a consistent initial condition for a Dirichlet condition on either end. 
We examine the effect of varying surface tension and the speed of the moving on the dynamics of the evolving sheet under stretching. 

The response of the moderately elastic sheet is markedly different from that of weak elasticity or Newtonian fluids.
Cummings et al. \cite{cummings2014extensional} modeled liquid crystal with weak elasticity, however this work focused primarily on the effect of an electric field on the liquid crystal. For liquid crystals with moderate elasticity, the elastic quality of the material is demonstrated well in Fig.~\ref{fig:minloc_comp}, which shows a recoil in the location of minima in a sheet with multiple waves. In the case of weak elasticity, the minima maintain their relative position in the sheet while undergoing stretching. In Fig.~\ref{fig:min_thickness}, we show that depending on the initial condition, the minimum sheet thickness can occur at almost any position in the sheet, from the very right end, to close to the left end. When varying the surface tension, we again see the elastic quality of the material; see Fig.~\ref{fig:one}. When varying the speed of the moving end, we see that for the same speed, the sheet with moderate elasticity thins slightly faster than in the case of weak elasticity; see Fig.~\ref{fig:two}. 

We also considered dynamics and mechanism for different initial wavenumbers in the sheet profile. We increase the number of waves in the initial condition, and observe the shape of the sheet as it undergoes stretching. We find, as might be expected, that the higher the surface tension, the more waves are lost from the initial shape under stretching. The amplitude of the waves has much less influence than the number of waves, as seen in Table \ref{tab:ampvwave}. At early times, depending on the number of sinusoidal waves, pressure either aids or opposes extensional flow, which changes the shape of the sheet and may result in the loss of a minimum or maximum at the moving end. When a maximum is lost from the moving end, and specifically when the moving end switches from a maximum to a minimum, we see a boundary layer form in the velocity profile. Fluid flows quickly out of the region at the end, and the sheet is unable to stretch for very long times  before numerics fail. This illustrates the more prominent role that pressure plays in determining the shape of the sheet with moderate elasticity; see, for example, Fig.~\ref{fig:allterms_5}. 

The menisci that develop in the thickness profiles when using Robin boundary conditions for moderate elasticity are reminiscent of the profiles found by several previous authors \cite{jones2006effect, jones2005dynamics,heryudono2007single,MakiBraun08,JossicLefevre09,AydemirBreward2010,Allouche17,mehdaoui2021numerical} for the aqueous layer of the tear film during a blink. Specifically, BC Case III (Robin condition at the moving end) yields profiles comparable to those of the tear film during the upstroke of a blink. BC Case IV (Robin condition at the fixed end) is similar to the meniscus corresponding to the lower lid during the upstroke when the upper lid would be moving away from it.  

We note that weak and moderate elasticity limits were considered for a nematic liquid crystal film on a substrate by Lin et al. \cite{lin2013note}  Those authors found that a larger scaling for the elastic terms (only) introduced an additional term in the single nonlinear PDE for the thickness $h$; the new term was diffusion-like and is similar to the effect of gravity in Newtonian films.\cite{ODB97}  In our work, there is no substrate for the free film, and compared to the weakly elastic limit, we made both the elastic and surface tension parameters larger.  As a result, we scaled the pressure to be larger, and the new balance gave us two PDEs, one each for the film thickness $h$ and axial velocity $u$, as is typical for extensional flow.\cite{cummings2014extensional}  The results reported here clearly show elastic behavior, and likely more obviously than the model found by Lin et al.\cite{lin2013note}

There are some limitations with our model in terms of computing for a longer time interval for for a wider range of parameter values. Both the finite differences and spectral methods work well up until $t=3$, and for some parameter values, much longer than that; however, to run other cases for longer time may require domain decomposition or an adaptive method to adequately resolve regions of small thickness.\cite{bertozzi94} Scenarios that result in beads on a string \cite{clasen2006beads,sostarecz2004beads} involve much more extension and longer computation times; the deformation we see here is less severe, and it is unclear what would result in our case.

We shall continue developing models for the lipid layer of the tear film in the eye. This will require using more realistic parameter values, and modifying the sheet's end speed to be more realistic.\cite{heryudono2007single,AydemirBreward2010,deng2013model,deng2014heat}  We have a model in hand with a shear-dominated aqueous layer added to the lipid layer in the spirit of previous works.\cite{peng2014evaporation, bruna2014influence, stapf2017duplex, zubkov2012coupling}  Much work remains to connect the dynamics of such models to the observed patterns of the lipid layer in the tear film.

\section*{Acknowledgments}
This work was supported by National Science Foundation grants DMS 1909846, DMS 2206127 and DMS 1815613. The content
is solely the responsibility of the authors and does not necessarily represent the official
views of the funding source.  


\newpage
\appendix

\section{Derivations}

\subsection{Ericksen-Leslie equations}
The Ericksen-Leslie equations describe the flow of nematic liquid crystals; here the $'$ denotes a dimensional quantity.
\begin{subequations}
\label{eq:ELwhole}
\begin{align}
    \frac{\partial }{\partial x'_i}\left( \frac{\partial W'}{\partial \theta_{x'_i}}\right)-\frac{\partial W'}{\partial \theta}+\tilde{g}'_i\frac{\partial n_i}{\partial \theta}&=0, \label{ELeq1} \\
    -\frac{\partial \pi'}{\partial x'_i}+\tilde{g}'_k\frac{\partial n_k}{\partial x'_i}+\frac{\partial \tilde{t}'_{ik}}{\partial x'_k}&=0,  \label{ELeq2}\\
    \frac{\partial v'_i}{\partial x'_i}&=0. \label{ELeq3}
\end{align}
\end{subequations}

The first equation corresponds to conservation of energy; the second to conservation of momentum, and the third to conservation of mass. Expanded, these equations become
\begin{align}
    \frac{\partial }{\partial x'}\left( \frac{\partial W'}{\partial \theta_{x}}\right)+\frac{\partial }{\partial z'}\left( \frac{\partial W'}{\partial \theta_{z}}\right)-\frac{\partial W'}{\partial \theta}+\tilde{g}'_x\frac{\partial n_x}{\partial \theta}+\tilde{g}'_z\frac{\partial n_z}{\partial \theta}=0,\\
    -\frac{\partial \pi'}{\partial x'}+\tilde{g}'_x\frac{\partial n_x}{\partial x'}+\tilde{g}'_z\frac{\partial n_z}{\partial x'}+\frac{\partial \tilde{t}'_{xz}}{\partial z'}+\frac{\partial \tilde{t}'_{xx}}{\partial x'}=0,\\
    -\frac{\partial \pi'}{\partial z'}+\tilde{g}'_x\frac{\partial n_x}{\partial z'}+\tilde{g}'_z\frac{\partial n_z}{\partial z'}+\frac{\partial \tilde{t}'_{zx}}{\partial x'}+\frac{\partial \tilde{t}'_{zz}}{\partial z'}=0,\\
    \frac{\partial v'}{\partial x'}+\frac{\partial w'}{\partial z'}=0,
\end{align}

where
\begin{align}
    \tilde{g}'_i=&-\gamma_1N'_i-\gamma_2e'_{ik}n_k, \hspace{35pt} e'_{ij}=\frac{1}{2}\left(\frac{\partial v'_i}{\partial x'_j}+\frac{\partial v'_j}{\partial x'_i}\right) , \\
    N'_i=&\;\dot{n}'_i-\omega'_{ik}n_k, \hspace{65pt} \omega'_{ij}=\frac{1}{2}\left(\frac{\partial v'_i}{\partial x'_j}-\frac{\partial v'_j}{\partial x'_i}\right),\hspace{35pt}\pi'=\;p'+W',\\
    W'=& \;\frac{1}{2}\bigg[K_1(\nabla' \cdot {\bf n})^2+K_2({\bf n}\cdot \nabla' \wedge {\bf n})^2+K_3(({\bf n}\cdot \nabla') {\bf n})\cdot (({\bf n}\cdot \nabla') {\bf n})\bigg],\\
    \tilde{t}'_{ij}=&\;\alpha_1'n_kn_pe'_{kp}n_in_j+\alpha'_2N'_in_j+\alpha'_3 N'_jn_i+\alpha'_4e'_{ij}+\alpha'_5e'_{ik}n_kn_j+\alpha'_6e'_{jk}n_kn_i .
\end{align}
Summation over the repeated indices is understood, and $\dot{n}_i$ denotes the convective derivative of the ith component of ${\bf n}$. The quantities defined above are described in Table \ref{tab:more_param}.
We solve the governing equations subject to the boundary conditions that follow. We list the boundary conditions for the top surface, $z'=\frac{1}{2}h'(x',t')$; those for the bottom surface, $z'=-\frac{1}{2}h'(x',t')$, are defined in the same way. The normal stress condition is
\begin{equation}
    \hat{n}'\cdot \sigma' \cdot \hat{n}' = -\gamma' \kappa' \hat{n}'\quad\text{at}\;z=\frac{1}{2}h'(x',t'), 
\end{equation}
where $\hat{n}'=(-h'_{x'}/2,1)/\sqrt{1+\left(h'_{x'}/2\right)^2}$ is the unit vector normal to the top surface, and $\kappa' = -\nabla\cdot \hat{n}'$ is the curvature of the top surface, and $\sigma'_{ij} = -p'\delta_{ij} + \theta_i\theta_j + \tilde{t}'_{ij}$. The definition of $\sigma'$ is taken from Lin et al., \cite{lin2013note} and contains an additional term from the form given in Cummings et al. \cite{cummings2014extensional} The tangential stress condition is
\begin{align}
    \hat{n}'\cdot \sigma' \cdot t' = 0\quad\text{at}\;z'=\frac{1}{2}h'(x',t'),
\end{align}
where $t'=(1,h'_{x'}/2)/\sqrt{1+\left(h'_{x'}/2\right)^2}$ is the unit vector tangent to the top surface. 
The kinematic boundary condition is 
\begin{align}
    w'=\frac{1}{2}\left(h'_{t'}+u'h'_{x'}\right)\quad\text{at}\;z'=\frac{1}{2}h'(x',t').
\end{align}
Finally, the anchoring boundary condition, in the absence of an electric field, is
\begin{align}
    \theta&=\theta_B\quad\text{at}\;z'=\frac{1}{2}h'(x',t').
\end{align}

\begin{table*}
\caption{\label{tab:more_param} Parameters and variables used in the model. \cite{stewart2019static}  }
\begin{ruledtabular}
\begin{tabular}{ll}
Quantity   & Description\\
\hline
${\bf v}'=(u',0,w')$&velocity field of the flow\\
${\bf n}= (\sin\theta,0,\cos\theta)$& director field\\
$\theta(x,z,t)$&angle the director angle makes with the z-axis\\
$p'$           & pressure  \\
$W'$           & bulk energy density  \\
$\pi'=p'+W'$& modified pressure\\
$\tilde{t}'_{ij}$& extra stress tensor (viscous stress)\\
$\sigma$ & stress tensor\\
$\alpha'_i,\,i=1,...,6$ & Leslie viscosities (Newtonian: $\mu'=\alpha'_4/2$, all other $\alpha_i=0$)\\
$\gamma'_1=\alpha'_3-\alpha'_2$  & rotational/twist viscosity  \\
$\gamma'_2=\alpha'_6-\alpha'_5$  & torsion coefficient \\
$K_1,\;K_2,\;K_3$           & elastic constants representing splay, twist, and bend respectively       \\
$\omega'_{ij}$ & vorticity tensor \\
$e'_{ij}$ & rate of strain tensor\\
$N_i$ & co-rotational time flux of the director \bf{n}
\end{tabular}
\end{ruledtabular}
\end{table*}

\subsection{Scalings for weak elasticity}
The scalings for weak elasticity are 
\begin{align*}
    x'&=Lx, \hspace{35pt} z'=\delta Lz,\hspace{35pt}t'=\frac{L}{U}t,\hspace{35pt}\gamma'= \frac{\mu U}{\delta}\gamma,\\
    u'&=Uu, \hspace{32pt} w'=\delta Uw,\hspace{28pt} h'=\hat{h}h,\hspace{35pt} \hat{N} = \frac{K}{\mu U \delta L}, \\
    p'&= \frac{\mu U}{L}p, \hspace{20pt} W'=\frac{K}{\delta^2 L^2}W,\hspace{15pt}  \alpha_i'=\mu\alpha_i.
\end{align*}

Nondimensionalizing with these scalings yields
\begin{widetext}
\begin{align}
    \hat{N}\frac{\p}{\p x}\left(\frac{\partial W}{\p \theta_x}\right)+\hat{N}\frac{\p}{\p z}\left(\frac{\partial W}{\p \theta_z}\right)-\hat{N}\frac{\partial W}{\p \theta}+ g_x\frac{\p n_x}{\p \theta}+ g_z\frac{\p n_z}{\p \theta}&=0,\\
    -\delta^2\frac{\partial p}{\partial x}-\delta\hat{N}\frac{\partial W}{\p x}+\delta g_x\frac{\p n_x}{\p x}+\delta g_z\frac{\p n_z}{\p x}+\delta\frac{\p t_{xx}}{\p x}+\frac{\p t_{xz}}{\p z}&=0,\\
    -\delta\frac{\partial p}{\partial z}-\hat{N}\frac{\partial W}{\p z}+ g_x\frac{\p n_x}{\p z}+ g_z\frac{\p n_z}{\p z}+\delta\frac{\p t_{zx}}{\p x}+\frac{\p t_{zz}}{\p z}&=0.
\end{align}
\end{widetext}
Then, the leading order system of equations is
\begin{align}
    h_t+(hu)_x&=0,\\
    \frac{F(\theta_B)}{G(\theta_B)}(hu_x)_x+\frac{\gamma}{2}hh_{xxx}&=0.
\end{align}
where
\begin{align}
    G(\theta_B)=&\;\alpha_1-2\alpha_2+2\alpha_3+8+2\alpha_5+2\alpha_6-\alpha_1\cos(4\theta_B)\nonumber\\
    &-2\cos(2\theta_B)(\alpha_2+\alpha_3-\alpha_5+\alpha_6),\label{eq:G}\\
    F(\theta_B)=&\;\alpha_1(-\alpha_2+\alpha_3+8+2\alpha_5+2\alpha_6)-\alpha_2(8+\alpha_5+3\alpha_6)\nonumber\\&+\alpha_3(8+\alpha_5+\alpha_6)+32+\alpha_5(16+2\alpha_5+4\alpha_6)+\alpha_6(16+2\alpha_6)\nonumber\\ &-2\cos(2\theta_B)(\alpha+4+\alpha_5+\alpha_6)(\alpha_2+\alpha_3-\alpha_5+\alpha_6)\nonumber\\ &-\cos(4\theta_B)\Big[\alpha_1\alpha_2-\alpha_1\alpha_3+(\alpha_2+\alpha_3)(\alpha_5-\alpha_6)\Big]\label{eq:F}.
\end{align}
In the Newtonian case, all viscosities are zero ($\alpha_i=0,\;i=1,\cdots6$), and $F(\theta_B)/G(\theta_B)=4$. 

\subsection{Deriving the equations for moderate elasticity}
\label{sec:ModElasticity}

To consider the case of moderate elasticity, we rescale the inverse Ericksen number, the pressure, the surface tension as follows, while keeping the other scalings the same.
\begin{align*}
    \hat{N}=\frac{K}{\mu U L},\quad p'=\frac{\mu U}{\delta L}p,\quad \gamma'=\frac{\mu U}{\delta^2}\gamma.
\end{align*}

Using the scaling for moderate elasticity, the nondimensionalized governing equations become
\begin{align}
    \hat{N}\frac{\p}{\p x}\left(\frac{\partial W}{\p \theta_x}\right)+\hat{N}\frac{\p}{\p z}\left(\frac{\partial W}{\p \theta_z}\right)-\hat{N}\frac{\partial W}{\p \theta}+\delta g_x\frac{\p n_x}{\p \theta}+\delta g_z\frac{\p n_z}{\p \theta}&=0,\\
    -\delta\frac{\partial p}{\partial x}-\hat{N}\frac{\partial W}{\p x}+\delta g_x\frac{\p n_x}{\p x}+\delta g_z\frac{\p n_z}{\p x}+\delta\frac{\p t_{xx}}{\p x}+\frac{\p t_{xz}}{\p z}&=0,\\
    -\delta\frac{\partial p}{\partial z}-\hat{N}\frac{\partial W}{\p z}+\delta g_x\frac{\p n_x}{\p z}+\delta g_z\frac{\p n_z}{\p z}+\delta^2\frac{\p t_{zx}}{\p x}+\delta\frac{\p t_{zz}}{\p z}&=0.
\end{align}
We consider the case with no electric field, so we set ${\bf E= 0}$.
Now we asymptotically expand the dependent variables in powers of $\delta=\hat{h}/L$. For example, 
\[u = u_0(x,z,t)+\delta\, u_1(x,z,t)+\delta^2\,u_2(x,z,t)+ \cdots,\]
and we do the same for $\theta,\,v,\,p,$ and $h$. 
We substitute these into the governing equations and boundary conditions, and then collect like powers of $\delta$. 
Then at order 1,
\begin{align}
\frac{1}{2} \big(\alpha_2+\alpha_3-\alpha_5+\alpha_6+2 \alpha _1 \cos{2 \theta_0}\big)\sin{2\theta_0}u_{0z}\theta_{0z}\nonumber-\hat{N}\theta_{0z} \theta_{0zz}\\
    +\frac{1}{2}\left[2+(\alpha_5-\alpha_2) \cos^2{\theta_0}+(\alpha_3+\alpha_6)\sin^2{\theta_0}+\frac{1}{2}\alpha_1 \sin^2{2\theta_0}\right]u_{0zz}&=0,\label{eq1}\\
-\hat{N}\theta_{0z} \theta_{0zz}&=0, \label{eq2}\\
\hat{N} \theta_{0zz}&=0,\label{eq3}\\
u_{0x}+w_{0z}&=0,\label{eq4}\\
-\hat{N}\theta_{0z}^2&=0,\label{eq5}\\
\frac{1}{2}\left[2+(  \alpha_5-\alpha_2)\cos^2{\theta_0}+(\alpha_3+\alpha_6)\sin^2{\theta_0}+\frac{1}{2}\alpha_1\sin^2{2\theta_0}\right]u_{0z}\nonumber\\-\frac{1}{2}\hat{N}\theta_{0z}^2h_{0x}-\hat{N}\theta_{0z}\theta_{0x}&=0,\label{eq6}\\
w_0-\frac{1}{2}\big(h_{0t}+u_0h_{0x}\big)&=0,\label{eq7}\\
\theta_0-\theta_B&=0\label{eq8}.
\end{align}
Starting from the top, the equations represent momentum in the x and z-components respectively, energy, and continuity, followed by the boundary conditions on the top surface: normal and tangential stress, kinematic, and anchoring.
Solving the order 1 equations, we obtain
\begin{align}
    &\theta_0=\theta_B,\\
   & u_0=u_0(x,t),\\
   & w_0=w_0(x,z,t)=-u_{0x}z,\\
  &  h_{0t}+(u_0h_0)_x=0, \label{eq:massb}
\end{align}
where (\ref{eq:massb}) is the mass balance equation. At order 1, we do not have an equation for $h$, so to close the system, we continue on to order $\delta$. After making the above substitutions, the order $\delta$ equations are
\begin{align}
\frac{1}{8}\Big[8+\alpha_1-2(\alpha_2+\alpha_3+\alpha_5+\alpha_6)
-2(\alpha_2+\alpha_3-\alpha_5+\alpha_6)\cos{2\theta_B}\nonumber\\-\alpha_1\cos{4\theta_B}\Big]u_{1zz} -p_{0x}&=0,\label{eq9}\\
p_{0z}&=0, \label{eq10}\\
\hat{N}\theta_{1zz}&=0,\label{eq11}\\
u_{1x}+w_{1z}&=0,\label{eq12}\\
p_{0}+\frac{\gamma}{2}h_{0xx}&=0,\label{eq13}\\
-\frac{1}{2}(\alpha_1\cos{2\theta_B}-\alpha_5+\alpha_6)\sin{2\theta_B}u_{0x}\nonumber\\ +\frac{1}{4}\Big[4-2(\alpha_2-\alpha_5)\cos^2{\theta_B}+2(\alpha_3+\alpha_6)\sin^2{\theta_B}+\alpha_1\sin^2{2\theta_B}\Big]u_{1z}&=0,\label{eq14} \\
w_1-\frac{1}{2}(h_{1t}+u_1h_{0x}+u_0h_{1x})&=0,\label{eq15}\\
\theta_1 &=0.\label{eq16}
\end{align}
Solving, we obtain
\begin{align}
p_0(x,t)&=-\frac{\gamma}{2}h_{0xx}, \\
    u_1(x,z,t)&=\frac{p_{0x}}{F(\theta_B)}\left(\frac{z^2}{2}-\frac{h_0}{2}z\right)-\frac{ A(\theta_B,u_{0x})}{B(\theta_B)}z+K(x,t),\\
    \theta_1(x,z,t)&= 0,
\end{align}
where $K(x,t)$ is as yet unknown. While we have found some higher order terms, we must proceed to order $\delta^2$ to find an equation for $h$.
We make the above substitutions, in addition to the substitution $w_{1zz}=-u_{1xz}$, obtained from differentiating the the continuity equation. At this order, the equations are too long to be profitable displayed in their entirety, so we summarize the steps. First, we use $z$-momentum and the normal stress condition to determine 
\begin{align}
p_1(x,z,t)=&\;G_1(\theta_B)u_{0x}+H_1(\theta_B)\gamma(2z-h_0)h_{0xxx}-\frac{\gamma}{2}h_{1xx},
\end{align}
where
\begin{align*}
    G_1(\theta_B)=&\;\frac{1}{4} \big[-8 - \alpha_1 - 2 \alpha_5 - 2 \alpha_6 - 
    2 (\alpha_1 + \alpha_5 + \alpha_6) \cos
      2 \theta_B - \alpha_1 \cos
      4 \theta_B\big] \\&+ \frac{(-\alpha_1 - \alpha_2 - \alpha_3 - \
\alpha_5 - \alpha_6 - \alpha_1 \cos
      2 \theta_B) (-\alpha_5 + \alpha_6 + \alpha_1 \cos
      2 \theta_B) \sin2 \theta_B^2}{
 8 + \alpha_1 - 2 \alpha_2 + 2 \alpha_3 + 2 \alpha_5 + 
  2 \alpha_6 - 
  2 (\alpha_2 + \alpha_3 - \alpha_5 + \alpha_6) \cos
    2 \theta_B - \alpha_1 \cos4 \theta_B},\\
    \\
    H_1(\theta_B)=&\; \frac{(-\alpha_1 - \alpha_2 - \alpha_3 - \alpha_5 - \alpha_6 \
- \alpha_1 \cos2 \theta_B)  \sin
  2 \theta_B }{2 \big[8 + \alpha_1 - 2 \alpha_2 + 2 \alpha_3 + 
   2 \alpha_5 + 2 \alpha_6 - 
   2 (\alpha_2 + \alpha_3 - \alpha_5 + \alpha_6) \cos
     2 \theta_B - \alpha_1 \cos4 \theta_B\big]}.
\end{align*}
Then, substituting $p_1$ and others into the x-momentum equation, we solve for $u_{2zz}$ and integrate across the sheet. From the tangential stress condition we get the solvability condition, which leads to 
\begin{align}
   &\frac{B_2(\theta_B)}{A_2(\theta_B)}\left(h_0u_{0x}\right)_x+\frac{C_2(\theta_B)}{A_2(\theta_B)}\gamma(h_0^2h_{0xxx})_x+4\gamma h_0h_{1xxx}=0, \label{eq:axforce}
   \end{align}
   where
   \begin{align}
\quad A_2(\theta_B)=&\;8+\alpha_1-2\alpha_2+2\alpha_3+2\alpha_5+2\alpha_6,\nonumber\\ & -2(\alpha_2+\alpha_3-\alpha_5+\alpha_6)\cos{2\theta_B}-\alpha_1\cos{4\theta_B},\\
   \label{eq:BigB_2}
   B_2(\theta_B)=&-4 \Big[2 \alpha_1 \alpha_2 + 2 \alpha_1 \alpha_3 - 
    2 \alpha_1 \alpha_5 + 2 \alpha_1 \alpha_6\Big] \cos{6 \theta_B} \nonumber\\ &- 
 4 \Big[-64 - 16 \alpha_1 - \alpha_1^2 + 16 \alpha_2 + 
    2 \alpha_1 \alpha_2 - 16 \alpha_3 - 2 \alpha_1 \alpha_3 - 
    32 \alpha_5 \nonumber\\ 
    &- 4 \alpha_1 \alpha_5  + 6 \alpha_2 \alpha_5 - 
    2 \alpha_3 \alpha_5 - 4 \alpha_5^2 - 32 \alpha_6 - 
    4 \alpha_1 \alpha_6 + 2 \alpha_2 \alpha_6 - 
    6 \alpha_3 \alpha_6\nonumber\\
    & - 8 \alpha_5 \alpha_6 - 4 \alpha_6^2 + 
    2 \big(\alpha_2 + \alpha_3 - \alpha_5 + \alpha_6\big) \big(\alpha_1 + 
       2 [4 + \alpha_5 + \alpha_6]\big) \cos{2 \theta_B}\nonumber\\ 
       &+ 2 \big(\alpha_1 [\alpha_2 - \alpha_3] - [\alpha_2 + \alpha_3] [
\alpha_5 - \alpha_6]\big) \cos{4 \theta_B} + \alpha_1^2 \cos{
      8 \theta_B}\Big],\\
   \label{eq:BigC_2} 
   C_2(\theta_B)=&-8\big(\alpha_2+\alpha_3+\alpha_1\cos{2\theta_B}\big)\sin{\theta_B}.
\end{align}
Note that the undetermined function $K(x,t)$ from $u_1$ does not appear in the solvability condition. However, if take the continuity equation from order $\delta$, integrate with respect to $z$, and apply the kinematic boundary conditions, we find
\begin{align}
 h_{1t}+u_0h_{1x}+\big(h_{0}K(x,t)\big)_x-\frac{\gamma}{6A_2(\theta_B)}\big(h_0^3h_{0xxx}\big)_x&=0, \label{eq:forK}
\end{align}
where $A_2(\theta_B)$ is as defined above. \\
To close the system, we assume that
\begin{align}
    h(x,t)=h_0(x,t)+\delta^2h_2(x,t)+O(\delta^3).\label{eq:h_expansion}
\end{align}
In other words, there is no correction to $h$ at order $\delta$. Then (\ref{eq:massb}), (\ref{eq:axforce}), and (\ref{eq:forK}) simplify to
\begin{align}
    h_{0t}+(u_0h_0)_x&=0,\\B_2(\theta_B)\left(h_0u_{0x}\right)_x+C_2(\theta_B)\gamma(h_0^2h_{0xxx})_x&=0,\\
    \big(h_{0}K(x,t)\big)_x-\frac{\gamma}{6A_2(\theta_B)}\big(h_0^3h_{0xxx}\big)_x&=0. \label{eq:tofindK}
\end{align}
So we have three equations with three unknowns: $u_0(x,t),\;h_0(x,t),$ and $K(x,t)$. 
We solve for $K(x,t)$ by integrating Eq. (\ref{eq:tofindK}) with respect to $x$. We determine the constant of integration by integrating $u_1$ through the depth; no net flux along the film due to $u_1$ results in 
\begin{align}
    K(x,t)&=\frac{\gamma}{2}\frac{h_0^2h_{0xxx}}{3A_2(\theta_B)},\\
    u_1(x,z,t)&=\frac{p_{0x}}{F(\theta_B)}\left(\frac{z^2}{2}-\frac{h_0}{2}z\right)-\frac{ A(\theta_B,u_{0x})}{B(\theta_B)}z+\frac{\gamma}{2}\frac{h_0^2h_{0xxx}}{3A_2(\theta_B)}.
\end{align}
Then, to find the axial velocity $u_0$ and the sheet thickness $h_0$, we can solve the coupled system
\begin{align}
    h_{0t}+(u_0h_0)_x&=0,\\
    \left(h_0u_{0x}\right)_x+\hat{\gamma}(h_0^2h_{0xxx})_x&=0,
\end{align}
where
\begin{align}
    \hat{\gamma}=\frac{C_2(\theta_B)}{B_2(\theta_B)}\gamma.\label{eq:gamma}
\end{align}
In the paper we take $\hat{\gamma}=\gamma$.

\bibliography{refs}

\end{document}